\documentclass[%
 reprint,
superscriptaddress,
 amsmath,amssymb,
aps,
prd,
]{revtex4-2}

\usepackage{graphicx}
\usepackage{dcolumn}
\usepackage{bm}
\usepackage{xcolor} 
\usepackage{hyperref} 
\usepackage{relsize}
\hypersetup{
    colorlinks=true,                
    breaklinks=true,                
    urlcolor= black,                
    linkcolor= blue,                
    bookmarksopen=false,
    filecolor=black,
    citecolor=blue,
    linkbordercolor=red
}
\usepackage[T1]{fontenc}  
\usepackage[export]{adjustbox} 
\usepackage{ntheorem} 

\usepackage{booktabs}
\usepackage{pgfplots}
\usepackage{pgfplotstable}

\usepackage{tabularx}

\AtBeginDocument{\hypersetup{pdfborder={0 0 1}}}
\newcommand{\specialcell}[2][c]{%
  \begin{tabular}[#1]{@{}c@{}}#2\end{tabular}}

\DeclareGraphicsExtensions{.pdf,.png}

\theoremseparator{:} 

\newtheorem*{hyp*}{$H_\protect\hypnumber$} 

\newcommand{\hypnumber}{}

\begin{document}

\preprint{APS/123-QED}



\title{\textbf{Inferring Neutron-Star Properties from Post-merger Gravitational-wave Spectra with Neural Networks}}


\author{Dimitrios Pesios}
\email{dipesios@auth.gr}
\affiliation{Department of Physics, Aristotle University of Thessaloniki, GR-54124 Thessaloniki, Greece}




\author{Nikolaos Stergioulas}
\email{niksterg@auth.gr}
\affiliation{Department of Physics, Aristotle University of Thessaloniki, GR-54124 Thessaloniki, Greece}%


\date{\today}

\begin{abstract}

We present a proof-of-concept study of the inverse problem of inferring neutron-star properties directly from the post-merger gravitational-wave spectrum of equal-mass binary neutron star mergers. Using noise-free spectra  for equal-mass mergers obtained from numerical relativity simulation catalogs, we train and compare three artificial-neural-network regression models and two algebraic multivariate linear-regression baselines to predict three source properties: the stellar mass, $M$, and the quadrupolar tidal deformability, $\kappa_2^\tau$, and the local slope of the mass--radius relation, $dR/dM$.  Because the inverse mapping is nonlinear and cannot be obtained by analytically inverting the direct neural-network model, we construct dedicated inverse surrogates and train the neural-network models with a two-stage procedure in which residuals from an initial pass are used to define sample weights for a second pass, together with regularization through dropout, Gaussian-noise injection, and early stopping. We find that the neural-network models consistently outperform the linear baselines, demonstrating that the inverse relation between post-merger spectra and source properties is more effectively captured by nonlinear surrogates than by formal algebraic inversion. The best overall performance is achieved by an ensemble of single-task neural networks, while a multi-task network gives comparable accuracy for the prediction of the mass--radius slope, and a mixture-of-experts architecture provides useful indications about the relative importance of different spectral regions. We further show that the best model reproduces empirical relations between the dominant post-merger frequency and tidal deformability, and also recovers representative equation-of-state-dependent mass--tidal-deformability trends, indicating that the predicted parameters remain physically consistent beyond pointwise regression accuracy. Although the present analysis is restricted to a limited catalog of idealized, noise-free spectra, the results demonstrate that neural-network surrogates can provide a promising complementary route for extracting neutron-star information from post-merger signals with future third-generation detectors.
\end{abstract}

\maketitle

\section{INTRODUCTION}

Neutron stars (NSs) are among the densest objects in the Universe, with central densities of order $10^{15}\;{\mathrm{g/cm^{3}}}$ \cite{haensel_potekhin_yakovlev_2007} that place their interiors well beyond the reach of terrestrial experiments. Determining the equation of state (EOS) that governs matter at such densities is therefore one of the central goals of NS physics \cite{abac2025scienceeinsteintelescope,bauswein_exploring_2016,dietrich2021interpreting,abbott2019properties,abbott2018gw170817,iacovelli2023nuclear,friedman_astrophysical_2020,baiotti_binary_2017,baiotti_gravitational_2019,hild_s_et_al_sensitivity_2011,ascenzi_neutron-star_2024}, and gravitational-wave (GW) observations of binary neutron star (BNS) mergers have emerged as a uniquely powerful probe of it. While the inspiral phase constrains the EOS through the tidal deformability of cold, beta-equilibrated matter, the \textit{post-merger} signal carries complementary information about hot, dense matter in regimes that inspiral measurements cannot access \cite{bauswein_exploring_2016,dietrich2021interpreting,bauswein_equation_2020,haster_inference_2020}. Extracting this information is a central science target for third-generation (3G) ground-based detectors such as the Einstein Telescope and Cosmic Explorer \cite{abbott_observation_2016, luck2022third,Abbott_2017,reitze2019cosmic, zhang_gravitational_2022, evans_m_et_al_horizon_2021,borhanian_listening_2024,ackley_k_et_al_neutron_2020,branchesi_m_et_al_science_2023, takami_constraining_2014,10.1093_mnras_staf1147}, whose improved high-frequency sensitivity is expected to render the post-merger spectrum observable for the first time.

Connecting an observed post-merger spectrum to the physical properties of 
its source is, however, an \textit{inverse problem}: one is given the 
frequency-domain signal and asked to infer the parameters that produced it. 
Traditional approaches to this problem fall into two broad families. The 
first relies on empirical or quasi-universal relations between a small 
number of spectral features---most commonly the dominant post-merger 
frequency $f_\mathrm{peak}$---and bulk neutron-star properties such as 
compactness, radius, or the threshold mass for prompt collapse 
\cite{shibata_constraining_2005,takami_constraining_2014,
bauswein_equation_2020}. The second addresses inference through Bayesian 
parameter estimation against waveform templates or phenomenological 
spectral models \cite{haster_inference_2020,breschi_constraints_2022,
PhysRevD.102.043011,Wijngaarden_2022,clark_observing_2016,
iacovelli_nuclear_2023,vretinaris2025robustfastparameterestimation,
Banagiri_2020,jain_improving_2024,criswell_hierarchical_2023,
Huez_2026,mitra2025prospectconstrainingeosneutron,
panther2025carelesswhisperspopulationsubthreshold}; see 
\cite{christensen2022parameter,carson_future_2019,baiotti_gravitational_2022} 
for reviews and complementary perspectives.

\begin{figure}[ht!]
    \includegraphics[width=\linewidth]{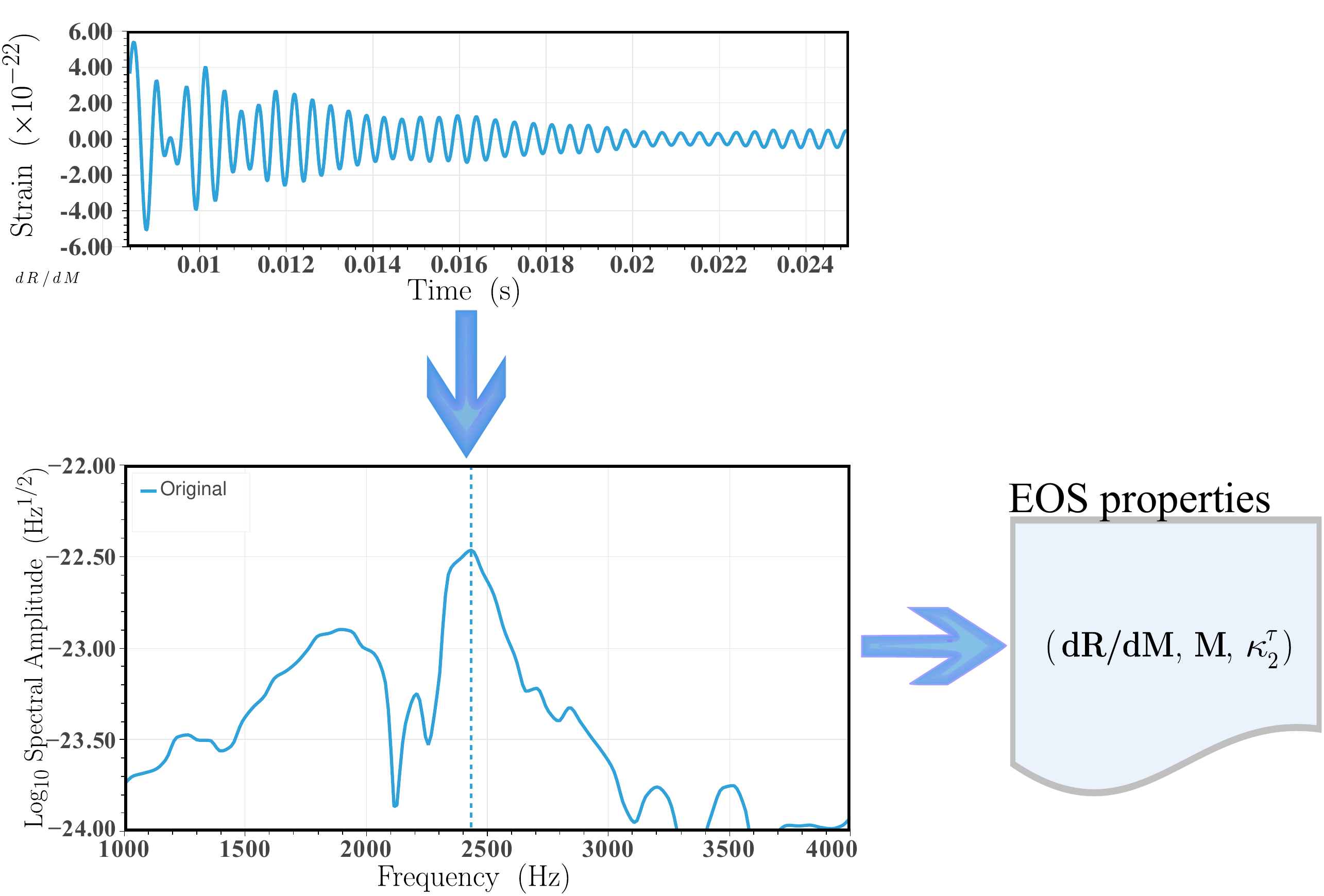}
        \caption{Schematic of our inverse-surrogate approach. The post-merger BNS time-series waveform is converted to its frequency-domain effective amplitude spectrum, which is then passed to a trained neural network that predicts the local mass--radius slope $dR/dM$, the component mass $M$, and the quadrupolar tidal deformability $\kappa_2^\tau$ of the (equal-mass) neutron stars. A BNS merger using the BHBlp EOS is shown for illustration.}
    \label{fig:first_image}
\end{figure}

Our aim here is complementary: we ask whether a non-Bayesian \textit{inverse surrogate}, trained directly on numerical-relativity spectra, can regress multiple continuous NS properties from the full post-merger spectrum without committing to a specific phenomenological template. We do not seek to replace Bayesian PE. Rather, the present study is intended as a pilot investigation: given the limited number of available numerical-relativity simulations and the still broad EOS uncertainty left open by current inspiral observations, we ask whether machine-learning-based inverse mappings already produce competitive and physically consistent predictions on the current training catalog. A positive answer would motivate a more ambitious implementation in the future, once larger simulation sets concentrated in a narrower EOS parameter space (constrained by more accurate inspiral measurements with 3G detectors) become available. Our general approach is depicted in Fig.~\ref{fig:first_image}.

The NS properties we target are the individual (equal-mass) component mass $M$, the quadrupolar tidal deformability $\kappa_2^\tau$, and the local slope of the mass--radius relation $dR/dM$ at the given $M$. Together these three quantities probe both a bulk source property ($M$) and two EOS-sensitive quantities ($\kappa_2^\tau$ and $dR/dM$) that encode the stiffness and local curvature of the $M$--$R$ relation in the relevant mass range. Typical NS masses lie between about $1.1$ and $2.2\,\mathrm{M_{\odot}}$, with radii estimated in the range $10$--$13\,\mathrm{km}$ \cite{branchesi_m_et_al_science_2023}. All three target properties can, in principle, be constrained by a combination of pre- and post-merger information \cite{wijngaarden_probing_2022, puecher_unraveling_2023}. Here, we restrict attention to post-merger spectra alone, so that the resulting surrogate can later be combined with inspiral-based inference as an independent cross-check. Since the inspiral primarily constrains the cold, low-to-intermediate density EOS, a discrepancy between inspiral and post-merger inferences would provide a handle on effects that appear only in the post-merger regime, such as phase transitions at high densities or possible beyond-GR phenomena like spontaneous scalarization. Beyond-GR scenarios \cite{miller2019new,arimoto_gravitational_2021,PhysRevD.8.3308,abbott2021tests,will2014confrontation,sathyaprakash2009physics,maggiore2000gravitational} lie outside the scope of the present study, but the inverse-surrogate framework we develop is agnostic to the underlying theory and could in principle be retrained on non-GR simulation catalogs.

Several works have proposed direct constraints on the EOS from post-merger signals. In \cite{huxford_accuracy_2024}, a thorough analysis is provided for pinpointing the NS radius with next-generation networks of ground-based GW detectors. In \cite{vretinaris2025robustfastparameterestimation}, the authors use an analytic waveform model and empirical relations to predict the prior ranges for post-merger frequencies for PE, based on the methodology in \cite{easter_detection_2020}. In \cite{Banagiri_2020}, a phase-agnostic likelihood model is developed for PE, demonstrating the ability to set upper limits on certain remnant properties. In \cite{jain_improving_2024}, the incorporation of post-merger remnant information is explored and how it can enhance the inference of NS properties. In \cite{criswell_hierarchical_2023}, a hierarchical Bayesian method is presented for refining the EOS using data from an ensemble of BNS post-merger remnants, irrespective of signal-to-noise ratios (SNRs). In \cite{Huez_2026}, the authors present a comprehensive Bayesian analysis of the full GW spectrum of BNS mergers, demonstrating that even a single detection near the post-merger signal-to-noise threshold can constrain the maximum mass and density of neutron stars to within $\sim 6\%$ and $\sim 10\%$ respectively. In \cite{mitra2025prospectconstrainingeosneutron}, the authors employ the setup of \cite{criswell_hierarchical_2023} to incorporate a broader set of EOS proxy parameters, demonstrating that upcoming A+ detector sensitivities could constrain neutron star radii to within $\sim 0.55\text{--}1$ km. Another work \cite{panther2025carelesswhisperspopulationsubthreshold} proposes a hierarchical Bayesian framework to constrain the EOS by stacking sub-threshold post-merger GW signals from a population of BNS mergers, demonstrating that aggregating approximately 50 non-detections at A+ sensitivities can resolve $f_\mathrm{peak}$ to within $\sim 50$ Hz, effectively recovering astrophysical information from signals otherwise lost due to noise.

Further works concentrate on the feasibility of restricting the EOS of NSs in future scenarios. In \cite{ecker_listening_2024}, it is demonstrated that the ``long ring-down'' of BNS mergers encodes information about the high-density EOS, and that measuring this signal in future detectors could improve constraints at extreme densities. In \cite{breschi_kilohertz_2019}, a time-domain model was developed for post-merger gravitational waves used to constrain the properties of ultra-dense matter in NS, with its respective follow-up in \cite{breschi_kilohertz_2024}. In \cite{tsokaros_masking_2024}, it is investigated how certain physical processes can mask the imprints of the equation of state in BNS merger signals, complicating their interpretation. In \cite{Soultanis_2022}, an analytic frequency-domain model is presented, which accounts for various oscillation modes of the remnant and its time evolution, to perform detection and inference of NS properties. In \cite{tringali_morphology-independent_2023}, a morphology-independent method is presented to determine the fate of the remnant, and consequently to gain insights on the EOS. In \cite{kochankovski2024hyperonsneutronstarmergers}, a study with hyperonic vs nucleonic EOS models is presented, along with their impact on the post-merger BNS signal, concluding that there is a shift in the dominant spectrum frequency $f_\mathrm{peak}$ upward by approximately 150 Hz, providing a potential observational signature of hyperons in dense, hot NS merger remnants, see also \cite{Blacker_2024}. The long-term evolution of post-merger oscillations was initially studied in \cite{2018PhRvL.120v1101D,2020PhRvD.101f4052D}, showing that convectively-driven inertial modes could contribute to the GW spectrum. Their detectability was assessed in \cite{2024PhRvD.109d3045S}, while \cite{guerra2026treatmentthermaleffectsequation} reinforced these results using tabulated finite-temperature EOS.

A substantial body of work has used machine learning for post-merger GW analysis, including classification of merger outcomes \cite{puecher_machine-learning_2024}, parametrization of post-merger signals \cite{whittaker_using_2022}, sparse-dictionary approaches \cite{llorensmonteagudo2025clawdiadictionarylearningframework}, K-nearest-neighbor time-domain modeling at fixed EOS \cite{soultanis2024gravitationalwavemodelneutronstar}, and an artificial neural network (ANN)-based \textit{forward} modeling of the post-merger spectrum from source properties \cite{Pesios2024}; see \cite{Cuoco_2025} for a recent review. CLAWDIA \cite{llorensmonteagudo2025clawdiadictionarylearningframework}, in particular, yields physically interpretable representations and targets data-sparse regimes, with classification primarily driven by $f_\mathrm{peak}$ and achieving encouraging results at modest SNR for next-generation detectors like ET and NEMO.

In Pesios et al. \cite{Pesios2024}, the direct precursor of the present work, a methodology for predicting the BNS post-merger spectra was presented, employing an ANN regression-based scheme and using as input a three-component vector of physical properties: the inverse gradient of radius versus mass $dR/dM$ at the specific mass $M$ in the mass--radius relation, the individual neutron star mass $M$ itself, and the induced tidal deformability $\kappa_2^\tau$. This work had as a starting point the work of Easter et al. \cite{PhysRevD.100.043005}, who used Multivariate Linear Regression (MLR) and the compactness $C$ instead of $dR/dM$. In \cite{Pesios2024}, the dataset was extended to 87 waveforms and it was demonstrated that even an imbalanced dataset can be used to train the model. This preceding work successfully utilized both ANNs and MLR to predict the BNS post-merger spectrum (output) given a three-component vector of NS properties (input), demonstrating superior performance of the ANN approach. In the remainder of this work, we will call this problem the \textit{direct problem}.

\begin{figure}[ht!]
    \centering
    \includegraphics[width=\columnwidth]{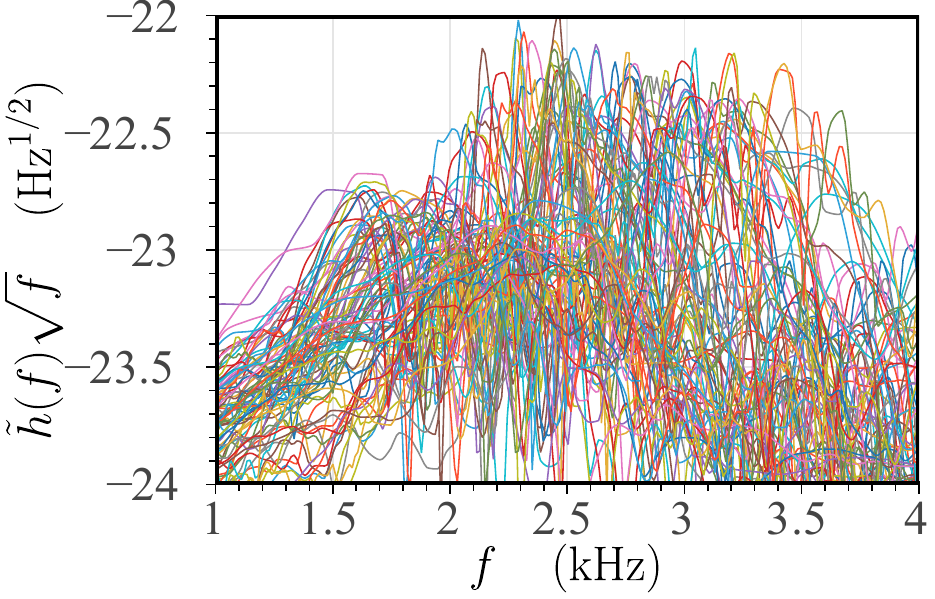}
    \caption{Collective plot of the effective amplitude spectra $\tilde{h}(f)\sqrt{f}$ of the 77 equal-mass BNS post-merger waveforms used to train and evaluate the inverse surrogates. All spectra are distance-normalized to 50~Mpc and are noise-free. In contrast to our previous work~\cite{Pesios2024}, the spectra are {\it not} aligned to a reference frequency.}
    \label{fig:misaligned_spectra_mesh}
\end{figure}

In the present work we address the corresponding \textit{inverse problem}: regressing $(dR/dM, M, \kappa_2^\tau)$  from the post-merger spectrum. The inverse mapping cannot be obtained by analytically inverting the trained forward ANN of \cite{Pesios2024}, because the nonlinear activations and dimensionality changes preclude a closed-form inverse; a dedicated training procedure is therefore required. We compare three ANN architectures: a multi-task network (MT-ANN), an ensemble of single-task networks (ST-ANNs), and a single-task mixture-of-experts (ST-MoE-ANN), against two algebraic baselines derived from multivariate linear regression, including a pseudo-inverse construction (PI-MLR) obtained by formally inverting the direct MLR of \cite{Pesios2024}. All neural models are trained with a two-stage residual-reweighting scheme, in which residuals from an initial pass are used to define sample weights for a second pass, together with dropout, Gaussian-noise injection, and early stopping to limit effective capacity on the 77-waveform training catalog. Beyond pointwise regression accuracy, we assess whether the predictions remain physically consistent by checking them against  quasi-universal $M_\mathrm{tot}f_\mathrm{peak}$--$\kappa_2^\tau$ relations \cite{vretinaris2020,chakravarti2020} and against representative EOS-specific $M$--$\kappa_2^\tau$ trends. Among the proposed architectures, the ensemble of single-task ANNs (ST-ANNs) yields the best overall results, both compared with the algebraic baselines and in reproducing the empirical fitting relations.

To our knowledge, this is the first study to regress multiple continuous neutron-star properties directly from the full post-merger gravitational-wave spectrum using dedicated inverse surrogates, rather than classifying the EOS or the remnant, or performing inference dominated by inspiral information. The present pilot study is deliberately restricted to idealized, noise-free, equal-mass spectra. A follow-up study will address detectability under realistic 3G noise curves.

This work is organized as follows. In Section~\ref{section_ii} we introduce the datasets, the algebraic inverse-regression baselines, and the ANN architectures, together with the weighted two-stage training procedure and loss functions, as well as deep learning concepts such as invertibility of ANNs and ensemble learning mixture-of-experts (MoE) models. In Section~\ref{section_iii} we present the results of the five models, the chi-squared analysis of the MoE gating, and the validation against empirical and EOS-dependent relations. Finally, in Section~\ref{section_iv} we make some final remarks and draw conclusions about their performance.

\begin{table*}
\caption{
Training set of 77 equal-mass BNS post-merger waveforms. Columns give, for each EOS, the number of simulations, the source catalog, and the component masses.  }
\label{tab:table1}
\begin{ruledtabular}
\resizebox{\textwidth}{0.15\textheight}{%
\begin{tabular}{cccc}
{\bf EOS}  & {\bf Waveforms} & {\bf References} & {\bf Component Masses} (in $M_\odot$)\\
\hline
ALF2 & 10 &Rezzolla \& Takami \cite{PhysRevD.93.124051} and CoRe v2 \cite{Gonzalez_2022mgo}& \specialcell{1.2, 1.225, 1.25, 1.275, 1.3, 1.325,\\ 1.35, 1.3505, 1.375, 1.3755}\\ \hline
APR4 & 7 & Rezzolla \& Takami \cite{PhysRevD.93.124051} & 1.2, 1.225, 1.25, 1.275, 1.3, 1.325, 1.35 \\ \hline
BHBlp & 4 & CoRe v2 \cite{Gonzalez_2022mgo} & 1.25, 1.3, 1.35, 1.4 \\ \hline
BLh & 4 & CoRe v2 \cite{Gonzalez_2022mgo} & 1.3, 1.3325, 1.364, 1.4 \\ \hline
DD2 & 7 & CoRe v2 \cite{Gonzalez_2022mgo} & 1.2, 1.25, 1.3, 1.35, 1.364, 1.4, 1.5 \\ \hline
ENG & 1 & CoRe v2 \cite{Gonzalez_2022mgo} & 1.3495 \\ \hline
GNH3 & 7 & Rezzolla \& Takami \cite{PhysRevD.93.124051} & 1.2, 1.225, 1.25, 1.275, 1.3, 1.325, 1.35 \\ \hline
H4 & 13 &Rezzolla \& Takami  \cite{PhysRevD.93.124051} and CoRe v2 \cite{Gonzalez_2022mgo}& \specialcell{1.2, 1.225, 1.25, 1.275, 1.3, 1.325, 1.3495\\ 1.35, 1.3505, 1.3715, 1.3725, 1.3735, 1.3795}  \\ \hline
LS220 & 4 & CoRe v2 \cite{Gonzalez_2022mgo} & 1.2, 1.35, 1.364, 1.4 \\ \hline
MPA1 & 8 & Soultanis, Bauswein \& Stergioulas \cite{Soultanis_2022} & 1.2, 1.25, 1.3, 1.35, 1.4, 1.45, 1.5, 1.55 \\ \hline
SFHo & 2 & CoRe v2 \cite{Gonzalez_2022mgo} & 1.35, 1.364 \\ \hline
SLy & 10 &Rezzolla \& Takami \cite{PhysRevD.93.124051} and CoRe v2 \cite{Gonzalez_2022mgo}& \specialcell{1.2, 1.225, 1.25, 1.275, 1.3, 1.325,\\ 1.35, 1.351, 1.3575, 1.364} \\ 
\end{tabular}%
}
\end{ruledtabular}
\end{table*}

\section{PREREQUISITES} \label{section_ii}

This section is dedicated to introducing the necessary prerequisites for the rest of the manuscript and the relevant nomenclature that we are going to use, as well as the available datasets.

\subsection{Dataset descriptions} \label{section_iii_a}

\begin{figure*}[ht!]
    
    \begin{minipage}{\textwidth}
        \begin{minipage}{0.45\textwidth}
          \includegraphics[width=\linewidth,left]{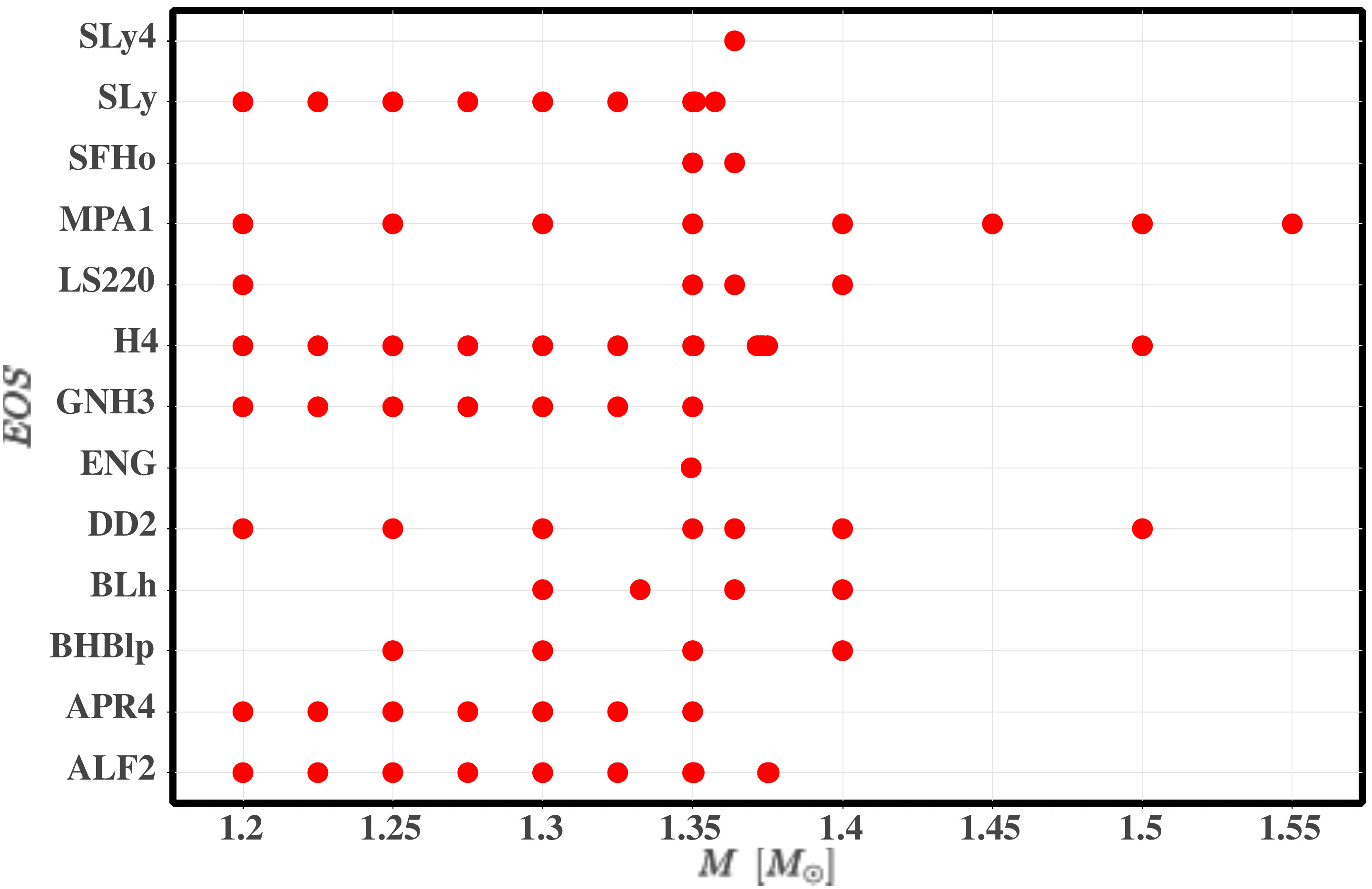}
        \end{minipage}
        \begin{minipage}{0.45\textwidth}
          \includegraphics[width=\linewidth,right]{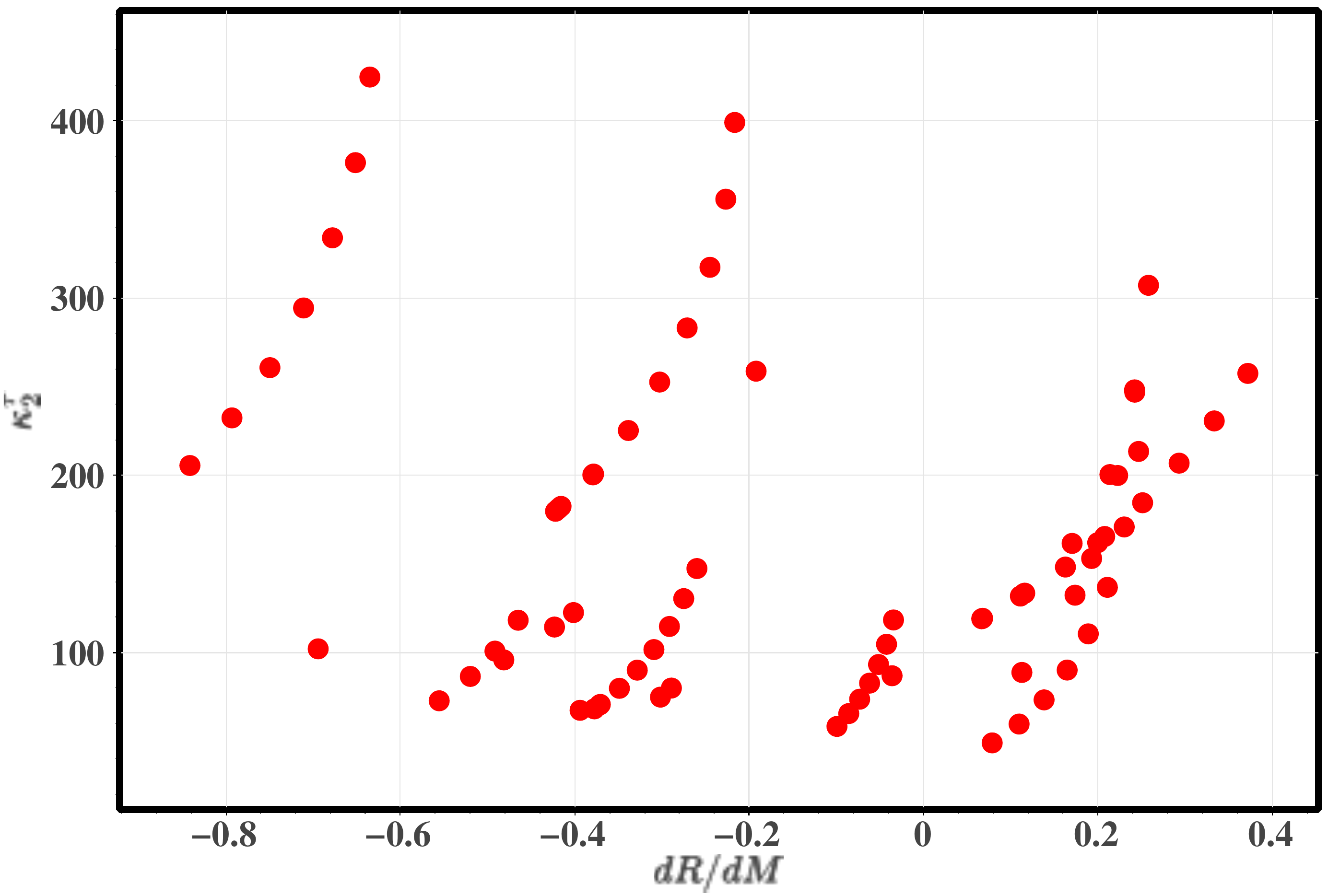}
        \end{minipage}%
    \end{minipage}
    
    \begin{minipage}{\textwidth}
        \begin{minipage}{0.45\textwidth}
          \includegraphics[width=\linewidth, left]{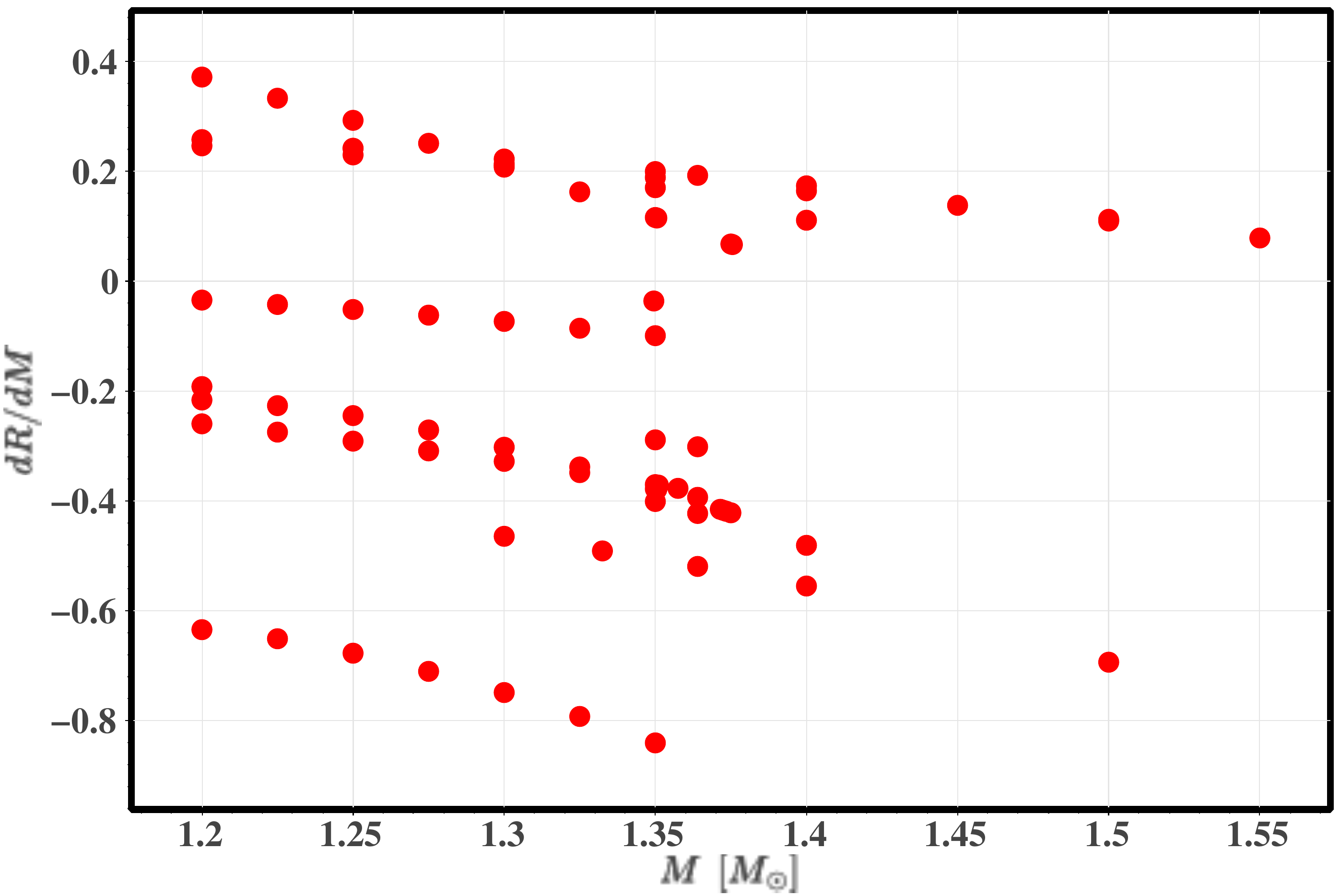}
        \end{minipage}
        \begin{minipage}{0.45\textwidth}
          \includegraphics[width=\linewidth, right]{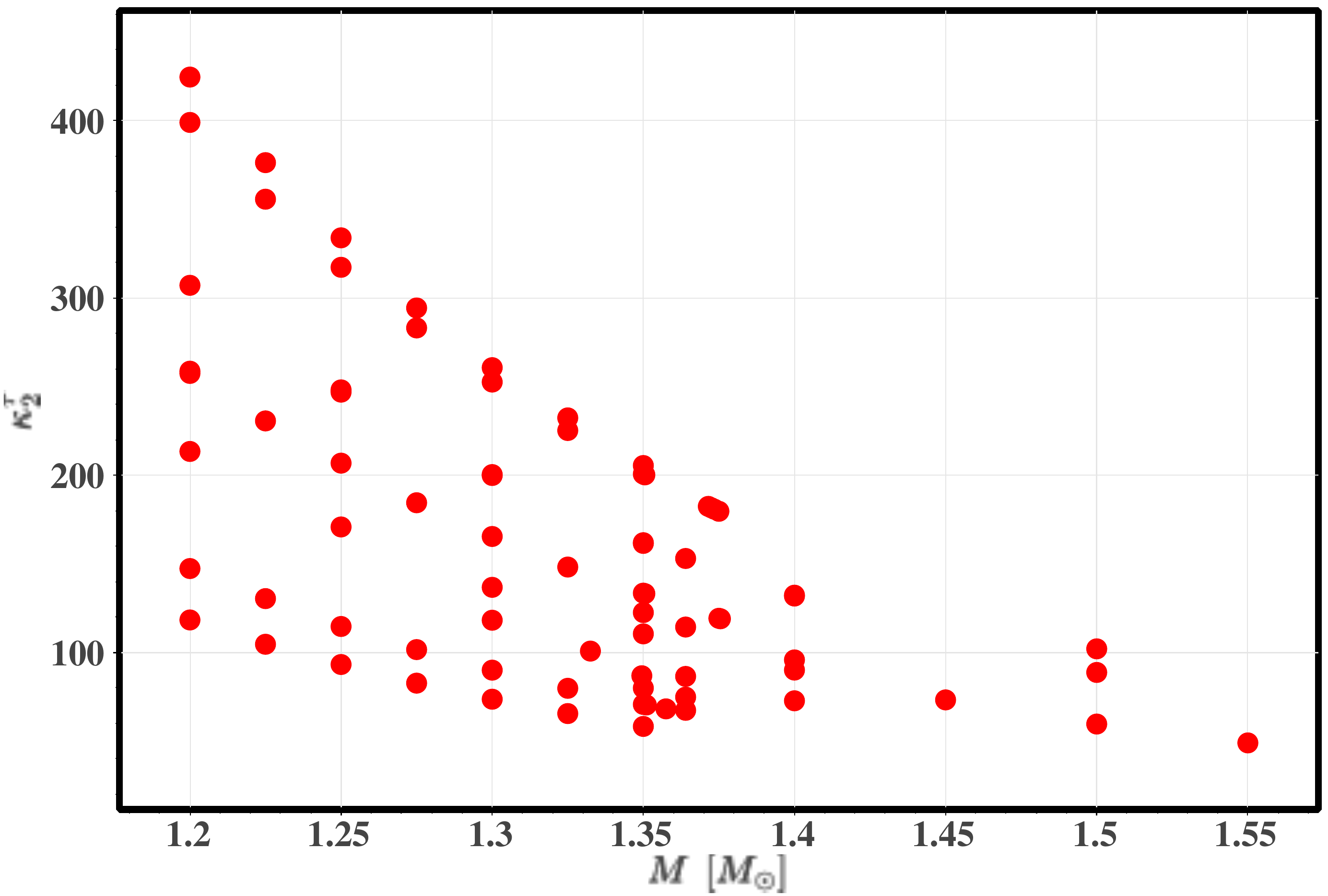}
        \end{minipage}
    \end{minipage}

    \caption{Distribution of the 77 training samples across the parameter space. \textit{Upper left}: component mass $M$ for each EOS, visualizing the dataset composition of Table~\ref{tab:table1}. \textit{Upper right}: tidal deformability $\kappa_2^\tau$ against the local mass--radius slope $dR/dM$. \textit{Lower left}: $dR/dM$ against $M$. \textit{Lower right}: $\kappa_2^\tau$ against $M$. Each point corresponds to a single post-merger spectrum. Three of the four panels display the target quantities that our inverse surrogates predict. Although the dataset is imbalanced across EOS families, it does span a physically meaningful range in each target variable.}
    
\label{fig:dataset_scatter_plots}.
\end{figure*}

The data set we employed in this work consisted of 77 Fourier domain representations (or spectra) of BNS post-merger signals in the astrophysical region of interest, which lies between $ [1-4] \; \mathrm{kHz} $. This dataset is a combination of simulated waveforms coming from CoRe v2 database \cite{Gonzalez_2022mgo}, and the Rezzola and Takami catalogue \cite{PhysRevD.93.124051}, while sevefal waveforms are from Soultanis et al. \cite{Soultanis_2022}, see Table \ref{tab:table1} and Fig. \ref{fig:dataset_scatter_plots} for a detailed description. All waveforms were normalized to a common distance of 50 Mpc and cut at their maximum amplitude to separate the inspiral and post-merger phases. In contrast with our previous work \cite{Pesios2024}, we feed every model with the original misaligned signals that have already been translated into a common frequency grid (meaning that they share the same frequency bins but their peak frequencies are at different locations) and they are shown in Figure \ref{fig:misaligned_spectra_mesh}. Compared to the previous set of 87 spectra used in \cite{Pesios2024}, we excluded 10 cases that are disfavored by the tidal deformability constraints of GW170817 \cite{abbott2018gw170817}, resulting in 77 waveforms.

It is evident that our dataset is imbalanced but still spanning a physically meaningful range, since an adequate number of values is covered in the respective vertical axes.

\subsection{Inverse linear regression}
\label{inverse_linear_regression}

Linear regression is one of the few supervised statistical and machine learning techniques that can be inverted, in that we can consider the output dataset as an input to a new model, the \textit{inverse} one, with a view to constructing the input dataset of the \textit{direct} model, that is, by exchanging regressors with regressands. The assumptions of attempting to do so, fall into the same ones of direct linear regression as far as multivariate analysis is concerned and they were briefly described in \cite{Pesios2024}. 

Thus, we can proceed by applying directly linear regression to our dataset or indirectly by using the coefficients of the model of our previous work manipulating the generalized inverse matrix, as we are going to see.

\subsubsection{Direct multivariate linear regression} 

Direct multivariate linear regression actually emanates from Ordinary Least-Squares (OLS) theory, reduced to multiple dimensions, and can be safely regarded as an optimization problem, where one attempts to minimize an objective function. This objective function corresponds to calculating the vertically projected distances of the $R^n$ points from a hyperplane, where $n$ is the dimensionality of the problem at hand.

In specific mathematical terms, multivariate linear regression is an extension of multiple linear regression allowing for more than one output variables, which may be correlated \cite{Izenman2008, foxweisberg}. The multivariate linear model is:
\begin{equation} \label{1_d}
    \underset{(n \times m)}{Y} = \underset{(n \times l+1)}{X} \underset{(l+1 \times m)}{B} + \underset{(n \times m)}{E} \; ,
\end{equation}
\noindent
where $Y$ is a matrix of $n$ observations on $m$ response variables (regressands), $X$ is a model (design) matrix with columns for $l + 1$ regressors, typically including a column of ones for the regression constant, $B$ is a matrix of regression coefficients, with one column for each response variable, and $E$ is a matrix of residuals, where the rows $E_i$, $i=1,\ldots,n$, have all mean 0 and the same error covariance matrix, and they are uncorrelated to each other.

\subsubsection{Pseudo-inverse multivariate linear regression (PI-MLR)}

The main equation in inverse linear regression is basically characterized by the same equation (as a specification) of direct linear regression, see Eq. \ref{1_d}. One can easily derive the specification of the inverse model equation using this equation of the direct model; however, we have to emphasize that the model itself, which coincides with the coefficient matrix $ B_\mathrm{inverted} \equiv \Gamma $, can be naturally derived from $ B_\mathrm{direct} \equiv B $ only if this matrix is invertible (square and nonsingular).

We denote the exchange as $ Y \longleftrightarrow{} X $, and our equation specification becomes: 
\begin{equation} \label{1}
    \underset{(n \times l+1)}{X} = \underset{(n \times m)}{Y} \underset{(m \times l+1)}{\Gamma} + \underset{(n \times l+1)}{H} \; ,
\end{equation}
\noindent
where $ H $ is the matrix of residuals of the inverse model, having the same properties with the corresponding $ E $ residual matrix of the direct model.

By performing the matrix algebraic manipulations in Eq. \ref{1_d}, and in the case of square coefficients matrix, we obtain:
\begin{equation} \label{2}
    \underset{(m \times m)}{\Gamma} = \underset{(m \times m)}{B^{-1}} \;\;\; \mathrm{and} \;\;\; \underset{(n \times m)}{H} = \underset{(n \times m)}{-E}\underset{(m \times m)}{B^{-1}} \; ,
\end{equation}
\noindent
since we considered $ m = l +1 $ for a square matrix and that the residuals of the direct model have been precalculated.



In general, the coefficients matrix $ B $ is not a square matrix. Nevertheless, inverse regression can still be formulated by employing the concept of the pseudo-inverse matrix, which is a generalization of matrix inversion for matrices with unequal dimensions. We usually denote this matrix using a plus sign (+) notation.

To accommodate the intercept (bias) term, the input vector is already augmented with a unity constant, effectively adding a row to the design matrix (which correspond to the intercepts of individual equations when we analyze the matrix equation into a system of equations) in multivariate regression, since a specific approach would differentiate the matrix equation of the inverse model, as it was previously stated.

To be more precise, we consider that the design matrix $ X $, which plays the role of the regressands matrix in inverse linear regression, of direct linear regression contains ones in a column-wise fashion and particularly in the first row. Thus, if $ A $ is an arbitrary matrix, its pseudo-inverse is defined as:
\begin{equation} \label{3}
    A^{+} = A'{(AA')^{-1}} \; ,
\end{equation}
\noindent
where $ A' $ is the transpose matrix of $ A $. Here the pseudo-inverse is used solely to obtain a formal algebraic inverse of the direct linear model. The PI-MLR model is included solely as an algebraic baseline obtained by formally inverting the direct MLR and is not intended as a statistically optimal inverse estimator. Using this notation, that is assuming the rows of matrix $ A $ are linearly independent, Eq. \ref{3} is valid. If the intercepts were represented row-wise, then $ A^{+} $ would be given by a slightly different, but equivalent, algebraic expression.

The existence and uniqueness of this matrix can be mathematically guaranteed, and hence, a formal algebraic inverse can be derived under the stated rank assumptions.

In this point, and by manipulating Eq. \ref{1_d}, it is obvious that the equation specification of the inverse linear regression model can be directly derived from the direct one, using the following matrix equation:
\begin{equation} \label{4}
    X = YB'(BB')^{-1} - EB'(BB')^{-1} \; ,
\end{equation}
Or, stated differently, taking into account Eq. \ref{1}, we obtain:
\begin{subequations}
    \begin{equation} \label{5a}
        \underset{(m \times l+1)}{\Gamma} = \underset{(m \times l+1)}{B^{+}} = B'(BB')^{-1} \; ,
    \end{equation}
    \begin{equation} \label{5b}
        \underset{(n \times l+1)}{H} = \underset{(n \times l+1)}{-EB^{+}} = -EB'(BB')^{-1} \; .
    \end{equation}
\end{subequations}

Thus, by applying the least squares theory to the direct linear regression problem and finding the estimator coefficients of $ B $, denoted by $ \hat{B} $, one can directly derive the inverse model from its direct counterpart, that is $ \hat{\Gamma} $, using the above set of equations.

In this work, the PI-MLR baseline is obtained by reusing the coefficient matrix $\hat{B}$ of the direct MLR model of \cite{Pesios2024} and computing its pseudo-inverse $\hat{\Gamma} = \hat{B}^{+}$ analytically, as described above. As a cross-check, we additionally fit a direct inverse MLR model on our dataset by applying OLS to Eq.~\ref{1} (i.e.\ by taking the exchange $Y \longleftrightarrow X$ relative to \cite{Pesios2024}). We refer to this simply as MLR and use it to validate the pseudo-inverse construction. One caveat should be kept in mind: $\hat{B}$ was originally fit on spectra that had been aligned to a common reference frequency using an empirical relation. This alignment cannot be reproduced at inference time in the inverse setting, since it would require prior knowledge of the neutron-star properties one is trying to predict. The practical consequences of this limitation for PI-MLR accuracy are discussed in Section~\ref{section_iii}.

\subsection{Regression methods using Artificial Neural Networks}
Artificial Neural Networks (ANNs) generally approximate the mapping between input and output variables that was given to them in the process of training \cite{HORNIK1989359,HORNIK1991251}. As universal approximators, these structures can derive complex mappings (mathematical functions) by adapting to a particular dataset and learn from data using a large set of training samples, so as to acquire the ability to generalize. The ability to generalize is considered to be the capacity of the model produced to accurately predict the properties of an unseen sample that was presented to them as input, as accurately as possible. 

If $ f_\mathrm{direct}: X \mapsto Y $  is the function approximated by the network, which in general is non-linear by nature, then in order to define the inverse model all we have to do is to invert the mapping between $ X $ and $ Y $, that is, by exchanging and retraining. So, our goal is to derive the function $ f_\mathrm{inverse}: Y \mapsto X $. 
This inverse mapping need not be unique — different parameter combinations can in principle produce similar post-merger spectra, so the inverse relation is one-to-many in general and learning a well-defined conditional inverse is nontrivial.

In many circumstances and depending on the application context, these structures are considered \textit{black boxes}. The outcome of this assumption is reflected on the complexity of the resulted function they represent. Consequently, it is intractable to mathematically derive the inverse model out of the direct one by manipulating equivalent algebraic relations. Unlike the linear regression case, the non-linear activations and dimensionality changes in deep neural networks preclude an analytic inversion of $f_\mathrm{direct}$ to obtain $f_\mathrm{inverse}$.


We employ the same or even different enhancements as in \cite{Pesios2024} to optimize the model production, such as Learning Rate Schedulers, Gaussian Noise, Dropout and input/output Scaling. Also, the logarithmic training representation of the input is a shared technique.

\subsection{Ensemble learning and input partitioning}

To improve and investigate predictive robustness, we employ ensemble learning techniques which comply well with the divide-and-conquer paradigm, in that individual models specialize in different parts of the input space. In other words, each expert learns to handle a specific subset of input data and specializes in it. This expertise of each expert in their own dataset, which is generally a subset of the overall dataset according to the application context, is called a \textit{subtask} in feature-partitioned MoEs terminology. Here, a subtask is a spectral partition.

However, MoE models exhibit serious interactions with respect to the bias-variance trade-off compared to simpler models. Since they are a combination of multiple models such as experts and gates, they become more sensitive to data splits leading to the so-called gating instabilities and over-parameterization issues \cite{mu2025, royer2023}. 

As far as this work is concerned, this is particularly significant when we use $k$-fold CV to estimate the performance of the model. When $k$ is small the model is trained on less data and the gating function may not fully specialize; when $k$ is large, such as in LOO-CV,  we have a low bias and very high variance and the gate behaves differently fold-to-fold taking unstable decisions. So, in order to gain the most of these kind of models, we have to select a moderate value of $k$ to keep a reasonable balance between folds, to reduce variance and obtain a reliable estimate of the performance. Consequently, the value of $k=4$ was chosen for our ST-MoE-ANN model, in contrast with the other models where LOO-CV was used.

\begin{figure}[ht!]
    \includegraphics[width=\linewidth]{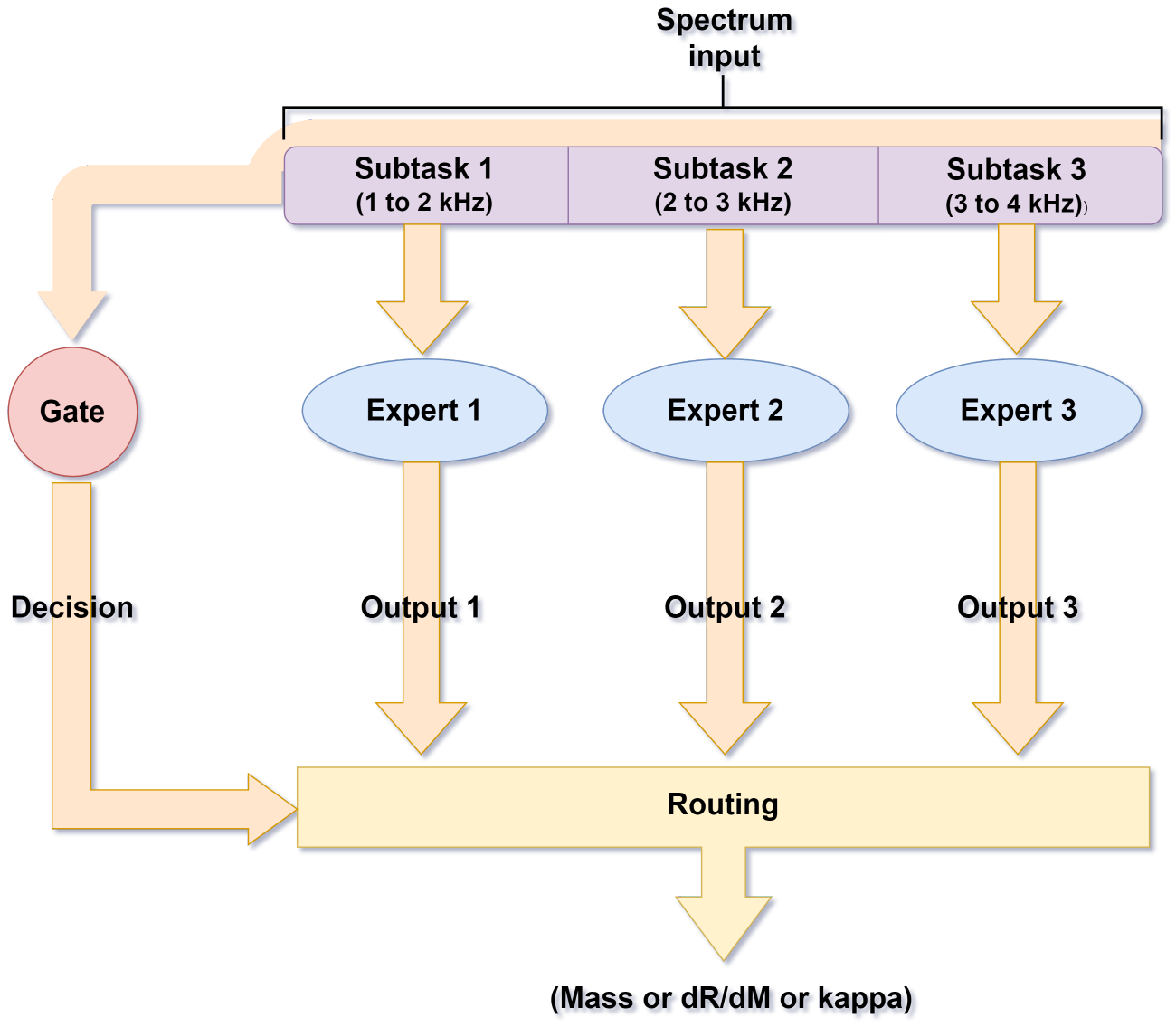}
    \caption{A schematic description of a mixture of experts model (MoE) with three expert models and a gate model. The input is divided into three spectral partitions (subtasks), so each expert is assigned 1/3 of input to gain specialization in it. The gate model, which is trained jointly with the expert models and takes the full spectrum as input, is responsible for selecting, via top-1 routing, the expert whose output is used by the overall MoE model.
    The architecture is instantiated three times, one for each NS property we want to predict.}
    \label{fig:MoE_drawing}
\end{figure}

In Figure \ref{fig:MoE_drawing} a typical architecture of an MoE model is presented, which comprises three expert models and one gate model. 

Each expert processes a different spectral partition. The gate network takes the full spectrum as input, produces three Softmax scores, and the expert with the highest score is selected (top-1 hard routing). Only the selected expert’s output is forwarded to the model output.

In this work in Section \ref{investigation_moe}, we are going to divide each input spectrum into three different equal-sized parts for three MoE models (as the one depicted in Figure \ref{fig:MoE_drawing}) that predict a single physical property of the triplet $ [ dR/dM, M, \kappa_2^\tau ] $ which describes a neutron star component of the binary system. Each part will have the same number of frequency bins and feed each expert of MoE models with its corresponding input. This way, in each MoE model the first expert will specialize in the low frequencies, the second one in the middle frequencies, and the third one in the high frequencies regime of frequency range of astrophysical interest, that is, $[1-4] \; \mathrm{kHz}$.

Another reason this approach might become useful in the future is that when 3rd generation detectors are constructed and operate, the visibility (related to 
the Signal-to-Noise Ratio of the signal) of the entire BNS post-merger signal will not be apparent at once. The SNR of a potentially detected post-merger signals are fundamentally and quantitatively linked to the detector's sensitivity curve through the matched-filter SNR integral  \cite{Thorne1987}. Firstly, when the main frequency peak will cease to be buried in noise, and as the sensitivity is further improved then, secondly, the other characteristics of the post-merger spectrum will become visible, as the lower frequency secondary peaks which lie one order of magnitude lower in amplitude. It would be useful to qualitatively demonstrate to which level improvements should happen with respect to this kind of models, since it can learn to prioritize the robust spectral regions.

\begin{figure}[ht!]
    \includegraphics[width=\linewidth]{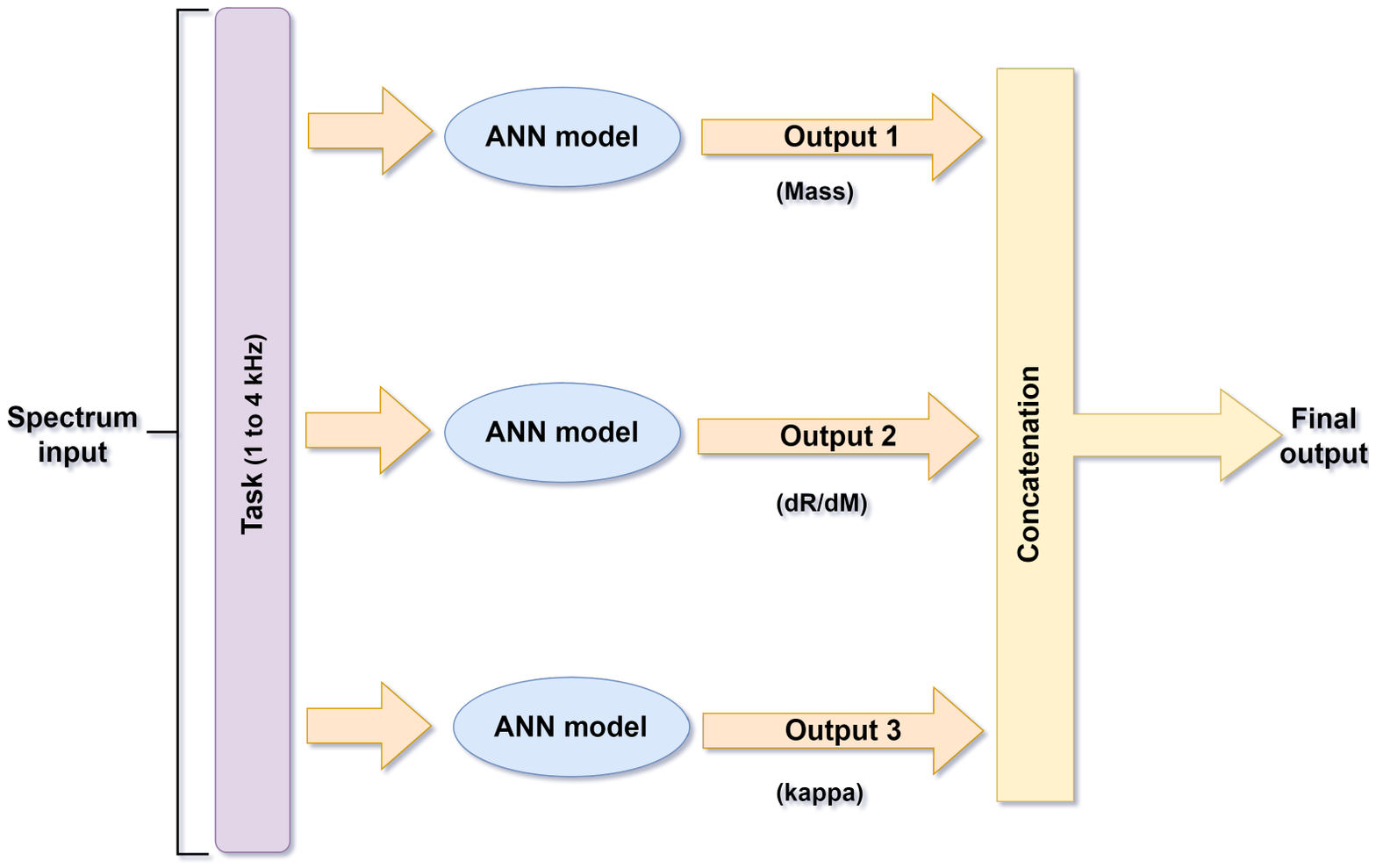}
    \caption{Schematic illustration of our overall Single-Task ANN model, consisting of an ensemble of three individual ANN models, each trained to predict one neutron-star property. Each model takes the full input spectrum and outputs a single value. The three outputs are then concatenated to form the final three-component prediction vector.}
    \label{fig:ST_ANN_drawing}
\end{figure}

On the contrary, in Figure \ref{fig:ST_ANN_drawing} we present a different single-task model-approach (ST-ANN) where the whole spectrum is fed into a single ANN without dividing it, and combine (concatenate) the result values into a single vector output. Although this approach is not formally considered an ensemble in strict terms, it can still be utilized to draw conclusions comparing it with the former.

Here, a relevant investigation will be conducted for identifying which part of spectrum most (if any) affects a specified neutron star property out of three (while the choice might not be unique and more than three in number) using a chi-squared statistical test.


\subsection{Weighted regression}
\label{weights_procedure}
Moreover, the aforementioned regression-oriented approaches, specifically for the deep learning ones, can be further enhanced through the incorporation of weighting schemes during model design, thereby improving their predictive capability.

Typically, a two-stage weighting procedure is employed to emphasize accurately predicted samples while penalizing poorly predicted ones. In the first stage, residuals for all samples are computed by performing an initial inference pass of the model. In the second stage, these residuals are inverted to derive the weights used in the subsequent training phase.

Badly predicted samples are penalized in the sense that their contribution to the loss is reduced via smaller residual-based weights, while well-predicted samples are correspondingly favored. This reflects a deliberate design choice and a trade-off between robustness and emphasis on regions of the training distribution that the model can represent reliably.

This approach enables the model to assign higher weights to samples with higher predictive accuracy, thereby yielding an overall improvement in its general predictive performance.
Concretely, the second-stage sample weights are defined as inverse residual-based weights,
\begin{equation}
w_i \propto \frac{1}{|e_i| + \epsilon},
\end{equation}
where $e_i$
 denotes the first-stage residual for sample 
i, and $\epsilon$ is a small positive constant introduced for numerical stability. In this way, samples with smaller first-stage errors receive larger weights and therefore contribute more strongly to the loss minimized during the second training stage.

In terms of the current work, we performed a range search for taking into account sample weights in the second stage, after normalizing them into a specified interval. We investigated normalized ranges $r$ of the form:
\begin{equation}
    r = \left(1, w_\mathrm{max}\right), \;\; \mathrm{where} \;\; w_\mathrm{max} \in \left( 2,4,8,16,32 \right)
\end{equation}

This way, a $w_\mathrm{max}$ value indicates how many more times a specific sample $i$ will be included in the training procedure, while our dataset of 77 observations/samples is traversed, compared with samples that are weighted close to one.

For MT-ANN model, where a three-component vector is predicted, the weights are calculated as averages of individual property first stage values.

The proposed two-stage weighting procedure is applied entirely within each leave-one-out cross-validation (LOO-CV) fold. For a given fold, the held-out sample is excluded from both stages of training. In the first stage, the model is trained solely on the remaining $N$-1 samples and residuals are computed exclusively for these training samples. The resulting residuals are then used to define sample weights, which are applied in the second-stage training on the same $N$-1 samples. The held-out sample is used only for final evaluation and does not contribute to the computation of residuals, weights, or model parameters at any point. All preprocessing steps follow the same fold-wise protocol.

\subsection{Root Mean Square Error index}


While we used Fitting Factors histograms in our previous work \cite{Pesios2024}, it is not relevant to use them again here. Instead, we are going to use a simpler metric to demonstrate the accuracy of our inverse regression methodology describing the prediction capacity of a single model, which is commonly known as Root Mean Square Error (RMSE). RMSE's have the advantage that when they are calculated on a specific variable, the result has the same dimensions (units) as that variable and therefore represents a typical error magnitude.

Here, given a regressand variable $ y $ and a finished $k$-fold cross-validation (CV) evaluation procedure, the overall RMSE of $ y \in \left( dR/dM, M, \kappa_2^\tau \right)$ for a specific model is indicated by a calculation of the predicted values $ y_\mathrm{predicted} $ against its original true values $ y_\mathrm{original} $, using the following formula:
\begin{equation}
    \mathrm{RMSE}(y) = \sqrt{\mathlarger{\sum}_{i=1}^{N}{\frac{( y^{(i)}_\mathrm{predicted} - y^{(i)}_\mathrm{original})^2}{N}}} \; ,
\end{equation}
\noindent
where the summation index $ i $ traverses all the observations and, consequently, $ N = 77 $ in our case, as we will describe in Subsection \ref{section_iii_a}.  The results for our case are depicted in Table \ref{tab:rmses}.

\subsection{Optimal Huber loss}
\label{optimal_delta}
Another methodological selection we made regarding the implementation of two of our ANN-based models (MT-ANN and ST-ANN) was introducing a different loss function than the ones we used in \cite{Pesios2024}. The reason behind this choice, as we see from the scatter plots Figures \ref{fig:scatter_plots_dl_mlr_drdm} to \ref{fig:scatter_plots_dl_kappa}, is that this loss function demonstrates limited sensitivity to outliers.  However, the choice of the transition parameter $\delta$ must be made carefully: if set too small, genuine signal is treated as outlier and down-weighted into a linear regime, potentially preventing the model from fitting the bulk of the data tightly.

Huber loss is actually a robust loss function that behaves like MSE for small errors and like MAE for large errors and it is strongly related to smoothing the L1 metric while training.

Considering $\delta$ as the transition value between quadratic and linear behavior, the loss obtained for sample $i$ is:
\begin{equation}
    \mathrm{Loss}_i= \begin{cases}
    \frac{1}{2}\left(E_i\right)^2, & \text { if }\left|E_i\right| \leq \delta, \\[1ex]
    \delta\left(\left|E_i\right|-\frac{1}{2} \delta\right), & \text{otherwise}
    \end{cases}
\end{equation}

\noindent
where $ E_i = V^{(i)}_\mathrm{predicted} - V^{(i)}_\mathrm{original} $ is the difference between the original value and the predicted one, that is, the residual $E_i$. 

Usually, the choice of the $\delta$ parameter is empirical and according to the application context. However, there is a rule-of-thumb that works for the most cases using Median Absolute Deviation (MAD), as a starting point. In this work, we calculate the $\delta$ value using the following formula:
\begin{equation}
    \delta = 1.5 \times \mathrm{factor} \times \mathrm{MAD}
\end{equation}
\noindent
where 
\begin{equation}
    \mathrm{MAD} = \mathrm{median}_i| V_\mathrm{original}^{(i)} - \mathrm{median}_j\left( V_\mathrm{original}^{(j)} \right)|
\end{equation}
\noindent
and investigate using the following $\mathrm{factor}$ values,
\begin{equation}
    \mathrm{factor} \in \left(1,2,3,4,5,6\right),
\end{equation}
\noindent
while $i$ traverses the set of our $77$ sample values, and $ V \in \left( dR/dM, M, \kappa_2^\tau \right)$. Here we use the MAD of the target variable as a convenient proxy  and treat the multiplicative factor as a tunable hyperparameter, searched over (1, 2, 3, 4, 5, 6) and selected by validation performance. This is a heuristic choice. Its justification is empirical rather than theoretical, and the factor sweep is what compensates for the mismatch between target scale and residual scale.

For the sake of our MT-ANN model $V$ is an average of these original values since we have a three-component vector in its output. This way we enforce more aggressive outlier suppression.

Huber loss can also be included in the training scheme as a weighted loss function following the prescription we earlier stated. In each stage, we compute the residuals by subtracting the predicted values with the original ones, and invert them to be included in the next stage. We stop this iterative procedure when our validation metric (Huber on a held-out set) barely improves after stage $j$, so we do not iterate further. This resembles the Iteratively Reweighted Least Squares (IRLS) statistical principle \cite{Green1984, Rusiecki2012}, but with a Huber loss function in a neural network-based approach.

\begin{table*}[ht!]
\centering
\begin{tabularx}{\textwidth}{X c c c}
\textbf{Training parameters} & \textbf{MT-ANN} & \textbf{ST-ANN} & \textbf{ST-MoE-ANN} \\
\hline
NN model used & Single Feed-forward & Multiple Feed-Forward & Multiple Mixture-of-Experts \\
Number of input nodes & 370 & 370 each & Experts: 123, 123, and 124; gate: 370 \\
Number of hidden layers & 3 & 3 & Experts/gate: 4 \\
Number of output nodes & 3 & 1 & 1 \\
Loss function used & Huber loss & Huber loss & MAE/MSE/MSE \\
Training batch size & 6 & 6 & 6 \\
Early-stopping epoch & 30 & 40 & 120 \\
LR scheduling scheme & ReduceLROnPlateau & ReduceLROnPlateau in each & Experts/gate: WarmUpDecay  \\
I/O scaling & SD/MaxAbs & SD/SD (except for dR/dM output) & SD/SD (except for dR/dM output) \\
k in k-fold CV evaluation & 77 & 77 & 4 \\
\end{tabularx}
\caption{Training parameters of the MT-ANN, ST-ANN, and ST-MoE-ANN models.}
\label{tab:training_params}
\end{table*}

\section{APPLICATION OF DIFFERENT REGRESSION METHODS FOR BNS POST-MERGER PARAMETER PREDICTION} \label{section_iii}

Here, in this section, we are going to present the experiments carried out using the various neural network models we mentioned above, starting from the simplest to the most complicated one, namely 1) the Multi-Task ANN model (\textit{MT-ANN}), 2) the Single-Task ANN models (\textit{ST-ANNs}), 3) the Single-Task Mixture-of-Experts models (\textit{ST-MoE-ANN}), as well as 4) the Pseudo-Inverse Multivariate Linear Regression model (\textit{PI-MLR}), which essentially emanates from its direct counterpart, and 5) the direct inverse Multivariate Linear Regression model (\textit{MLR}) which is basically used for comparison with its pseudo-inverse equivalent. 

\begin{figure*}[ht!]
    \begin{minipage}{\textwidth}
        \textbf{First stage}%
    \end{minipage}
    \begin{minipage}{\textwidth}
        \begin{minipage}{0.33\textwidth}
          \includegraphics[width=\linewidth,left]{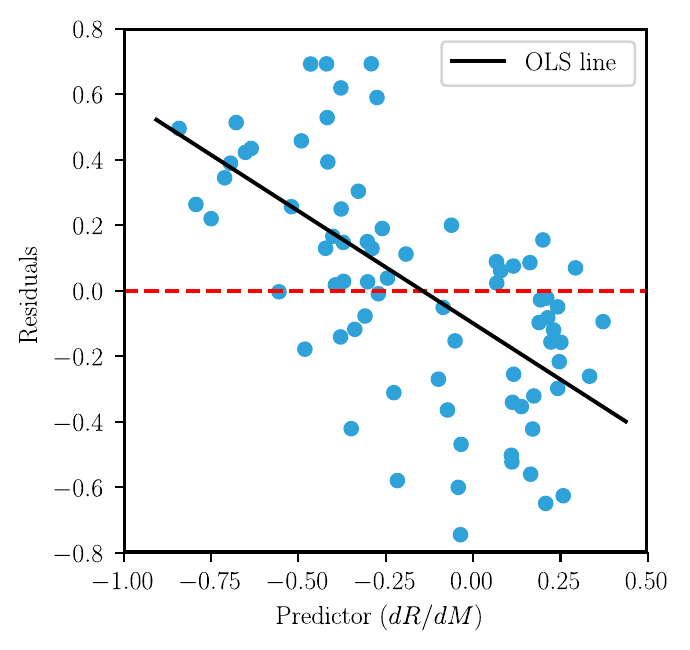}
        \end{minipage}
        \begin{minipage}{0.33\textwidth}
          \includegraphics[width=\linewidth,right]{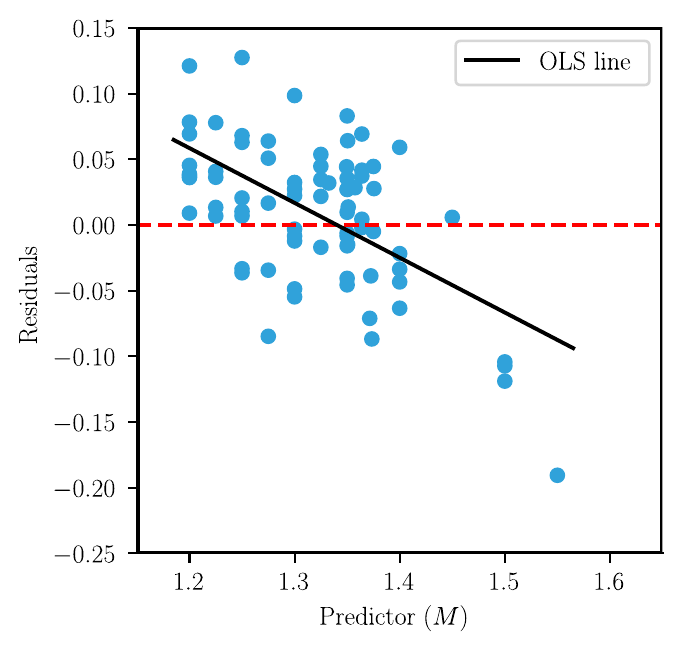}
        \end{minipage}%
        \begin{minipage}{0.33\textwidth}
          \includegraphics[width=\linewidth,right]{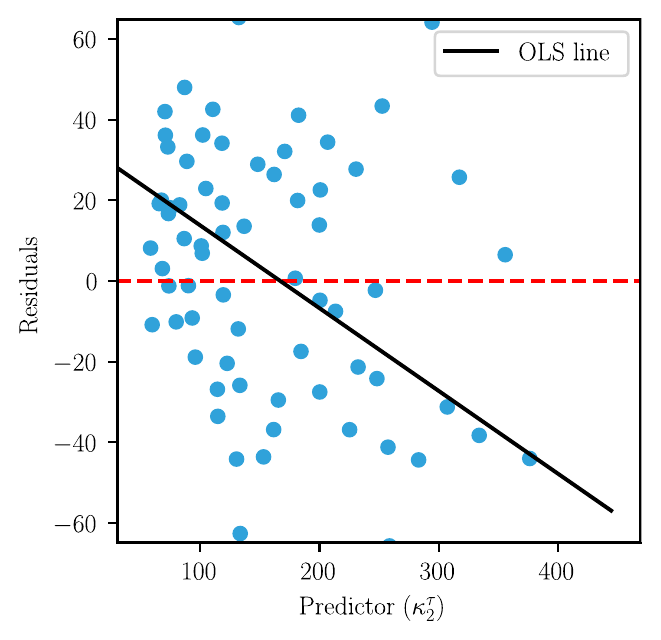}
        \end{minipage}%
    \end{minipage}
    \begin{minipage}{\textwidth}
        \textbf{Second stage}%
    \end{minipage}
    \begin{minipage}{\textwidth}
        \begin{minipage}{0.33\textwidth}
          \includegraphics[width=\linewidth,left]{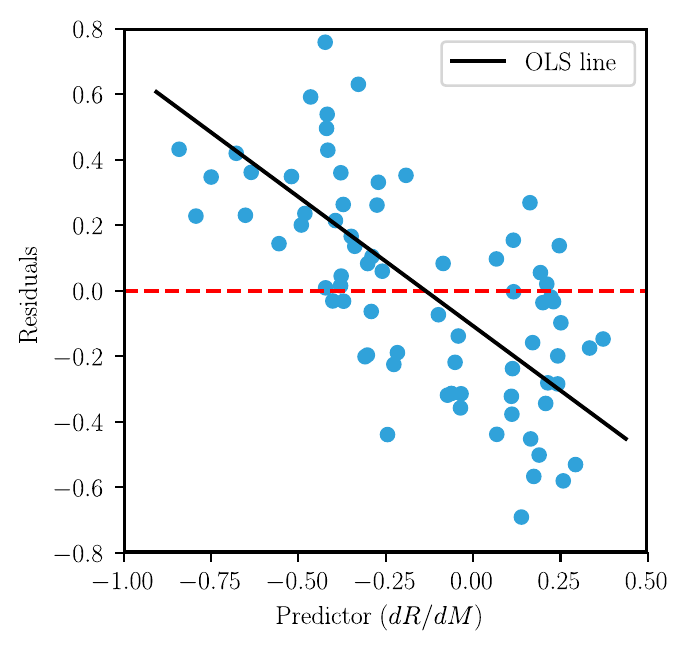}
        \end{minipage}
        \begin{minipage}{0.33\textwidth}
          \includegraphics[width=\linewidth,right]{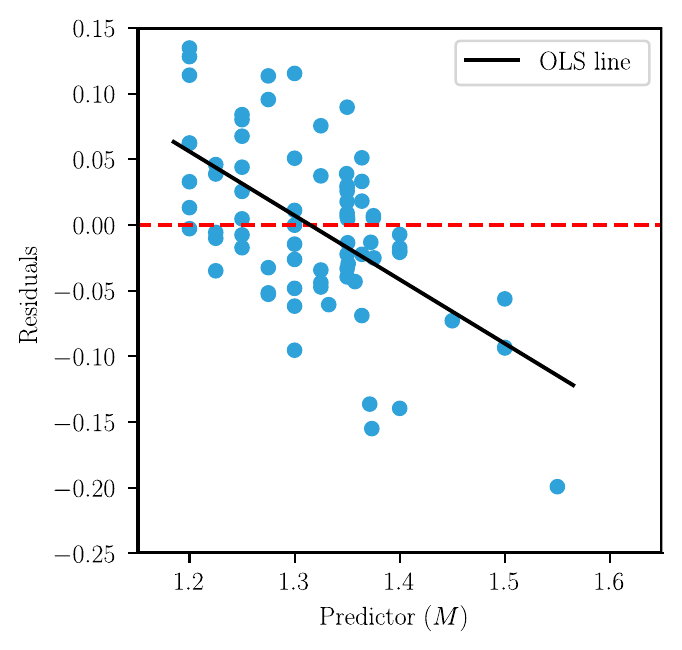}
        \end{minipage}%
        \begin{minipage}{0.33\textwidth}
          \includegraphics[width=\linewidth,right]{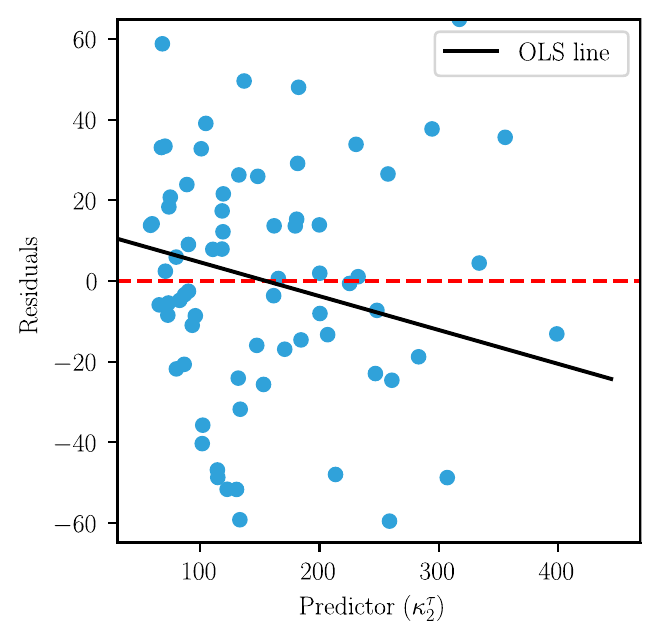}
        \end{minipage}%
    \end{minipage}
    
\caption{Scatter plots of the first-stage (upper panels) and second-stage (lower panels) residuals versus the neutron-star property predicted by the ST-ANN models. The black dashed line denotes an OLS trend line shown for visual guidance only. In the second stage, sample weights are assigned according to \(w_i \propto 1/(|e_i|+\epsilon)\), where \(e_i\) is the first-stage residual, so that samples with smaller residuals contribute more strongly to the weighted loss.}
\label{fig:residual_scatter_plots}.
\end{figure*}

All deep-learning models were trained in two stages, firstly determining the weights and, secondly, using them as input sample weights in the second stage to create stronger learners. Some typical residual vs. predictor scatter plots are depicted in Figure \ref{fig:residual_scatter_plots} for our best model for all three NS properties, demonstrating that residual trend information, with respect to property value, can be exploited to draw better results. 

In the second stage, we early stopped the training procedure when a number of epochs was reached. The stopping epoch was selected empirically from learning-curve diagnostics, identifying the range beyond which validation performance consistently deteriorated while training loss continued to decrease.
We also attempted to apply a weight procedure for our two MLR models, but the results were not satisfactory since the models would not improve.


\subsection{The models}

Our set of implemented models consists of three deep learning models (using artificial neural networks) that may comprise one or more combined neural networks, and two algebraic models based on multivariate linear regression constructed either indirectly (pseudo-inverse case) or directly (direct case).

We employed several enhancement techniques in our deep learning models, as in our previous work \cite{Pesios2024}, that yield better prediction results as we are going to see. We also had to increase the number of training parameters (both the number of layers and the number of nodes in each layer) in each network since the input now is much larger than the input of our previous work (370 instead of 3), and more correlations have to be determined by the various layers. An overview of the training parameters is presented in Table \ref{tab:training_params}. Moreover, all three DL models were trained using weighted regression.

\subsubsection{Multi-task ANN model}

The first deep learning model is an implemented ANN used for solving the problem of predicting multiple values using a regression-like scheme with 3 hidden layers. This type of set-up is typical of addressing the problem of prediction using neural networks when data can be represented in tabular form.

We feed it with 77 original spectra, each containing 370 values as much as the number of frequency bins, and predict a vector consisting of three values, that is BNS component mass $M$, the gradient of the individual NS radius versus the mass $dR/dM$, and induced tidal deformability $\kappa_2^\tau$.

We enhanced the training of this network using a \textit{Standard Scaler} in the input, by subtracting the average value of all samples and dividing with their standard deviation, an approach similar to that of our first paper. However, before doing that, a base-10 logarithmic representation of the input was calculated so that the network's weight adjustment can be more seamlessly achieved while training. On the other hand, we used a different scaler for the 3-value vector output, translating each component value so that the maximum absolute value in each individual component set equals 1. This scaler is widely known as \textit{MaxAbsScaler}, and was introduced as it empirically yielded faster and more accurate convergence, likely due to the specific distribution of the variable.

Additionally, during training, a Reduced Learning Rate (RLR) Scheduler was employed, starting from a learning rate value equal to 0.001 and gradually decreasing it to two orders of magnitude when there was no significant decrease in training validation loss, which was set as a mean squared value. The batch size was tuned to 6 samples, and the training procedure was completed using early stopping as mention earlier. We also employed a Huber loss function to overcome and eliminate to some extent outliers and achieve smoother training and validation loss curves. After performing the investigations we analyzed in Sections \ref{weights_procedure} and \ref{optimal_delta}, we took into consideration $w_\mathrm{max}=2$ for averaged weights, and $\mathrm{factor}=2$ for $delta$ optimal value in Huber loss function.

\begin{table}[th!]

\caption{Architecture of the multi-task ANN model, as reported by the \texttt{summary()} function of the TensorFlow library \cite{tensorflow2015-whitepaper}. The number of trainable parameters in each dense layer can be inferred from the numbers of nodes in the previous and current layers, together with the corresponding bias terms.}

\label{tab:simple_ann}%
\begin{ruledtabular}
\resizebox{\linewidth}{!}{%
\begin{tabular}{cccc}
Layer&Type&Shape&Activation\\
\hline
\#1&GaussianNoise(0.001)&(None, 370)&-\\
\#2&\textbf{Dense}&(None, 600)&Linear\\
\#3&GaussianNoise(0.05)&(None, 600)&-\\
\#4&Dropout(0.15)&(None, 600)&-\\
\#5&\textbf{Dense}&(None, 800)&Sigmoid\\
\#6&GaussianNoise(0.05)&(None, 800)&-\\
\#7&Dropout(0.15)&(None, 800)&-\\
\#8&\textbf{Dense}&(None, 800)&Sigmoid\\
\#9&GaussianNoise(0.05)&(None, 800)&-\\
\#10&Dropout(0.15)&(None, 800)&-\\
\#11&\textbf{Dense}&(None, 800)&Sigmoid\\
\#12&GaussianNoise(0.05)&(None, 800)&-\\
\#13&Dropout(0.15)&(None, 800)&-\\
\#14&\textbf{Dense}&(None, 3)&Linear\\
\end{tabular}
}
\end{ruledtabular}
\end{table}

To improve the generalization ability of the overall network, we introduced Gaussian noise in the input of each layer as well as dropping out nodes to some extent. The various parameters of this approach and the  architecture of the network is depicted in Table \ref{tab:simple_ann}. 

Evaluation was performed using Leave-One-Out Cross Validation (LOO CV) with the number of folds being equal to the cardinality of our dataset (77), as for the other models (except for the ST-MoE-ANN ones), a procedure that was briefly described in our previous work \cite{Pesios2024} too.

\subsubsection{Single-task ANN models}

In order to be able to assess the prediction capacity of the previous \textit{MT-ANN} network, we broke down the problem of predicting the output into three separate neural networks, where each one predicts a single component of the regressand vector while having the same regressors as input.

So, we constructed three ANNs that can be used as an ensemble, similar to the one described in Table \ref{tab:simple_ann}, except for the fact that the last output layer has (None, 1) shape, that is predicting a single value; either BNS component mass $M$, or inverse gradient of NS radius vs its mass $dR/dM$ or, lastly, tidal deformability $\kappa_2^\tau$. 


We used the same enhancements as in the \textit{MT-ANN} case, but we were slightly differentiated in input/output scaling of each network. To be more precise, the ANNs predicting mass $M$ and tidal deformability $\kappa_2^\tau$ were implemented using standard scaling (SD) both in input and output, a process that was reversed while predicting with the formulated model, while for inverse gradient of radius versus mass $dR/dM$ a standard scaler was used for input and a MaxAbsScaler for output. We emphasize again that the latter is justified by the fact that $dR/dM$ comprises sparse data and zero-centering would destroy sparsity.

Lastly, the investigations we analyzed in Sections \ref{weights_procedure} and \ref{optimal_delta}, indicated as best choice values for $w_\mathrm{max}$ and $delta$ factor, to be equal to 8 and 4, respectively.

\subsubsection{Single-task Mixture of Experts (MoE) models}

The third deep learning model we employed for predicting the specified neutron star properties is an ensemble of mixture-of-experts (MoE) models, one for each property out of three, and is depicted in Figure \ref{fig:MoE_drawing}, while we form each spectral partition so as to almost equally comprise the 1/3 of the input spectrum. In other words, we feed during training the first expert of each MoE model with 123 amplitude values equal to the number of frequencies of the lower range, the second expert with the corresponding 123 amplitude values of the middle range and, finally, the third expert with the last 124 amplitude values of the higher range, to predict a single value of the output vector.
Overall, across the three target properties, the ST-MoE setup comprises 9 experts and 3 gate networks.

\begin{table}[th!]
\caption{
Expert ANN architecture of one MoE model of the ST-MoE-ANN ensemble. The shape (None, 123) corresponds to the number of frequency bins in the first and second spectral partition, while for third one is 124 (handled by last expert) due to incomplete division. The Layer class was used instead of Dense class, equivalently.
}
\label{tab:expert_arch}%
\begin{ruledtabular}
\resizebox{\linewidth}{!}{%
\begin{tabular}{ccccc}
Layer&Type&Shape&Activation\\
\hline
\#1&\textbf{Layer}&(None, 123)&Linear\\
\#2&GaussianNoise(0.001)&(None, 800)&-\\
\#3&Dropout(0.15)&(None, 800)&-\\
\#4&\textbf{Layer}&(None, 800)&Sigmoid\\
\#5&GaussianNoise(0.05)&(None, 800)&-\\
\#6&Dropout(0.25)&(None, 800)&-\\
\#7&\textbf{Layer}&(None, 800)&Sigmoid\\
\#8&GaussianNoise(0.05)&(None, 800)&-\\
\#9&Dropout(0.25)&(None, 800)&-\\
\#10&\textbf{Layer}&(None, 800)&Sigmoid\\
\#11&GaussianNoise(0.05)&(None, 800)&-\\
\#12&Dropout(0.25)&(None, 800)&-\\
\#13&\textbf{Layer}&(None, 800)&Sigmoid\\
\#14&GaussianNoise(0.05)&(None, 800)&-\\
\#15&Dropout(0.15)&(None, 800)&-\\
\#16&\textbf{Layer}&(None, 1)&Linear\\
\end{tabular}
}
\end{ruledtabular}
\end{table}

\begin{table}[bh!]
\caption{
Gate ANN classifier architecture of one MoE model of the ST-MoE-ANN ensemble. The shape (None, 370) corresponds to the overall number of frequency bins in an individual spectrum.
}
\label{tab:gate_arch}%
\begin{ruledtabular}
\resizebox{\linewidth}{!}{%
\begin{tabular}{ccccc}
Layer&Type&Shape&Activation\\
\hline
\#1&\textbf{Layer}&(None, 370)&Linear\\
\#2&Dropout(0.001)&(None, 600)&-\\
\#3&\textbf{Layer}&(None, 600)&ReLU\\
\#4&Dropout(0.05)&(None, 600)&-\\
\#5&\textbf{Layer}&(None, 600)&ReLU\\
\#6&Dropout(0.05)&(None, 600)&-\\
\#7&\textbf{Layer}&(None, 600)&ReLU\\
\#8&Dropout(0.05)&(None, 600)&-\\
\#9&\textbf{Layer}&(None, 600)&ReLU\\
\#10&\textbf{Layer}&(None, 3)&Softmax\\
\end{tabular}
}
\end{ruledtabular}
\end{table}

The ANN architecture of an expert is described in Table \ref{tab:expert_arch} and the one of a gate is described in Table \ref{tab:gate_arch}. In order to implement the custom logic, we used the Layer keras class instead of Dense, in an equivalent fashion. We can see that the experts are essentially similar to the ANN architecture of Table \ref{tab:simple_ann}, consisting of 4 hidden layers having sigmoid activation functions, whereas the gate has the architecture of a classifier with 4 hidden layers and ReLU activation functions, notwithstanding SoftMax activation for the last output layer. 
The gate acts as a classifier over the three experts. Its Softmax scores are converted to a top-1 routing decision via argmax, and the selected expert alone is used to predict the target quantity.

We also employed warm-up learning rate scheduling, as in \cite{Pesios2024}, which gradually increases the LR (from a very low value) up to an epoch and then is decreased so that a smoother and better training process is accomplished. The overall training lasted 100 epochs, 15\% of which during warming up. Batch size was set to 6 samples after tuning, as in the other models. 

As regards the scaling we performed, it is similar to one described for the ST-ANN models. Lastly, this model was trained using $k=4$ instead of LOO-CV to avoid potential gating instabilities and variance surges. Lastly, we normalized input sample weights having as $w_\mathrm{max}$ equal to 16.

\subsubsection{Pseudo-Inverse Multivariate Regression model (PI-MLR)}
\label{sec:pi-mlr}

The first algebraic model we implemented takes advantage of Eq. \ref{5a}, as described in \ref{inverse_linear_regression}, where $B$ is the coefficients matrix of the direct model of our previous work, used to formulate the respective coefficients matrix $\Gamma$ of our inverse MLR model. All we have to do in order to calculate a triplet of neutron star properties is to multiply a column vector representing an individual spectrum with this matrix.

However, we have to note that coefficients matrix $B$ was produced after we aligned the BNS spectra (which is currently used as input) to formulate the direct model, something that leads to some implications when we make predictions using $\Gamma$ in the inverse case. Here, we have to align again the dataset of input spectra into a common reference frequency before feeding them one by one into the inverse model. This alignment should not be based on an empirical relation, because we are not eligible to use one, since the neutron star properties this relation takes into account are not known a priori (because they are the predicted values). On the contrary, we implement a perfect alignment and not a partial one.

If we were able to use the same empirical relation, we would have re-produced the partial alignment of the first step of the direct model, something that would yield much more equivalent results when predicting with the resulted PI-MLR inverse model.

In the future, when Equation-of-State is further restricted and determined to a great extent, the direct model of our previous work, as we have argued, will be much more precise since the optimized empirical relation will lead to perfect alignment, and, consequently, the PI-MLR model for the inverse case will be much more precise also.

\subsubsection{Multivariate Regression model (MLR)}

Apart from the PI-MLR model which takes advantage of the pseudo-inverse approach, we can use the well-known Ordinary Least Squares (OLS) methodology to fit the inverse linear model of Eq. \ref{2}, thereby estimating the prediction matrix $\Gamma$. That is, to deal with the inverse problem as a direct one to mainly make comparisons with the pseudo-inverse case.


\begin{figure*}
    \begin{minipage}{\textwidth}
        \begin{minipage}{\textwidth}
            \textbf{Multi-Task ANN}%
        \end{minipage}
        \begin{minipage}{0.9\textwidth}
          \includegraphics[width=\linewidth]{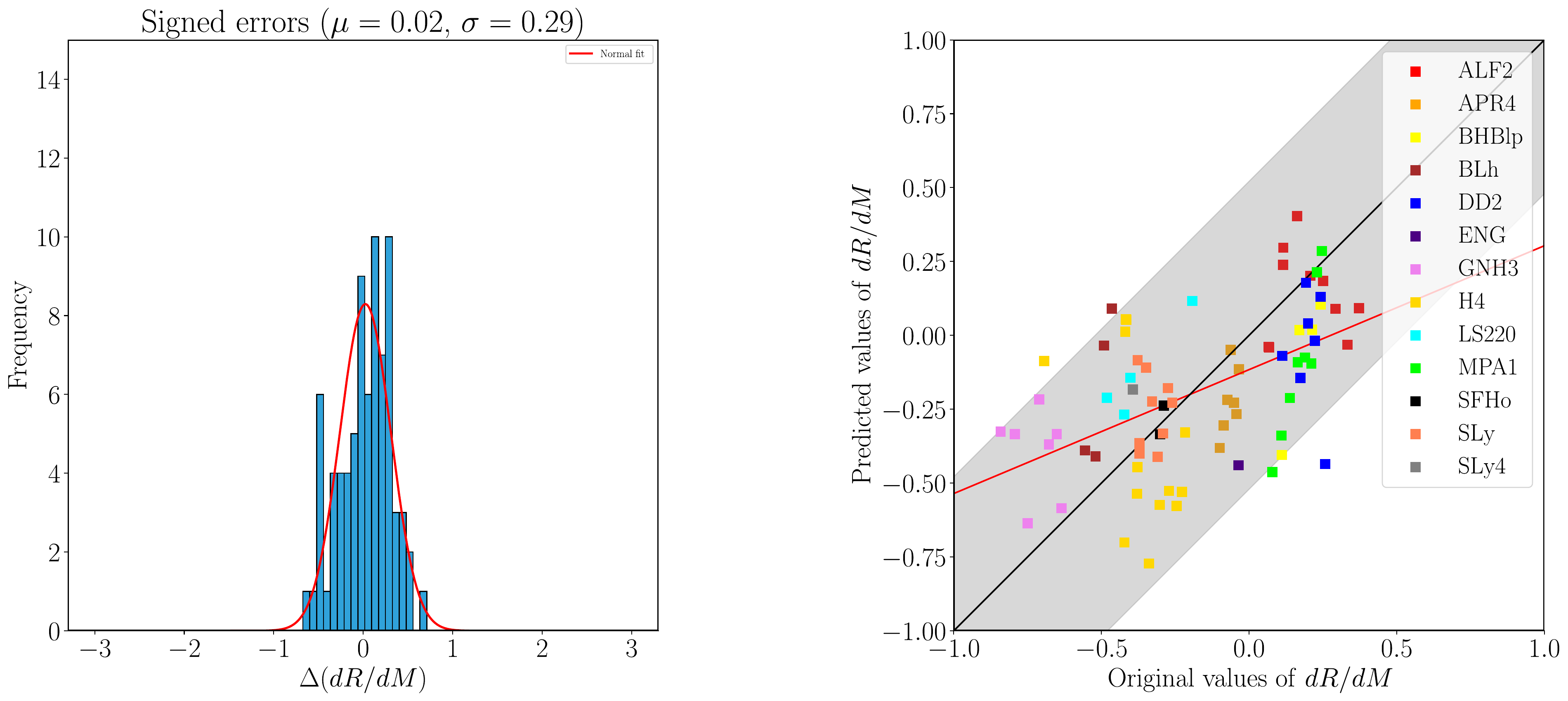}
        \end{minipage}\\%
        \begin{minipage}{\textwidth}
            \textbf{Single-Task ANN}%
        \end{minipage}
        \begin{minipage}{0.9\textwidth}
          \includegraphics[width=\linewidth]{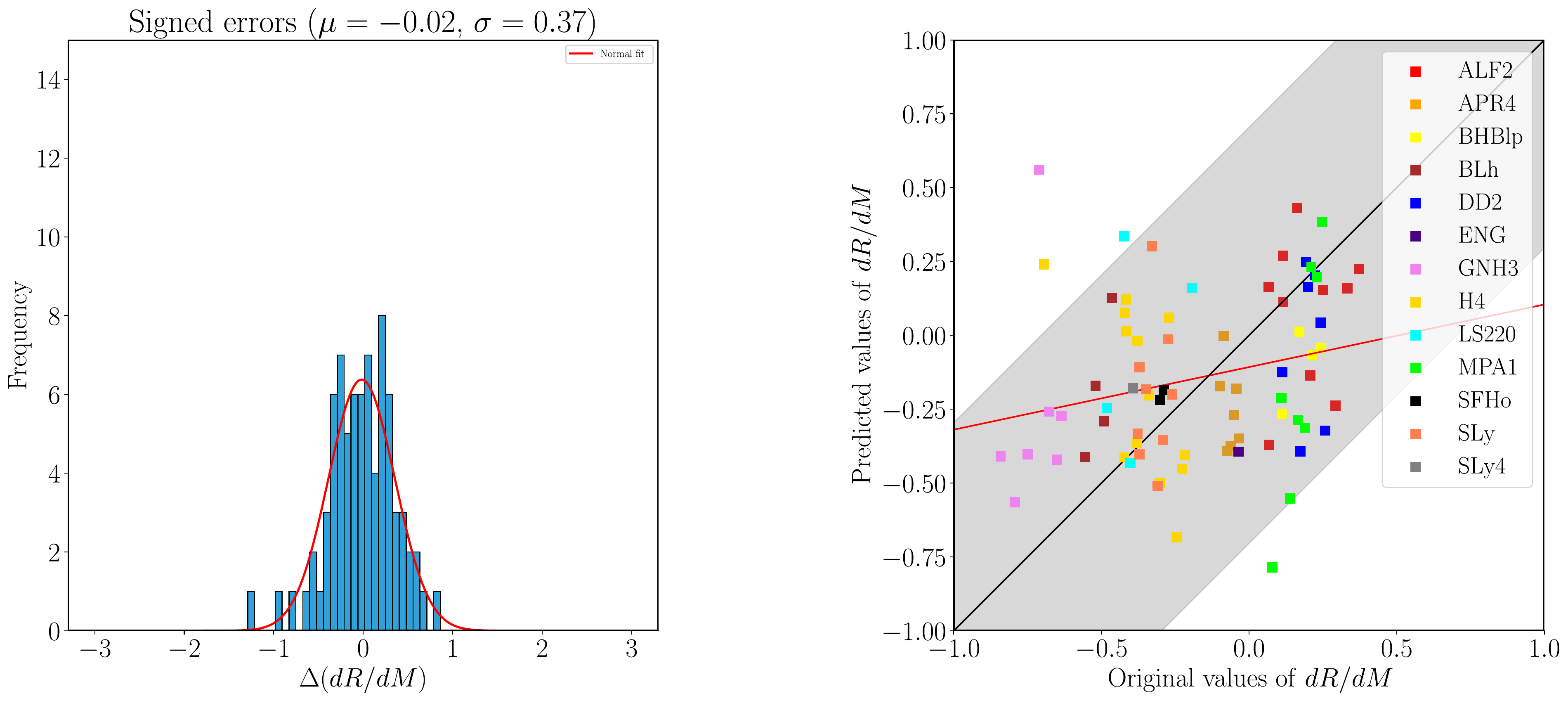}
        \end{minipage}\\%
        \begin{minipage}{\textwidth}
            \textbf{Single-Task MoE}%
        \end{minipage}
        \begin{minipage}{0.9\textwidth}
          \includegraphics[width=\linewidth]{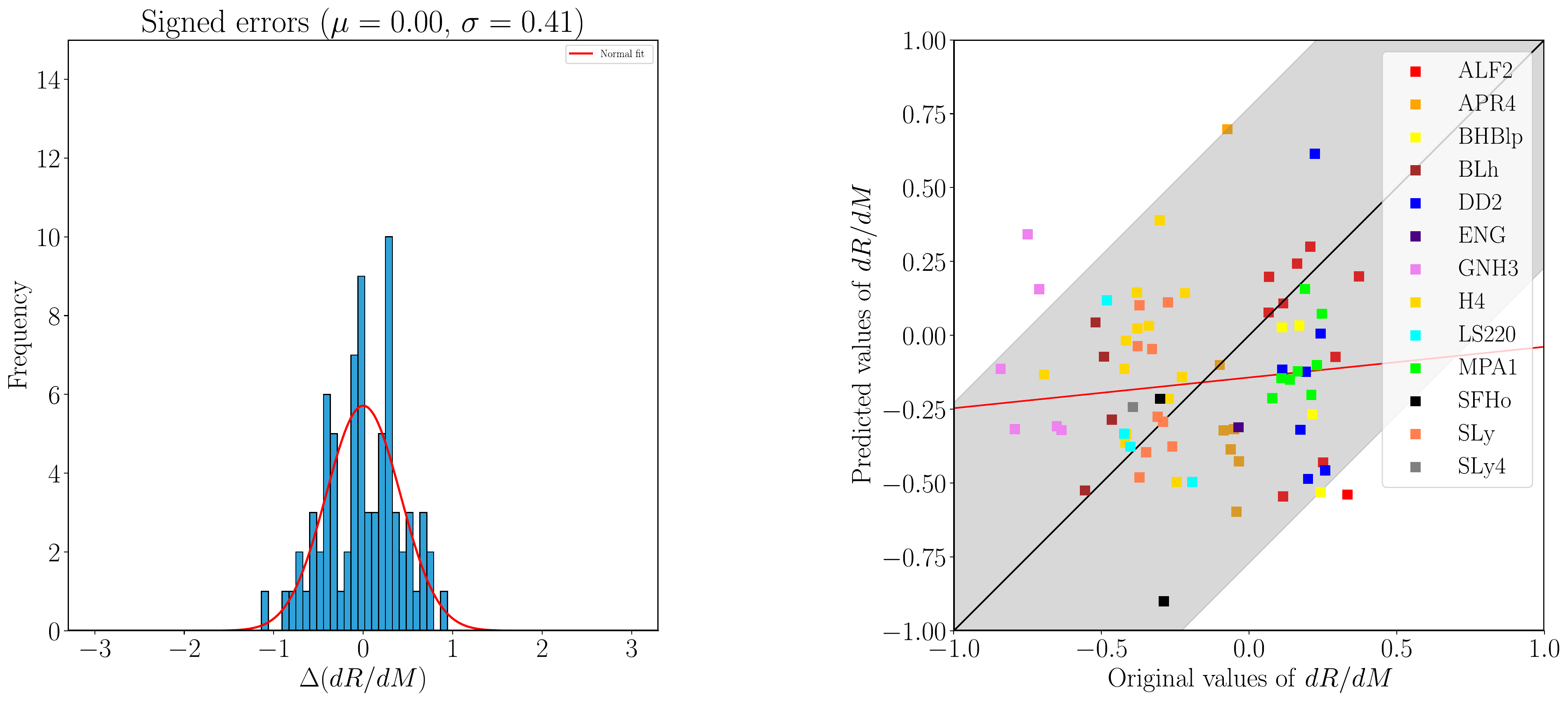}
        \end{minipage}%
    \end{minipage}
\caption{Prediction results for the three deep-learning models for $dR/dM$. Each panel shows a scatter plot of predicted versus original values together with the corresponding signed-error histogram. The black dashed line denotes the diagonal, the red line is the least-squares trend line, and the grey-shaded band indicates the central 90\% interval of the prediction errors. Among the three models, the MT-ANN achieves the best prediction accuracy for this target.}
\label{fig:scatter_plots_dl_drdm}
\end{figure*}

\begin{figure*}
    \begin{minipage}{\textwidth}
        \begin{minipage}{\textwidth}
            \textbf{Multi-Task ANN}%
        \end{minipage}
        \begin{minipage}{0.9\textwidth}
          \includegraphics[width=\linewidth]{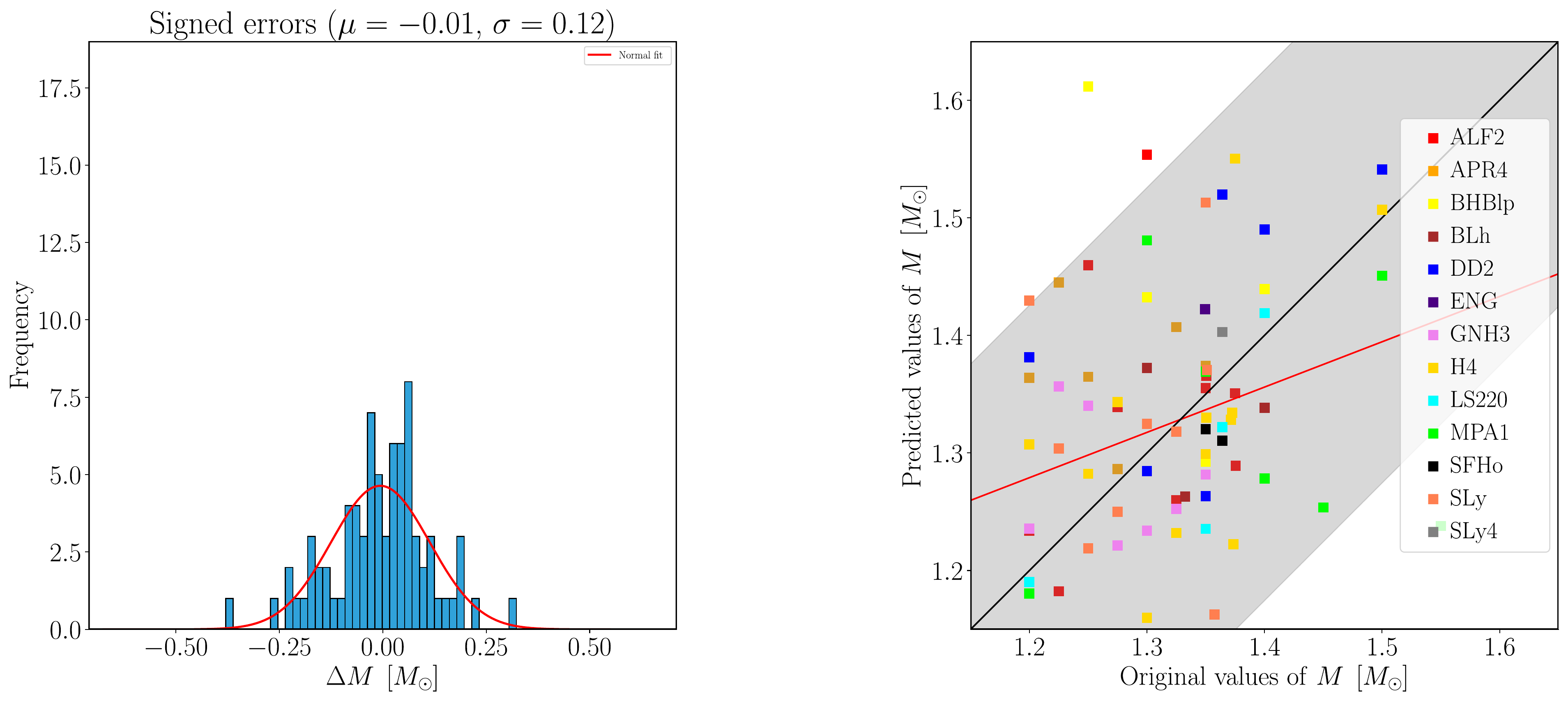}
        \end{minipage}\\%
        \begin{minipage}{\textwidth}
            \textbf{Single-Task ANN}%
        \end{minipage}
        \begin{minipage}{0.9\textwidth}
          \includegraphics[width=\linewidth]{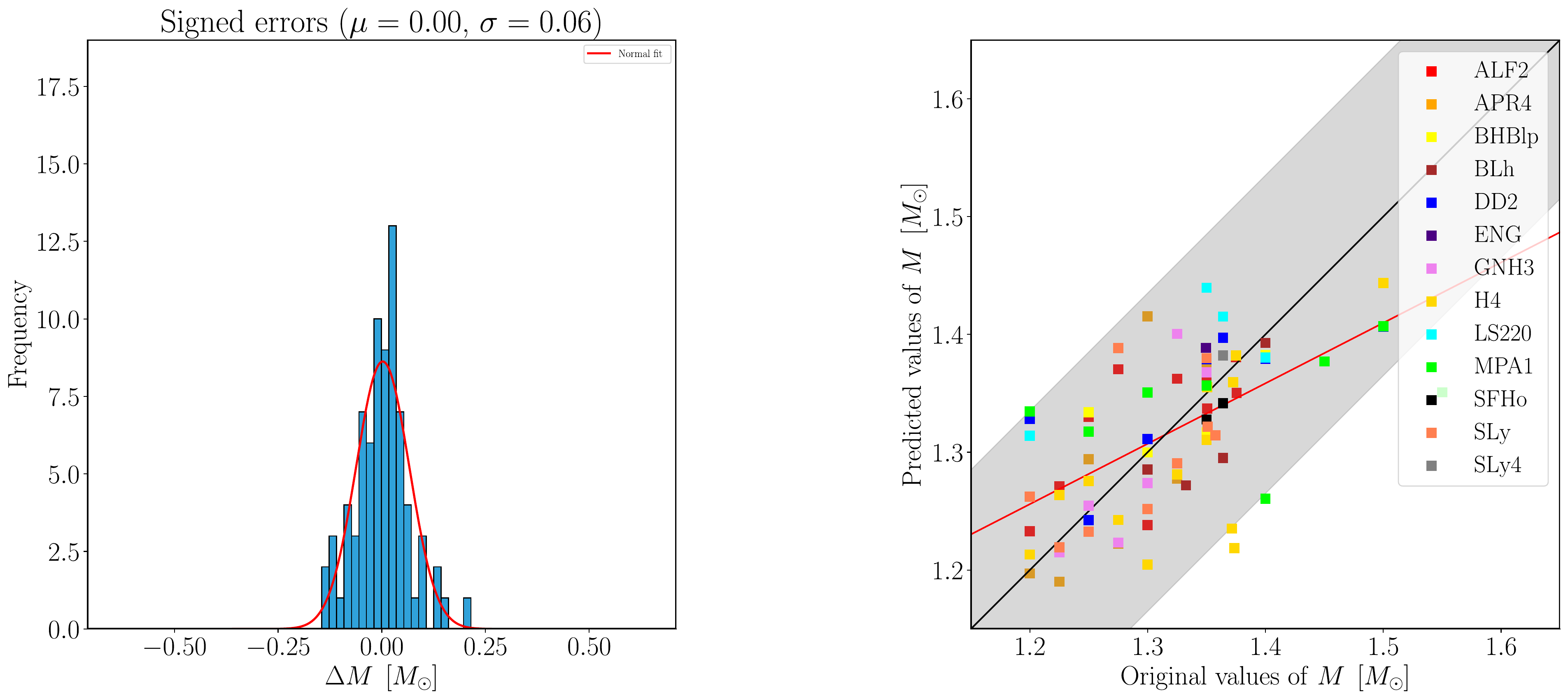}
        \end{minipage}\\%
        \begin{minipage}{\textwidth}
            \textbf{Single-Task MoE}%
        \end{minipage}
        \begin{minipage}{0.9\textwidth}
          \includegraphics[width=\linewidth]{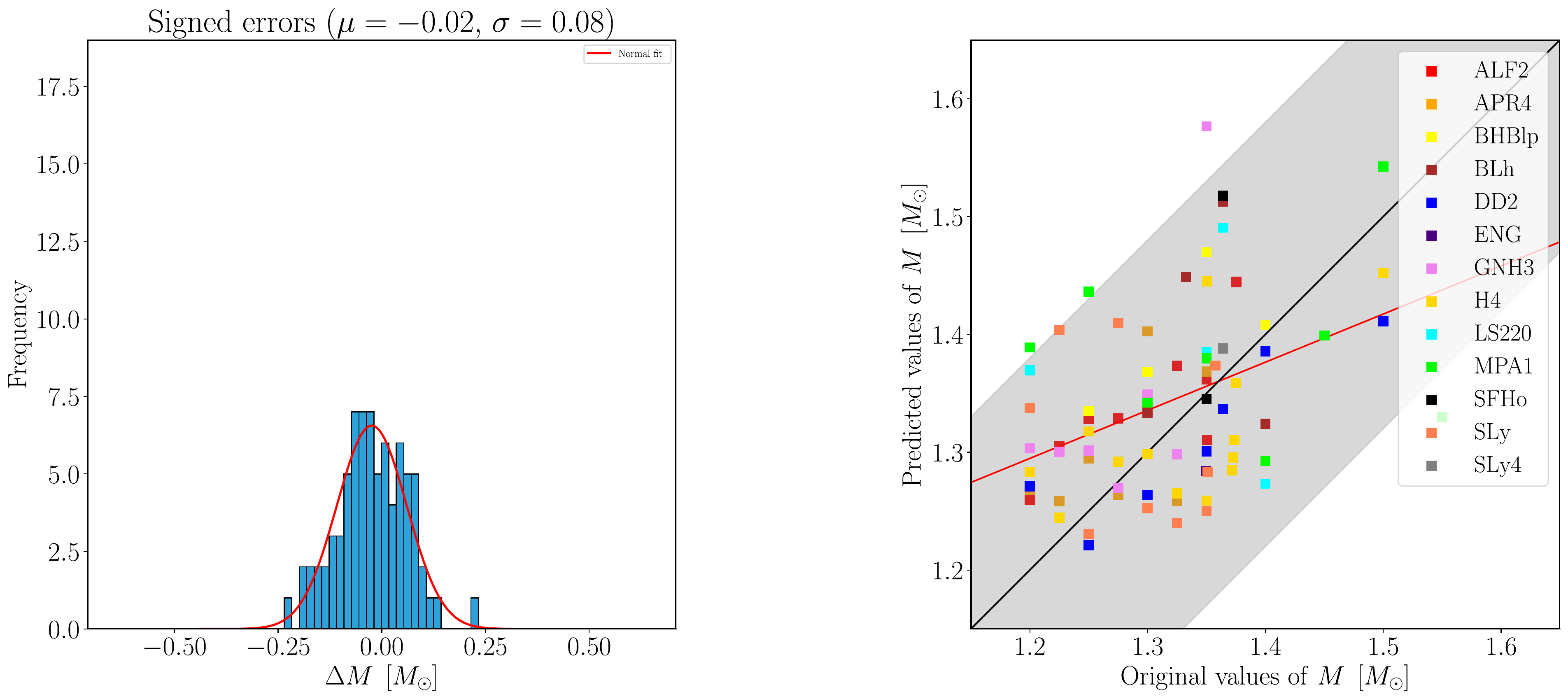}
        \end{minipage}%
    \end{minipage}
    \caption{Same as Fig. \ref{fig:scatter_plots_dl_drdm}, but for the Mass $(M)$ property.}
\label{fig:scatter_plots_dl_mass}
\end{figure*}

\begin{figure*}
    \begin{minipage}{\textwidth}
        \begin{minipage}{\textwidth}
            \textbf{Multi-Task ANN}%
        \end{minipage}
        \begin{minipage}{0.9\textwidth}
          \includegraphics[width=\linewidth]{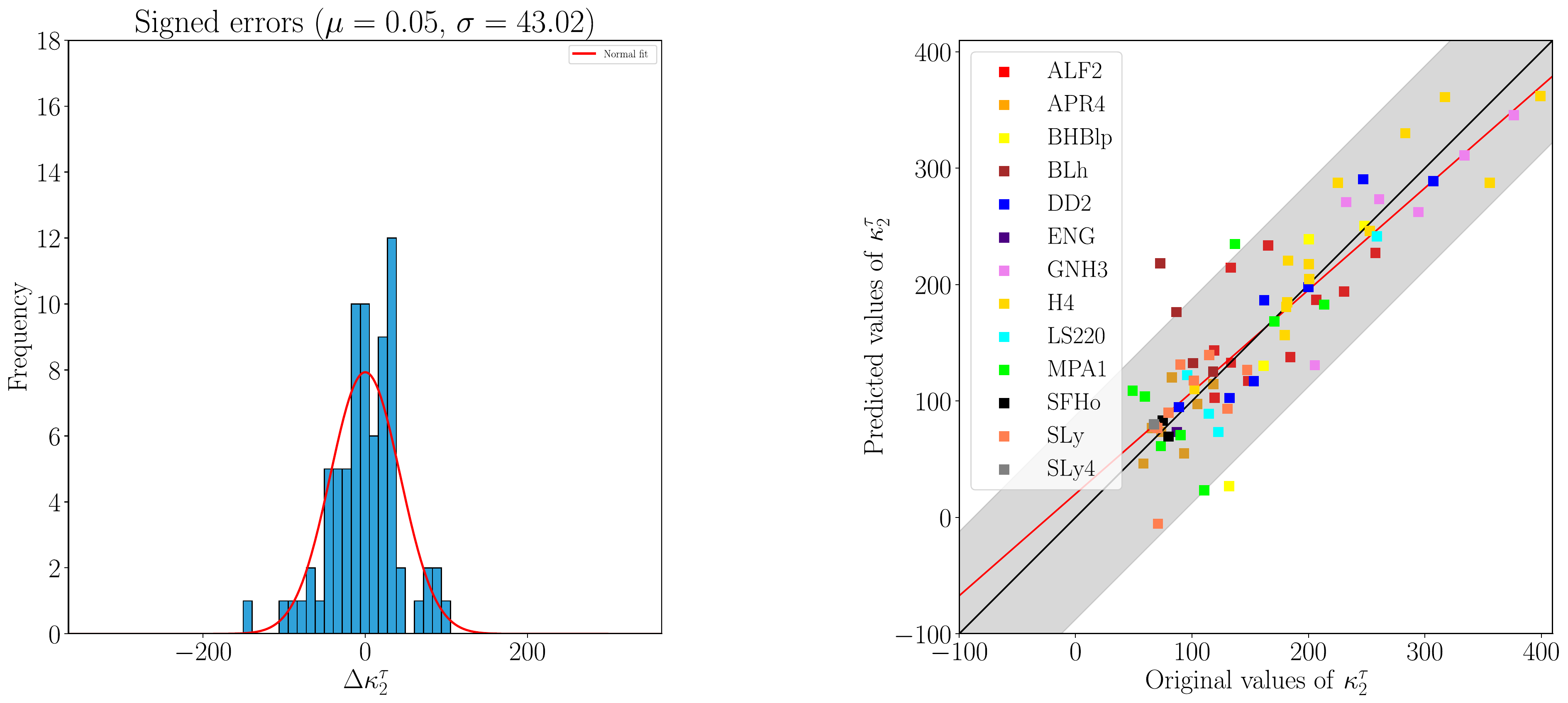}
        \end{minipage}\\%
        \begin{minipage}{\textwidth}
            \textbf{Single-Task ANN}%
        \end{minipage}
        \begin{minipage}{0.9\textwidth}
          \includegraphics[width=\linewidth]{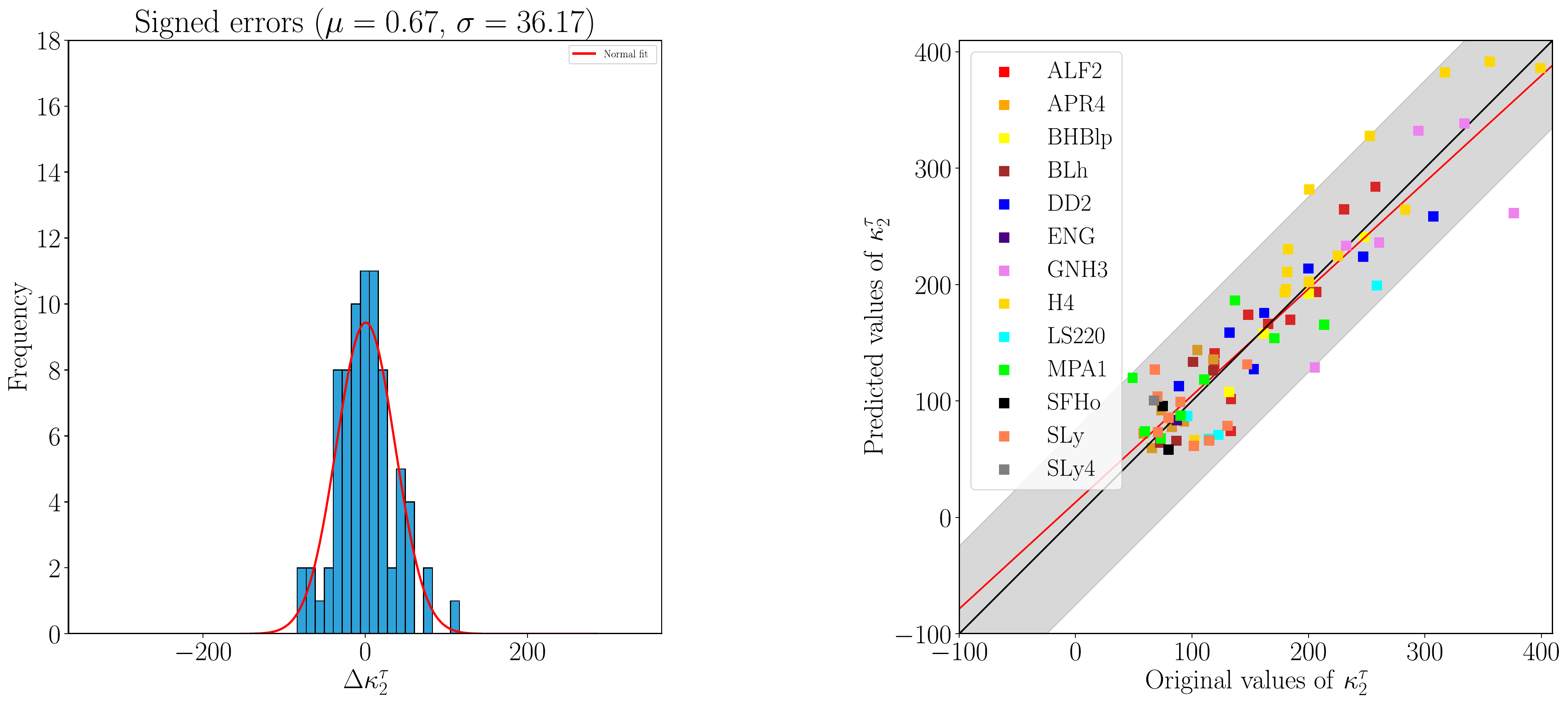}
        \end{minipage}\\%
        \begin{minipage}{\textwidth}
            \textbf{Single-Task MoE}%
        \end{minipage}
        \begin{minipage}{0.9\textwidth}
          \includegraphics[width=\linewidth]{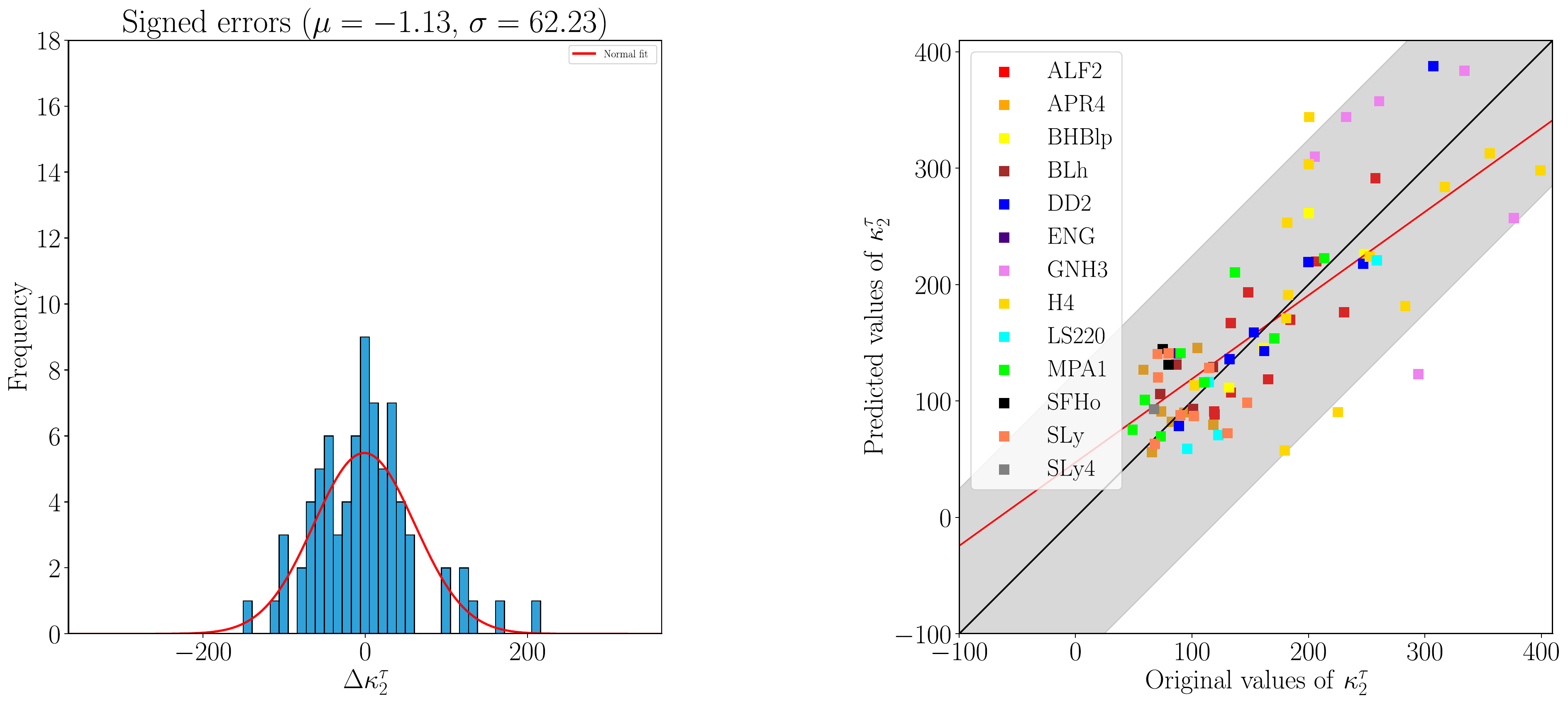}
        \end{minipage}%
    \end{minipage}
\caption{Same as Fig. \ref{fig:scatter_plots_dl_drdm}, but for the tidal deformability $\kappa_2^\tau$ property.}
\label{fig:scatter_plots_dl_kappa}
\end{figure*}

\subsection{Prediction accuracy and model comparison}
\label{comparisons}

\begin{figure*}
    \begin{minipage}{\textwidth}
        \begin{minipage}{\textwidth}
            \textbf{MLR}%
        \end{minipage}
        \begin{minipage}{0.9\textwidth}
          \includegraphics[width=\linewidth]{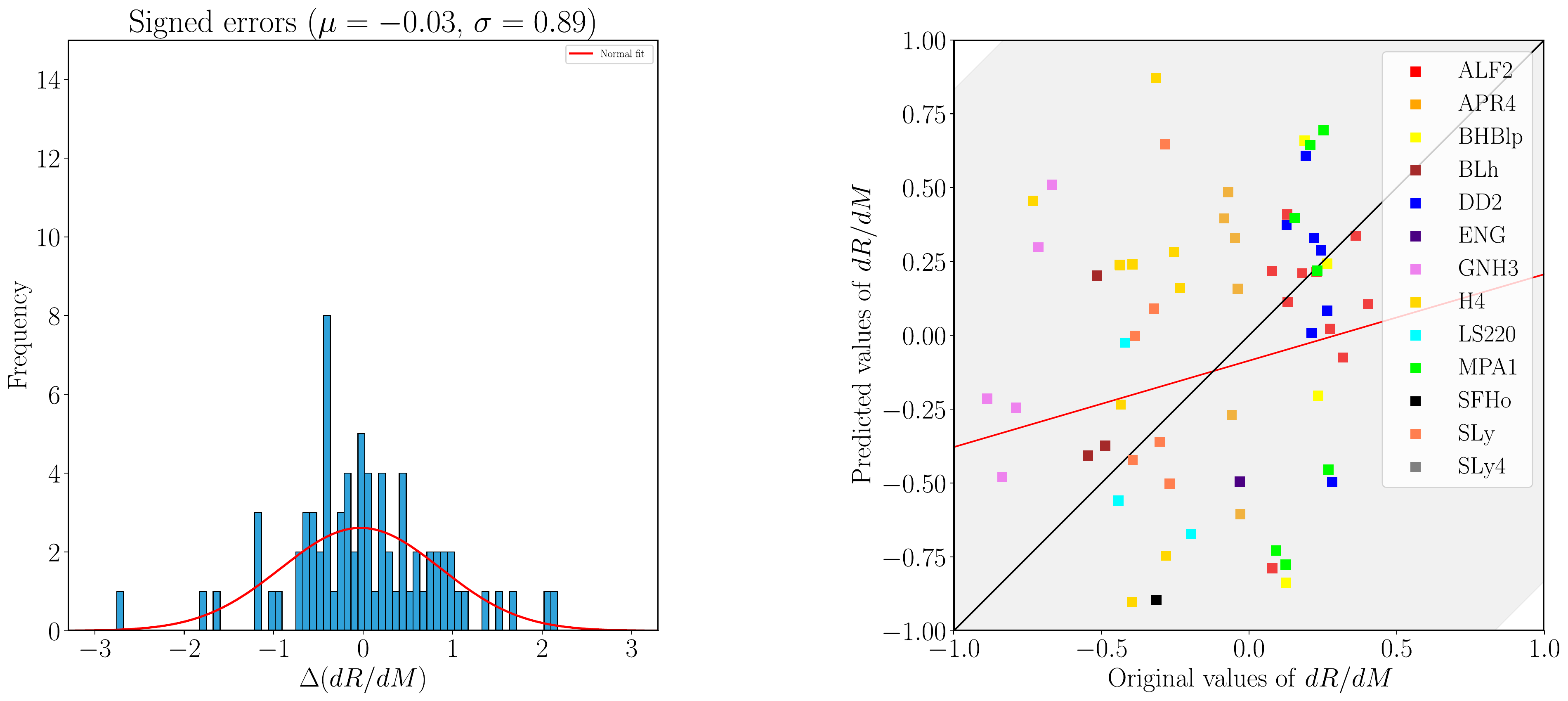}
        \end{minipage}\\%
        \begin{minipage}{\textwidth}
            \textbf{Pseudo-Inverse MLR}%
        \end{minipage}
        \begin{minipage}{0.9\textwidth}
          \includegraphics[width=\linewidth]{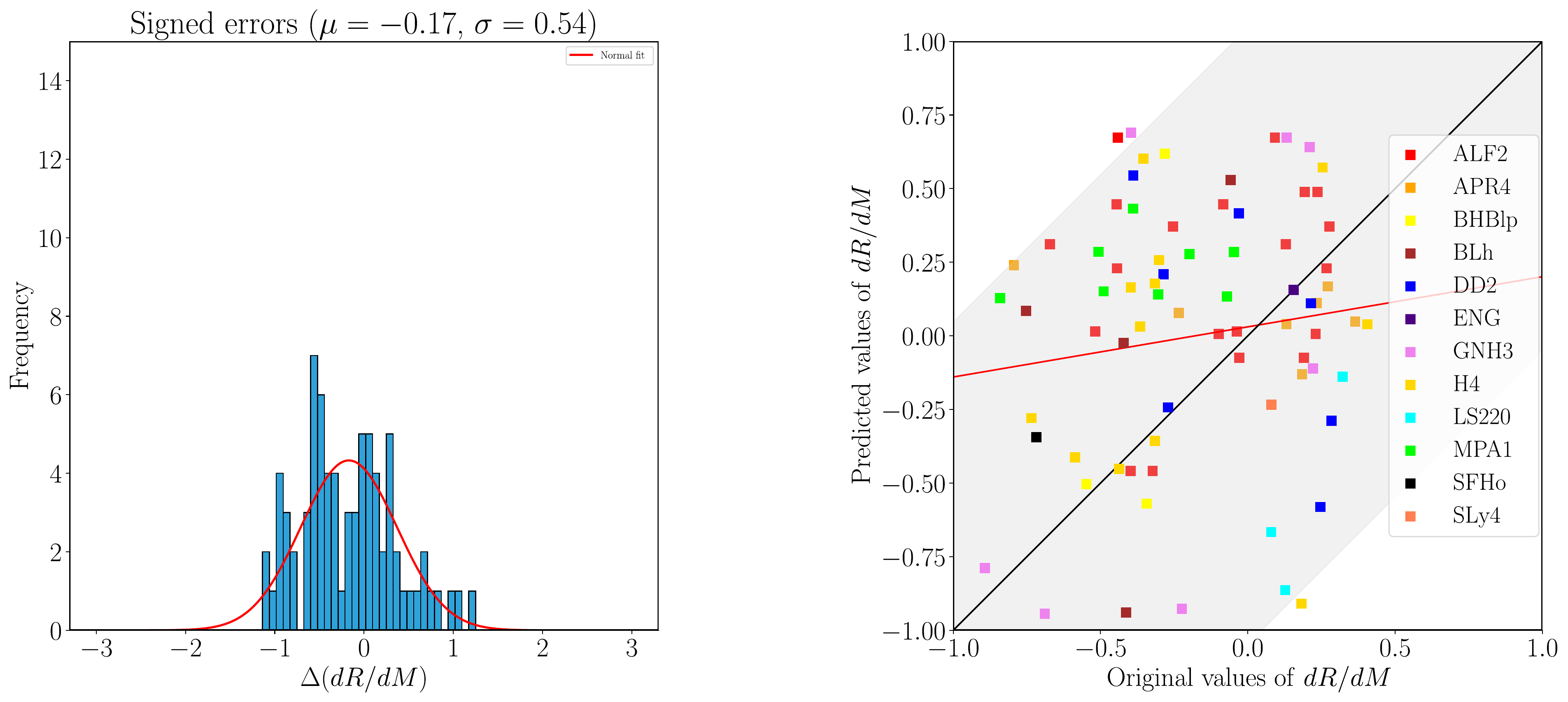}
        \end{minipage}\\%
    \end{minipage}
\caption{Same as Fig. \ref{fig:scatter_plots_dl_drdm}, but for the algebraic models.}
\label{fig:scatter_plots_dl_mlr_drdm}
\end{figure*}

\begin{figure*}
    \begin{minipage}{\textwidth}
        \begin{minipage}{\textwidth}
            \textbf{MLR}%
        \end{minipage}
        \begin{minipage}{0.9\textwidth}
          \includegraphics[width=\linewidth]{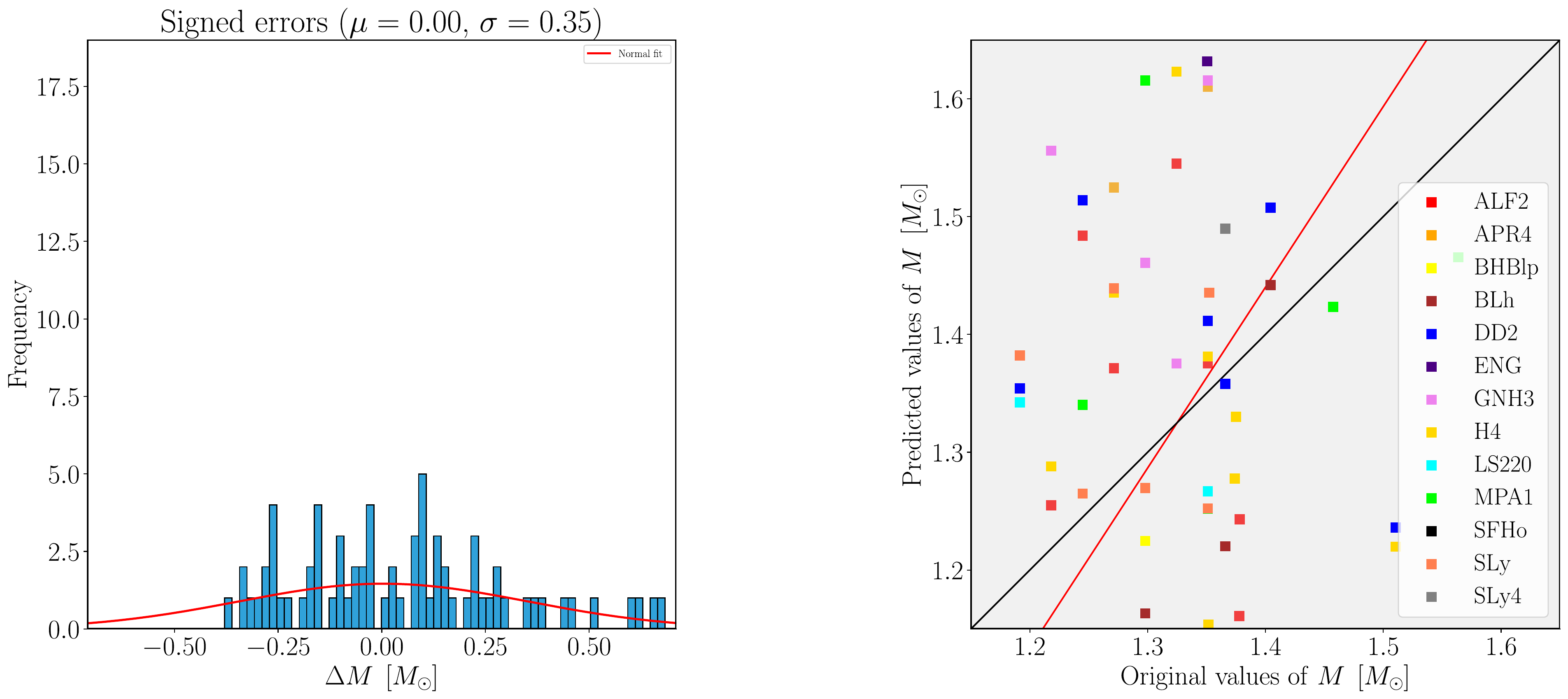}
        \end{minipage}\\%
        \begin{minipage}{\textwidth}
            \textbf{Pseudo-Inverse MLR}%
        \end{minipage}
        \begin{minipage}{0.9\textwidth}
          \includegraphics[width=\linewidth]{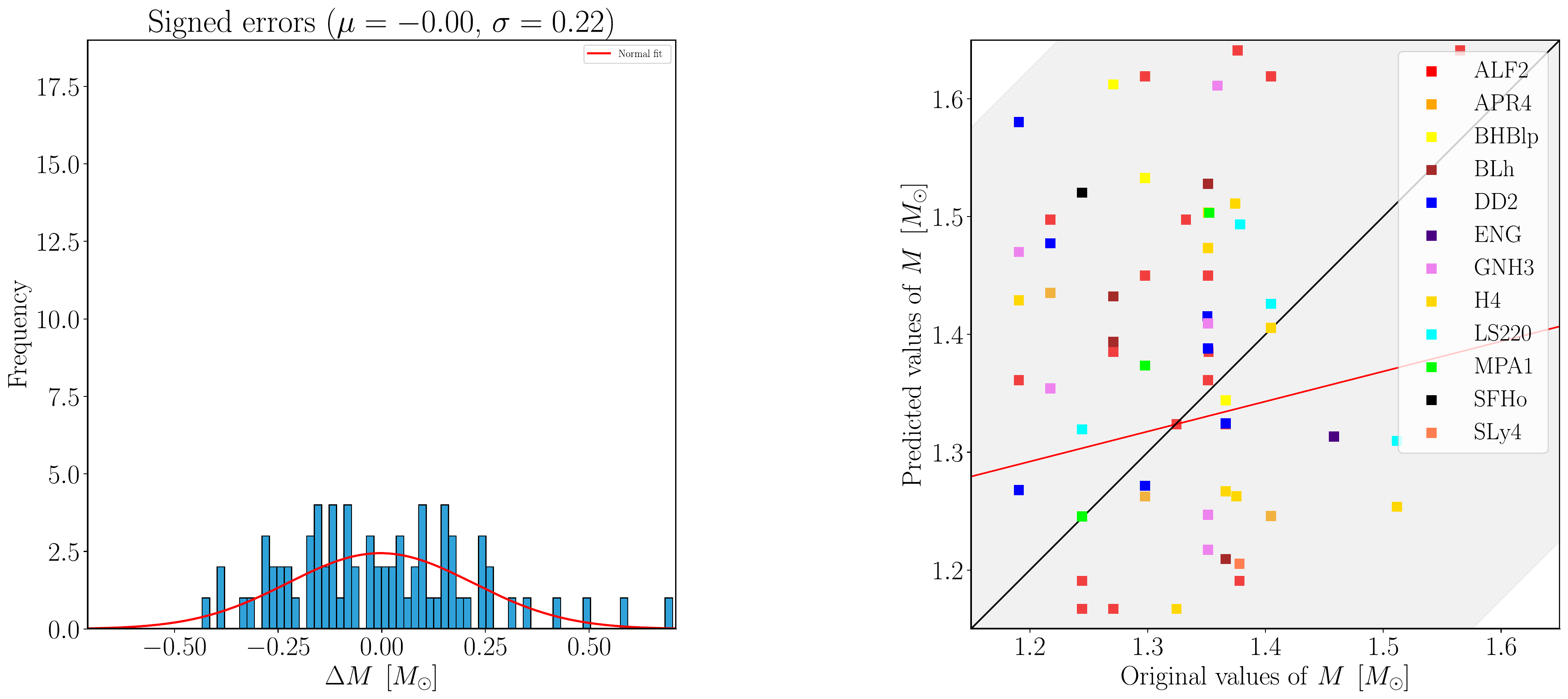}
        \end{minipage}\\%
    \end{minipage}
\caption{Same as Fig. \ref{fig:scatter_plots_dl_mass}, but for the algebraic models.}
\label{fig:scatter_plots_dl_mlr_mass}
\end{figure*}

\begin{figure*}
    \begin{minipage}{\textwidth}
        \begin{minipage}{\textwidth}
            \textbf{MLR}%
        \end{minipage}
        \begin{minipage}{0.9\textwidth}
          \includegraphics[width=\linewidth]{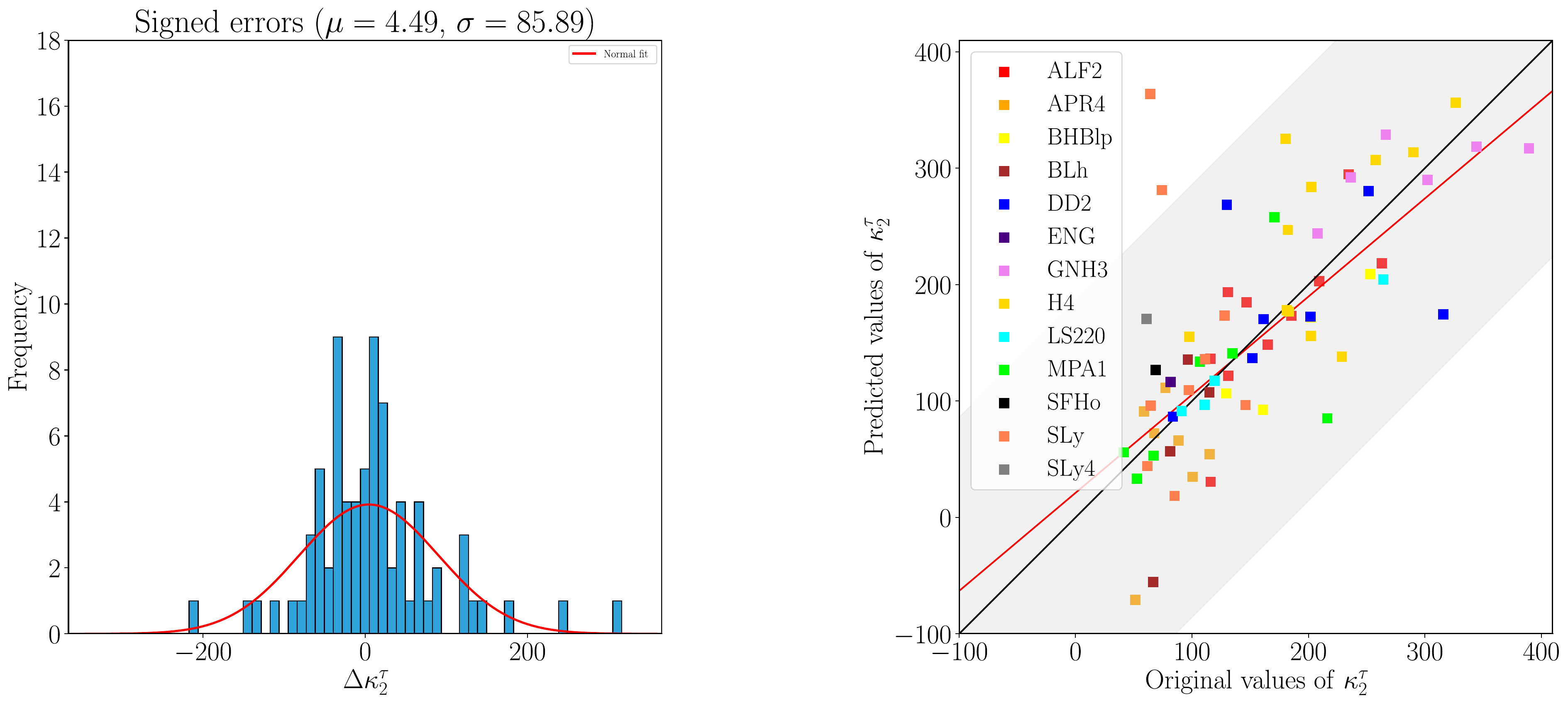}
        \end{minipage}\\%
        \begin{minipage}{\textwidth}
            \textbf{Pseudo-Inverse MLR}%
        \end{minipage}
        \begin{minipage}{0.9\textwidth}
          \includegraphics[width=\linewidth]{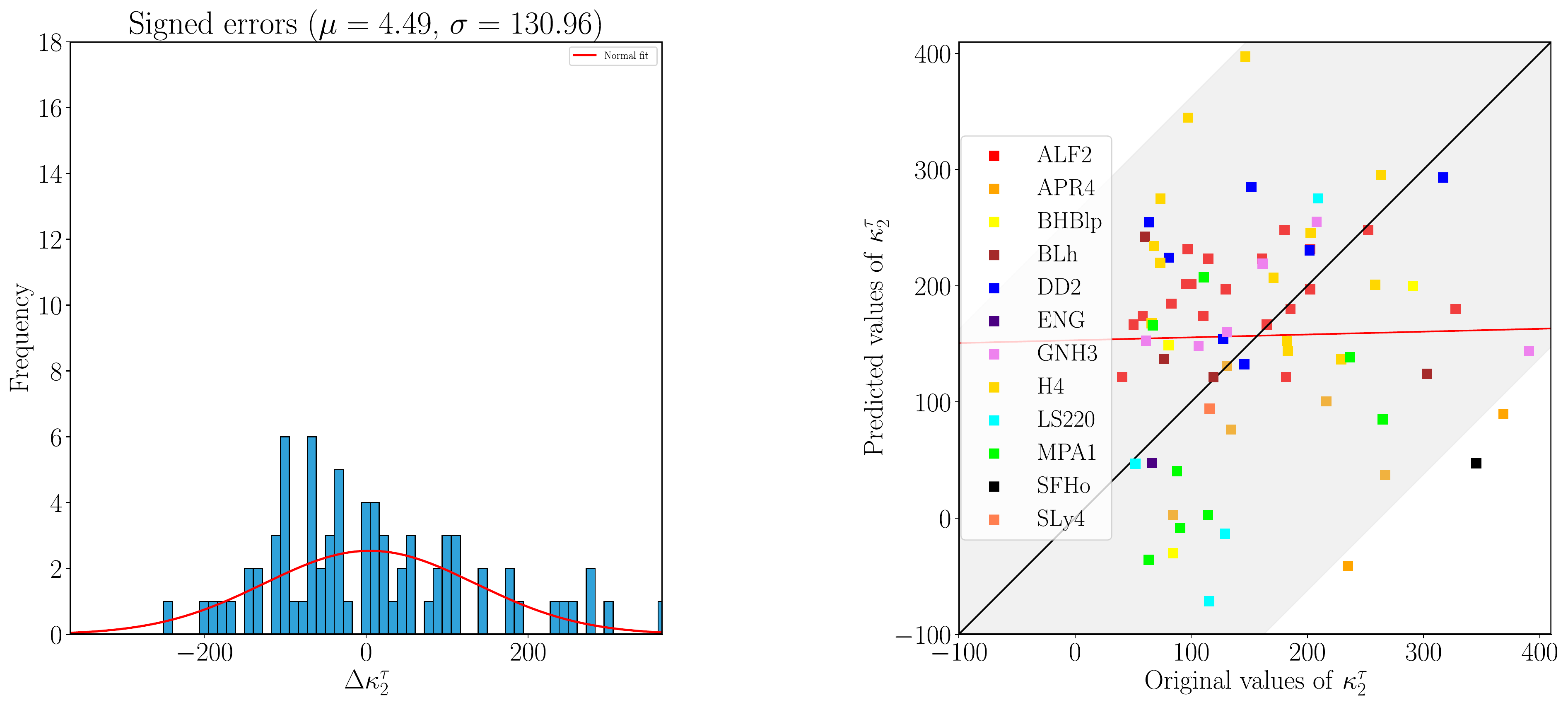}
        \end{minipage}\\%
    \end{minipage}
\caption{Same as Fig. \ref{fig:scatter_plots_dl_kappa}, but for the algebraic models.}
\label{fig:scatter_plots_dl_mlr_kappa}
\end{figure*}

Here, we present the prediction results as combined scatter plots and signed-error histograms for the two groups of models we implemented: the three deep-learning models in Figures \ref{fig:scatter_plots_dl_drdm} to \ref{fig:scatter_plots_dl_kappa}, and the two algebraic models in Figures \ref{fig:scatter_plots_dl_mlr_drdm} to \ref{fig:scatter_plots_dl_mlr_kappa}. In each scatter plot, the horizontal axis shows the original value of the predicted property, while the vertical axis shows the corresponding predicted value. The best predictions therefore lie close to the diagonal black dashed line. We also depict as a grey-shaded area the central 90\% interval of the prediction errors, bounded by the 5th and 95th percentiles. The red tilted line is a least-squares approximation to the overall trend of the data points.

The corresponding signed errors, defined as the difference between predicted and original values, are also shown as histograms in the same panels. For a given target quantity, the binning is kept fixed across models in order to enable direct visual comparison, although the bin width and number of bins differ between target quantities. The signed-error statistics used mainly to interpret the shape, spread, and bias of the prediction errors.

For quantitative comparison among models, we primarily rely on the root mean square error (RMSE), while the mean signed error $\mu$ and the standard deviation $\sigma$ of the signed errors are used as complementary descriptors of the error distributions shown in the histograms. For approximately unbiased predictions, RMSE is closely related to the standard deviation of the signed errors, since
\begin{equation}
    \mathrm{RMSE}^2 = \sigma^2 + \mu^2.
\end{equation}
Thus, when the mean signed error is small, RMSE and $\sigma$ take similar values. The histograms further indicate that, for the best-performing models, the signed-error distributions are reasonably close to Gaussian, so that the interval $(\mu-\sigma,\mu+\sigma)$ contains roughly 68\% of the values, in accordance with the empirical rule.
The RMSE values for all models and target quantities are reported in Table \ref{tab:rmses}.

Among the deep-learning models, the MT-ANN achieves the best performance for the prediction of $dR/dM$, with mean signed error $0.02$ and standard deviation $0.29$. For the mass $M$, the ST-ANN model gives the best predictions, with mean signed error $0.00\,M_\odot$ and standard deviation $0.07\,M_\odot$. The ST-ANN model also performs best for the dimensionless tidal deformability parameter $\kappa_2^\tau$, with mean signed error $0.67$ and standard deviation $36.17$. Overall, the ST-ANN ensemble provides the strongest performance across two of the three target quantities, while the MT-ANN remains the best model specifically for $dR/dM$.

The algebraic baselines perform less favorably than the ANN-based models. In particular, the direct inverse MLR model yields the largest RMSE for $dR/dM$ and $M$, while the PI-MLR model gives the largest RMSE for $\kappa_2^\tau$. This overall underperformance indicates that the inverse relation between post-merger spectra and neutron-star properties is not captured as effectively by linear surrogates as by the proposed nonlinear ANN models.

The signed-error histograms further show that the two MLR-based models behave qualitatively similarly, although for $M$ their error distributions deviate more clearly from an approximately Gaussian shape. This helps explain the larger mismatch between the standard deviation $\sigma$ of the signed errors and the corresponding RMSE values in those cases.

For the PI-MLR model specifically, the weaker performance relative to the ANN-based models is not only a consequence of limited predictive flexibility, but also reflects the fact that it inherits the alignment dependence of the direct model of \cite{Pesios2024}. In the inverse setting, this alignment procedure cannot be reproduced in the same way without prior knowledge of the target neutron-star properties, which places the PI-MLR baseline at an intrinsic disadvantage relative to dedicated inverse models trained directly on the original spectra.

\begin{table}[ht]
    \caption{Root mean square errors (RMSE indices) for predicting the corresponding NS property using our dataset and for each implemented model. RMSEs were calculated using LOO CV, except for ST-MoE-ANN group of models.}
    \label{tab:rmses}
    \centering
    \begin{tabularx}{\linewidth}{l l *{2}{>{\centering\arraybackslash}X}}
        \hline
        ~ & ~ & \textbf{RMSE indices} & ~ \\ \hline
        \textbf{Model} & $dR/dM$ & $M$ & $\kappa_2^\tau$ \\ \hline
        MT-ANN & 0.286 & 0.119 & 43.023 \\
        ST-ANNs & 0.371 & 0.064 & 36.177 \\
        ST-MoE-ANN & 0.414 & 0.087 & 62.244 \\
        PI-MLR & 0.472 & 0.178 & 90.961 \\
        MLR & 0.894 & 0.349 & 86.810 \\
    \end{tabularx}
\end{table}

\section{Gate-decision analysis of the ST-MoE models}
\label{investigation_moe}

The ST-MoE-ANN architecture offers a window into how the model distributes its routing decisions across the three spectral regions.  Because the gate selects exactly one expert per sample via top-1 hard routing, the aggregate routing statistics reveal which spectral partition the trained gate preferentially consults for each target property.  In our dataset, the dominant spectral peak $f_\text{peak}$ falls exclusively in Region~II ($2$--$3\,$kHz) or Region~III ($3$--$4\,$kHz), while Region~I ($1$--$2\,$kHz) contains only sub-dominant spectral features.  The routing patterns therefore allow us to characterize whether the gate tends to select the peak-containing region or one of the two remaining regions.

To quantify this, we first construct, for each target property, a $3 \times 3$ contingency table of gate decision versus true peak location (Table~\ref{tab:3x3}).  The column marginals are identical across all three properties $(0, 56, 21)$, confirming that the true peak distribution is determined by the spectra alone and is independent of the target variable.  The structurally empty first column (no peaks fall below $2\,$kHz) is a consequence of the expected frequency range of $f_\text{peak}$ for the EOS families in our dataset.

We then reduce each sample's gate decision to a binary outcome: whether the gate selected the region containing $f_\text{peak}$ (``peak'') or one of the two regions that do not (``non-peak'').  Under the null hypothesis that the gate routes independently of peak location, the expected non-peak rate is $p_0 = 2/3$, since two out of three regions do not contain the peak.  A significant departure from this rate would indicate that the gate has learned a directional routing preference related to $f_\text{peak}$.  We test:
\begin{itemize}
  \item[$H_0$:] The probability that the gate selects a non-peak region equals $p_0 = 2/3$ (no preference).
  \item[$H_1$:] The probability differs from $2/3$.
\end{itemize}
\noindent
using a two-sided binomial test (normal approximation), with test statistic:
\begin{equation}
  z = \frac{k - N p_0}{\sqrt{N\, p_0\,(1-p_0)}} \;,
  \label{eq:binomial_z}
\end{equation}
where $k$ is the observed number of non-peak gate selections out of $N = 77$ samples.  The results are reported in Table~\ref{tab:gate_routing}.

\begin{table}[!ht]
  \caption{$3\times3$ contingency tables of gate decision (rows) versus true peak location (columns) for each predicted quantity.  Column marginals are identical across properties, confirming that the true peak distribution is a property of the spectra alone.  Region~I is structurally empty (no peaks fall below $2\,$kHz).}
  \label{tab:3x3}
  \centering

  \smallskip
  \textbf{(I) $dR/dM$}\\[4pt]
  \begin{tabularx}{0.85\linewidth}{Xcccc}
    \hline
    Gate decision & Region I & Region II & Region III & Total \\
    \hline
    Region I  & 0 & 33 & 8  & 41 \\
    Region II & 0 & 9  & 4  & 13 \\
    Region III& 0 & 14 & 9  & 23 \\
    \hline
    Total     & 0 & 56 & 21 & 77 \\
    \hline
  \end{tabularx}

  \bigskip
  \textbf{(II) $M$}\\[4pt]
  \begin{tabularx}{0.85\linewidth}{Xcccc}
    \hline
    Gate decision & Region I & Region II & Region III & Total \\
    \hline
    Region I  & 0 & 11 & 9  & 20 \\
    Region II & 0 & 27 & 6  & 33 \\
    Region III& 0 & 18 & 6  & 24 \\
    \hline
    Total     & 0 & 56 & 21 & 77 \\
    \hline
  \end{tabularx}

  \bigskip
  \textbf{(III) $\kappa_2^\tau$}\\[4pt]
  \begin{tabularx}{0.85\linewidth}{Xcccc}
    \hline
    Gate decision & Region I & Region II & Region III & Total \\
    \hline
    Region I  & 0 & 23 & 12 & 35 \\
    Region II & 0 & 22 & 4  & 26 \\
    Region III& 0 & 11 & 5  & 16 \\
    \hline
    Total     & 0 & 56 & 21 & 77 \\
    \hline
  \end{tabularx}
\end{table}

\begin{table}[!ht]
  \caption{Gate routing preference for each target property.  The non-peak rate gives the fraction of samples for which the gate selected a region that does not contain $f_\text{peak}$.  Under $H_0$ (no preference), the expected non-peak rate is $2/3 \approx 66.7\%$.  The $z$-statistic and two-sided $p$-value are computed from Eq.~\eqref{eq:binomial_z}.}
  \label{tab:gate_routing}
  \centering
  \begin{tabularx}{\linewidth}{Xccccc}
    \hline
    Variable & Peak & Non-peak & Non-peak rate & $z$ & $p$-value \\
    \hline
    $dR/dM$         & 18 & 59 & 76.6\% & $+1.85$ & $0.064$ \\
    $M$             & 33 & 44 & 57.1\% & $-1.77$ & $0.077$ \\
    $\kappa_2^\tau$  & 27 & 50 & 64.9\% & $-0.32$ & $0.75\phantom{0}$ \\
    \hline
  \end{tabularx}
\end{table}

The three target properties exhibit qualitatively distinct routing behaviors.  For the mass--radius slope $dR/dM$, the gate selects a non-peak region in 76.6\% of cases, compared to the 66.7\% expected under random routing ($z = +1.85$, $p = 0.064$).  For the component mass $M$, the pattern reverses: the gate selects the peak-containing region 42.9\% of the time, above the 33.3\% random expectation ($z = -1.77$, $p = 0.077$).  For the tidal deformability $\kappa_2^\tau$, no significant preference is observed ($z = -0.32$, $p = 0.75$).

While neither the $dR/dM$ nor the $M$ result reaches the conventional $\alpha = 0.05$ significance threshold under the two-sided test, the \textit{directions} of the two departures are noteworthy and physically suggestive.  For $M$, a tendency toward peak tracking is consistent with the known sensitivity of $f_\text{peak}$ to the total binary mass, as reflected in the quasi-universal relations that involve
$M_{\rm tot} f_\text{peak}$ as a combined variable (see e.g. 
 \cite{vretinaris2020}). 

For $dR/dM$, the opposite tendency (routing away from $f_\text{peak}$) raises the possibility that sub-peak spectral features carry predictive information for the local curvature of the mass--radius relation that is complementary to the dominant frequency.

An important caveat must be noted regarding the interpretation of these routing statistics.  The gate network receives the \textit{full} $370$-bin spectrum as input and makes its routing decision based on the entire spectral shape, not solely on the content of the region it selects.  Consequently, when the gate chooses Region~I, this does not establish that Region~I alone is informative; rather, it indicates that, conditioned on the full spectral morphology, the gate determined that the Region~I expert would produce the best prediction.  Furthermore, top-1 hard routing is a discrete winner-take-all decision that does not reveal the margin by which one expert was preferred over another; near-ties and confident selections are treated identically.

As a result, two distinct interpretations of the observed routing patterns remain compatible with the data.  On one hand, the gate's preference for non-peak regions when predicting $dR/dM$ may reflect a genuine learned association between sub-peak spectral features and the mass--radius slope, indicating that the $1$--$2\,$kHz region carries independent predictive content not accessible through $f_\text{peak}$-based empirical relations.  On the other hand, the same pattern could arise from suboptimal gating: the gate may be making imperfect routing decisions, and the fact that the ST-MoE-ANN achieves worse overall RMSE than the ST-ANN (Table~\ref{tab:rmses}) is consistent with this possibility.  The present analysis cannot definitively distinguish between these two scenarios.

Nevertheless, the observation that routing patterns show opposite tendencies for dR/dM and M, while $\kappa_2^\tau$ shows no clear preference, is a robust descriptive finding and suggests that the gate does not behave identically across the three targets.
If the gate were simply routing randomly or poorly for all properties, one would not expect property-specific directional patterns.  This differentiation is consistent with the physical expectation that different neutron-star properties are encoded in different aspects of the post-merger spectral morphology.

The definitive test of whether non-peak spectral regions carry independent predictive information would require a controlled ablation study: comparing the RMSE of the ST-ANN model trained on the full $1$--$4\,$kHz spectrum against the same model trained on only the peak-containing $2$--$4\,$kHz band, and separately on only the sub-peak $1$--$2\,$kHz band.  A significant RMSE degradation when removing Region~I would directly demonstrate that the sub-peak region contributes information beyond what $f_\text{peak}$ and its immediate spectral environment provide, offering the strongest possible justification for full-spectrum inverse surrogates over $f_\text{peak}$-based empirical relations.  We defer this ablation study to future work.

\section{Consistency with empirical and EOS-dependent relations}

A useful way to assess the physical consistency of the predicted neutron-star properties is to compare them with relations already established in the literature. In particular, quasi-universal relations are especially valuable because they are approximately EOS-insensitive and therefore provide a stringent benchmark for any inverse model. Agreement with such relations indicates that the model does not merely reproduce individual target values, but also captures the underlying structure of the post-merger parameter space.

\begin{figure*}[ht!]
    \centering
    \includegraphics[width=0.75\textwidth]{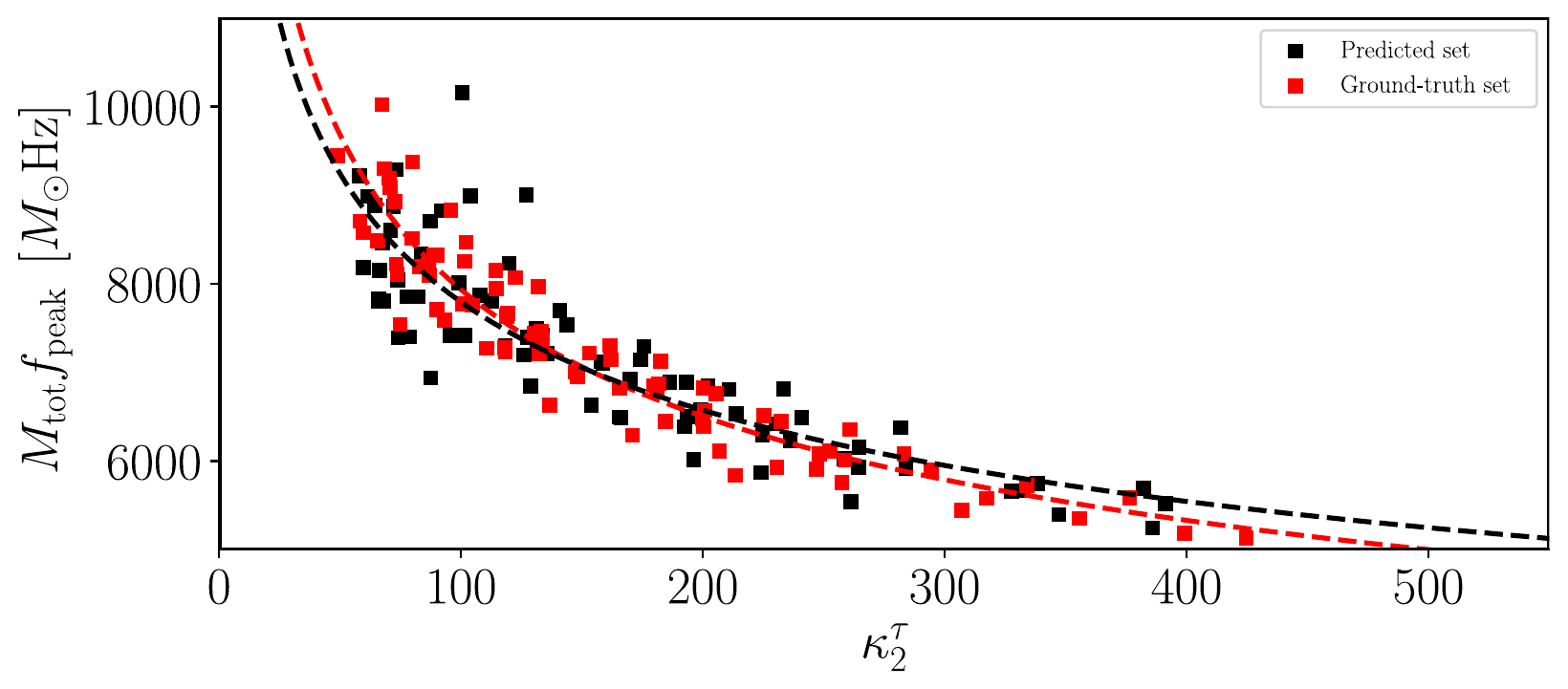}
    \caption{Validation of the inverse ST-ANN model in the $M_{\rm tot}f_{\rm peak}$--$\kappa_2^\tau$ plane. Predicted samples (black) and ground-truth samples (red) are shown together with the empirical fit of Vretinaris et al., illustrating that the network predictions recover the global trend followed by the reference data.}
    \label{fig:Mf2_vs_kappa_superposed}
\end{figure*}

As a first consistency test, we consider the empirical relation introduced by Vretinaris, Stergioulas  \& Bauswein \cite{vretinaris2020}, specifically their 41st fit, which we also used in our previous work for shifting and recovering the spectra in \cite{Pesios2024}. This relation differs from the one adopted in Easter et al. \cite{Easter2019} and provides a convenient EOS-agnostic reference for our predicted dataset. In Fig.~\ref{fig:Mf2_vs_kappa_superposed}, we compare the predicted values with the corresponding ground-truth dataset in the $M_{\rm tot}f_{\rm peak}$--$\kappa_2^\tau$ plane, together with the empirical fit. The level of agreement between the predicted and reference samples confirms that the predictions of the ST-ANN model are consistent with the global trend encoded by this quasi-universal relation.

\begin{figure}[ht!]
    \centering
    \includegraphics[width=\columnwidth]{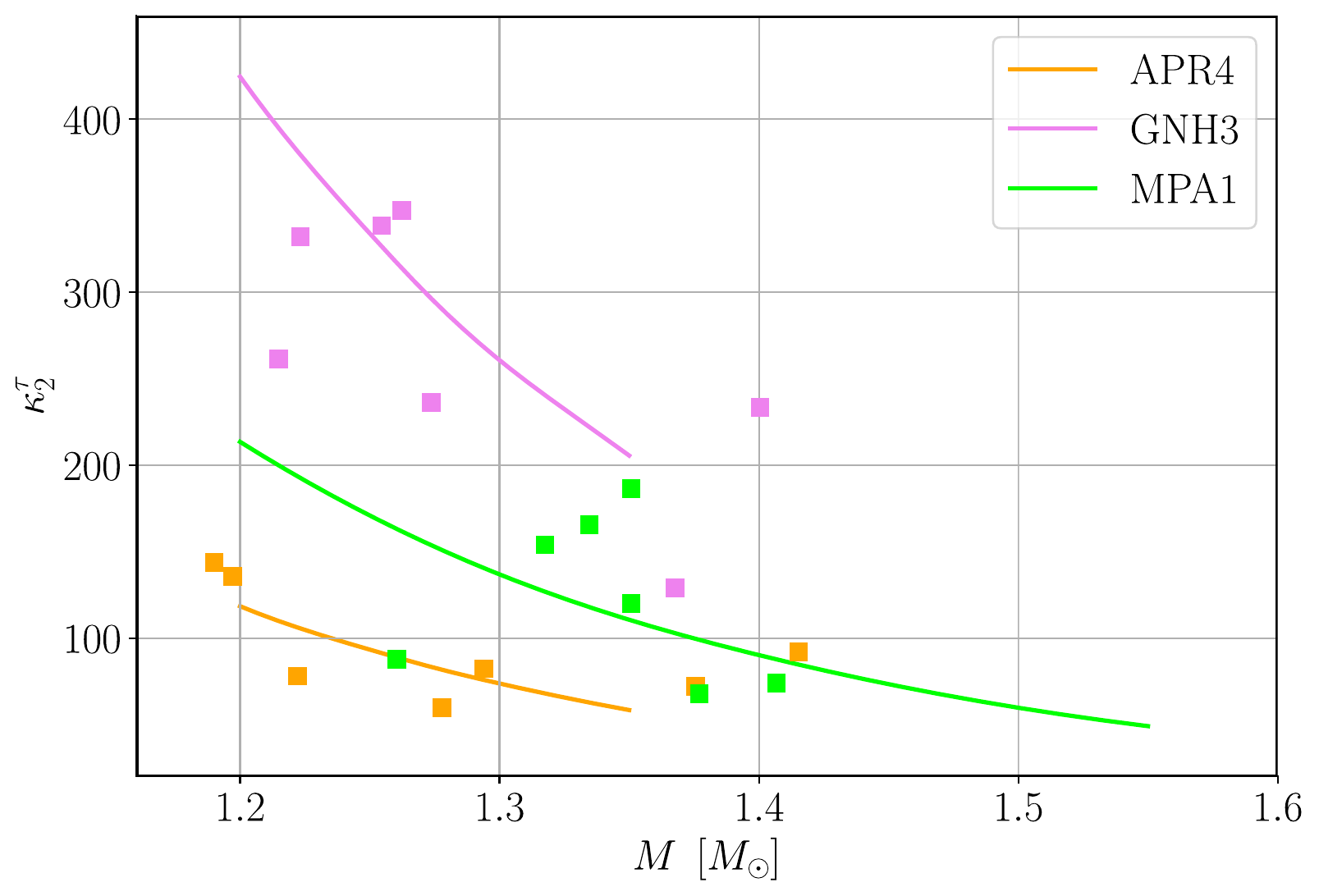}
    \caption{Validation of the inverse ST-ANN model in the $M$--$\kappa_2^\tau$ plane. Predicted values for three representative EOSs (APR4, GNH3, and MPA1) are compared with the corresponding ground-truth curves, showing promising prospects for EOS recovery from post-merger GW spectra, using ANNs.}
    \label{fig:comp_element_m_kappa}
\end{figure}

As a complementary consistency test, in Fig.~\ref{fig:comp_element_m_kappa} we examine whether the model also preserves EOS-dependent behavior. There, we compare the predicted $M$--$\kappa_2^\tau$ relation with the corresponding ground-truth curves for three representative EOSs, namely APR4, GNH3, and MPA1. This comparison is more demanding than the universal-relation test, since it probes the model's ability to recover EOS-specific structure rather than only the global behavior shared across different EOSs. 

The level of  agreement obtained in this case indicates that the predicted parameters remain physically consistent not only at the universal level, but also at the level of individual EOS families. A more detailed comparison with the corresponding ground-truth values is presented in Appendix~\ref{AppendixA}. This is a promising result for future work, where a substantially larger training set could enable our method to place direct constraints on the EOS from the post-merger GW spectrum, independently of the EOS information inferred from the inspiral phase. In the case that discrepancies with the EOS inferred from the inspiral phase are found, one could examine effects such as phase transitions or spontaneous a scalarization as possible explanations.

\section{CONCLUSIONS} \label{section_iv}

In this work, we presented five models that are used for predicting component neutron star properties out of the produced post-merger Binary Neutron Star spectrum, three of them being 
Deep Learning models and two based on Multivariate Linear Regression algebraic manipulations, such as OLS and pseudo-inverse concepts to compare with. 

The best model is the one where we used a individual ANN to predict each property separately in an ensemble fashion, namely ST-ANNs as described above, while the rest deep learning models exhibited comparable results. As for the MLR models, they do not seem to achieve similar behavior as the deep learning ones, with the PI-MLR being currently the worse, due to the fact that our previous work methodology is partially reversible for the particular model, due to the complications we stated in Section \ref{comparisons}, since 3rd generation detectors are under construction and we lack observations.

The superior performance of the ST-ANN framework indicates that the regression from post-merger spectra to neutron star properties is best captured by specialized models rather than shared representations. The degradation seen in Multi-Task learning suggests negative transfer, where the spectral features required for resolving mass may conflict with those needed for tidal deformability. Furthermore, the under-performance of the frequency-partitioned MoE model highlights the importance of global spectral context; while the dominant peak is crucial, the holistic spectral shape carries auxiliary information that is lost when the input is fragmented into sub-bands.

Our current work is a natural extension of our previous work \cite{Pesios2024}, which treated the NS properties as input to a regression-based scheme to predict the BNS spectra, by exchanging input and output. A natural concern is the large nominal number of trainable parameters compared with the 77 available training spectra. We emphasize, however, that the present study is a proof of concept, intended to test whether an ANN-based inverse surrogate is feasible and competitive with simpler algebraic baselines on the currently available catalog. Although the spectra are represented on a 370-bin frequency grid, they are generated by a restricted family of physically structured post-merger signals and are therefore expected to occupy a much smaller effective region of the input space than an arbitrary 370-dimensional dataset. In this setting, the network acts primarily as a flexible nonlinear regressor on structured inputs rather than as an unconstrained fit to generic high-dimensional data. We further limit effective model capacity through dropout, Gaussian noise injection, and early stopping, and we assess performance strictly in a fold-wise manner under cross-validation. We therefore do not claim that the present dataset is sufficient for definitive population-level generalization; rather, the results indicate that meaningful interpolation within the current simulation set is already possible, while larger future numerical-relativity catalogs will be needed to establish robustness more broadly.


Our method could be employed in the future as a way to obtain a fast parameter estimation for mass and tidal deformability, directly from the post-merger spectrum, for potentially discovered post-merger BNS signals using 3G detectors, such as Einstein Telescope or Cosmic Explorer.  It assumes that the source properties and the cold EOS properties up to densities encounter in pre-merger neutron stars are already well-constrained from the high-SNR inspiral part. The independent parameter estimation using the post-merger part would then serve as a probe for phase transitions and/or the thermal properties at central densities inaccessible to inspiral studies. 

Moreover, concerning the results obtained using the ST-MoE-ANN set models, which predict each property separately using a MoE model by examining each time a different part of the spectrum as a partition, there are indications that these properties are distinguishable enough by considering a shorter range of the spectrum which contains the $f_\mathrm{peak}$. As we have mentioned, the post-merger signal will not be visible at once, but gradually, when detectors will be improving their sensitivity curve and, therefore, someone can possibly draw some conclusions about the properties by observing the dominant feature, if it is distinguishable enough.


The choice of the three-component output vector $[dR/dM, M, \kappa_2^\tau]$ is not unique; as discussed in \cite{Pesios2024}, alternative or additional neutron star properties could serve as regression targets, and nonlinear terms could be incorporated to capture more complex relationships. Similarly, the post-merger dynamics can be influenced by a range of physical effects not considered here, including magnetic fields, viscosity, neutrino transport, phase transitions, and finite-temperature microphysics \cite{Kiuchi:2017zzg,Kiuchi:2015sga,Shibata:2017xht,Ciolfi:2019fie,Radice:2017zta,Giacomazzo:2014qba,Kiuchi:2014hja,PhysRevD.97.064016,blacker_constraining_2020,raithel_detectability_2024,prakash_detectability_2024,jacobi_effects_2023,most_emergence_2024,suarez-fontanella_gravitational_2024,fujimoto_gravitational_2023,rivieccio_gravitational-wave_2024,bauswein_identifying_2019,miravet-tenes_identifying_2024,vijayan_impact_2023,raithel_influence_2023,most_deconfinement_2020,bamber_post-merger_2024,espino_revealing_2024,prakash_signatures_2021,most_signatures_2019,fields_thermal_2023}, as well as possible deviations from General Relativity \cite{kedia_binary_2022,east_binary_2022,lam_binary_2024,kuan_binary_2023,shibata_coalescence_2014,staykov_differentially_2023,kuan_dynamical_2023,palenzuela_dynamical_2014,sagunski_neutron_2018,barausse_neutron-star_2013}. In this work we deliberately restricted the scope to a basic equal-mass hydrodynamics setting in order to isolate and validate the inverse regression methodology itself; extending the framework to incorporate these additional effects is a natural direction for future work once larger and more diverse simulation catalogs become available.

The present analysis uses noise-free spectra in order to establish the viability of the inverse mapping under ideal conditions; a dedicated follow-up study will address realistic detectability by injecting signals into simulated detector noise curves for specific next-generation instruments.

Ultimately, this work demonstrates that deep learning frameworks can bypass the computational bottlenecks of conventional parameter estimation, offering a viable pathway for real-time analysis of binary neutron star remnants in the next generation of gravitational-wave astronomy.

\begin{acknowledgments}

We are grateful to Grigorios Papigkiotis, Alexandra Koloniari and Georgios Vardakas for comments. 
This research work was supported by the Hellenic Foundation for Research and Innovation (HFRI) under the 3rd Call for HFRI PhD Fellowships (Fellowship Number: 6836). Virgo is funded through the European Gravitational Observatory (EGO), by the French Centre National de Recherche Scientifique (CNRS), the Italian Istituto Nazionale di Fisica Nucleare (INFN) and the Dutch Nikhef, with contributions by institutions from Belgium, Germany, Greece, Hungary, Ireland, Japan, Monaco, Poland, Portugal, Spain. Results presented in this work have been produced using the AUTh Computing Infrastructure and Resources. DP would like to acknowledge the support provided by the Scientific Computing Office throughout the progress of this research work and, especially, during its beginning.
\newline
\end{acknowledgments}

\appendix

\section{Additional validation plots and diagnostics}
\label{AppendixA}

\begin{figure*}[ht!]
    \begin{minipage}{\textwidth}
        \begin{minipage}{0.48\textwidth}
          \includegraphics[width=\linewidth,left]{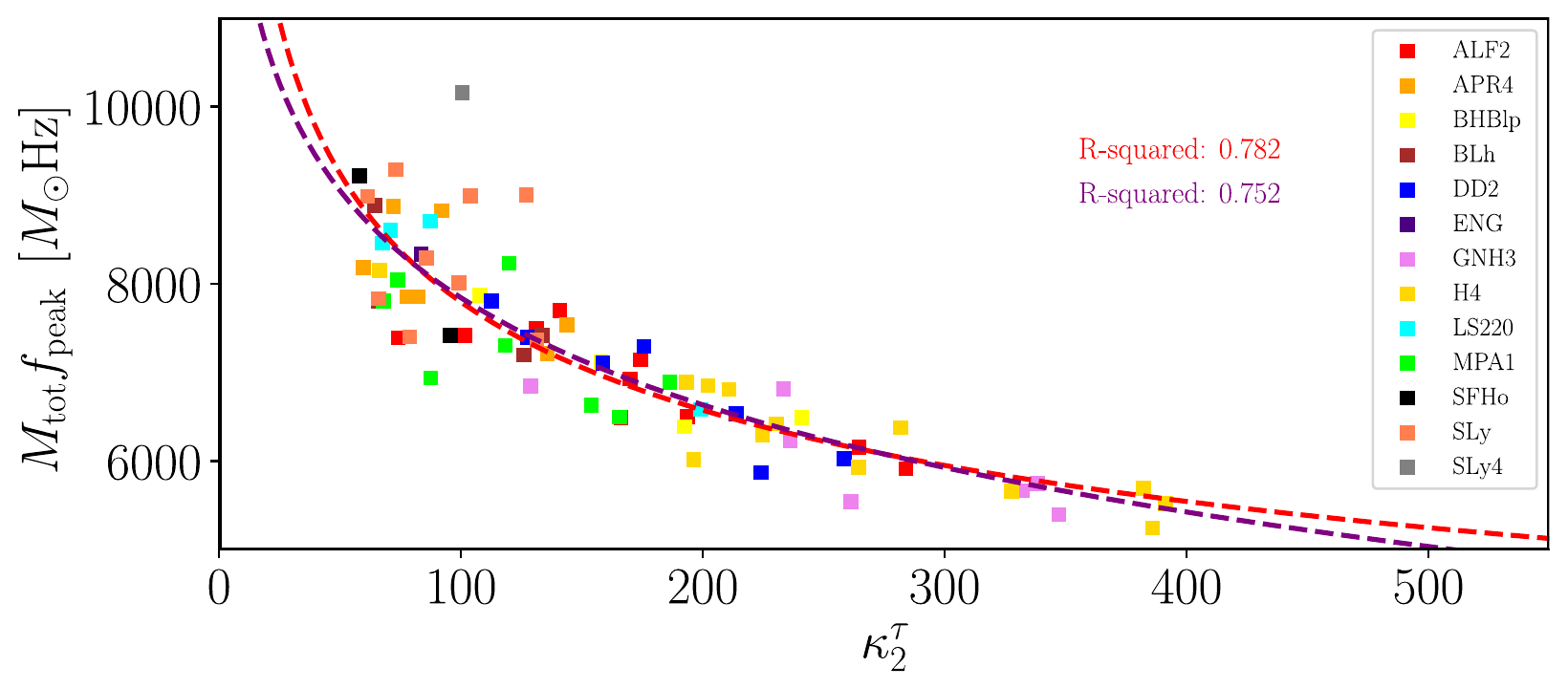}
        \end{minipage}
        \begin{minipage}{0.48\textwidth}
          \includegraphics[width=\linewidth,right]{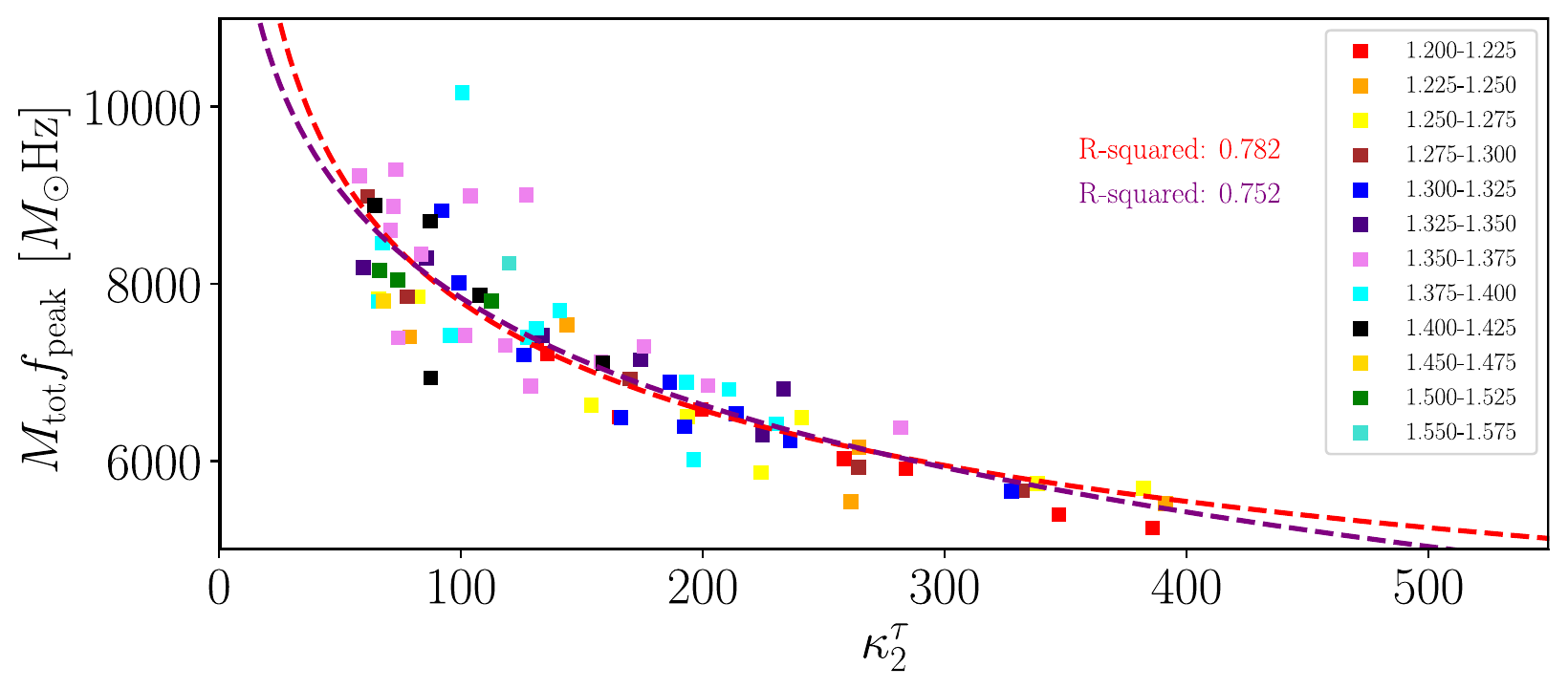}
        \end{minipage}%
    \end{minipage}
    
    \begin{minipage}{\textwidth}
        \begin{minipage}{0.48\textwidth}
          \includegraphics[width=\linewidth, left]{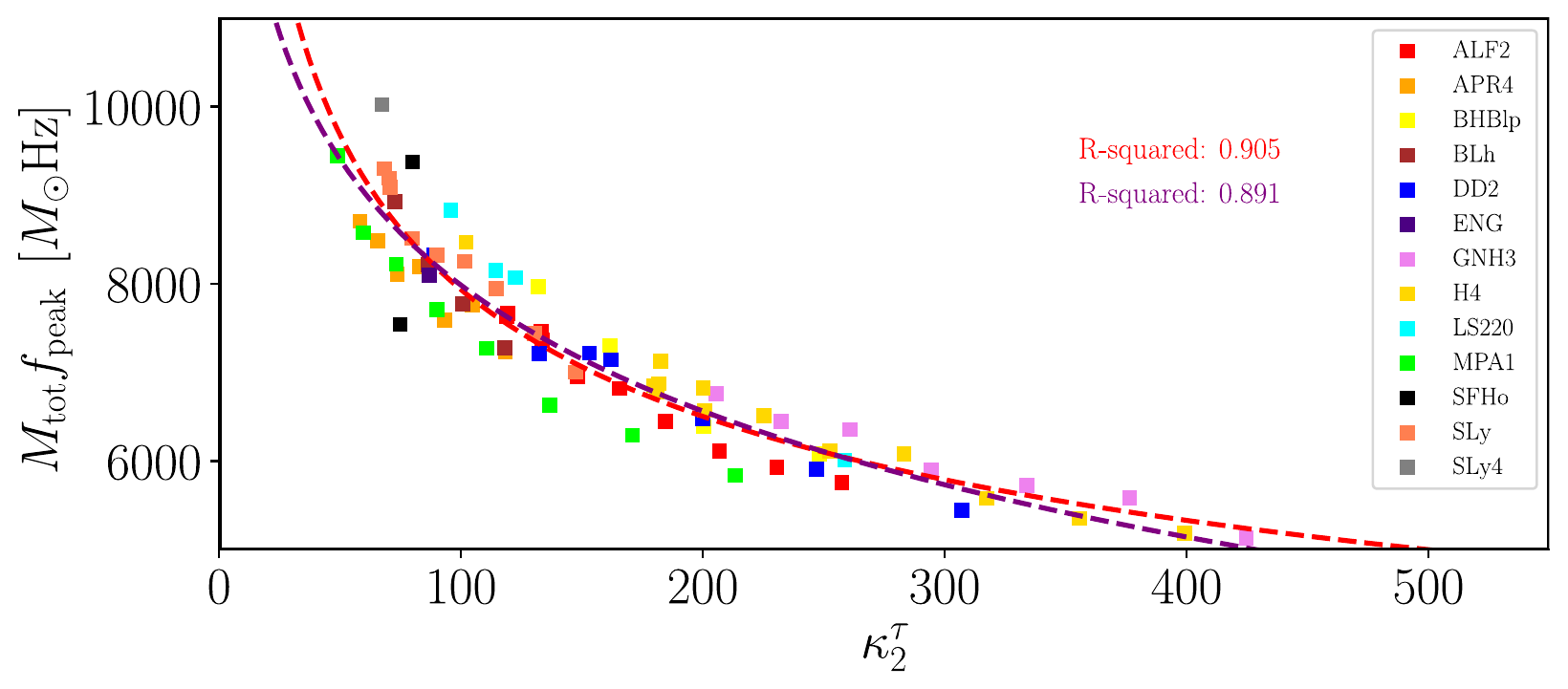}
        \end{minipage}
        \begin{minipage}{0.48\textwidth}
          \includegraphics[width=\linewidth, right]{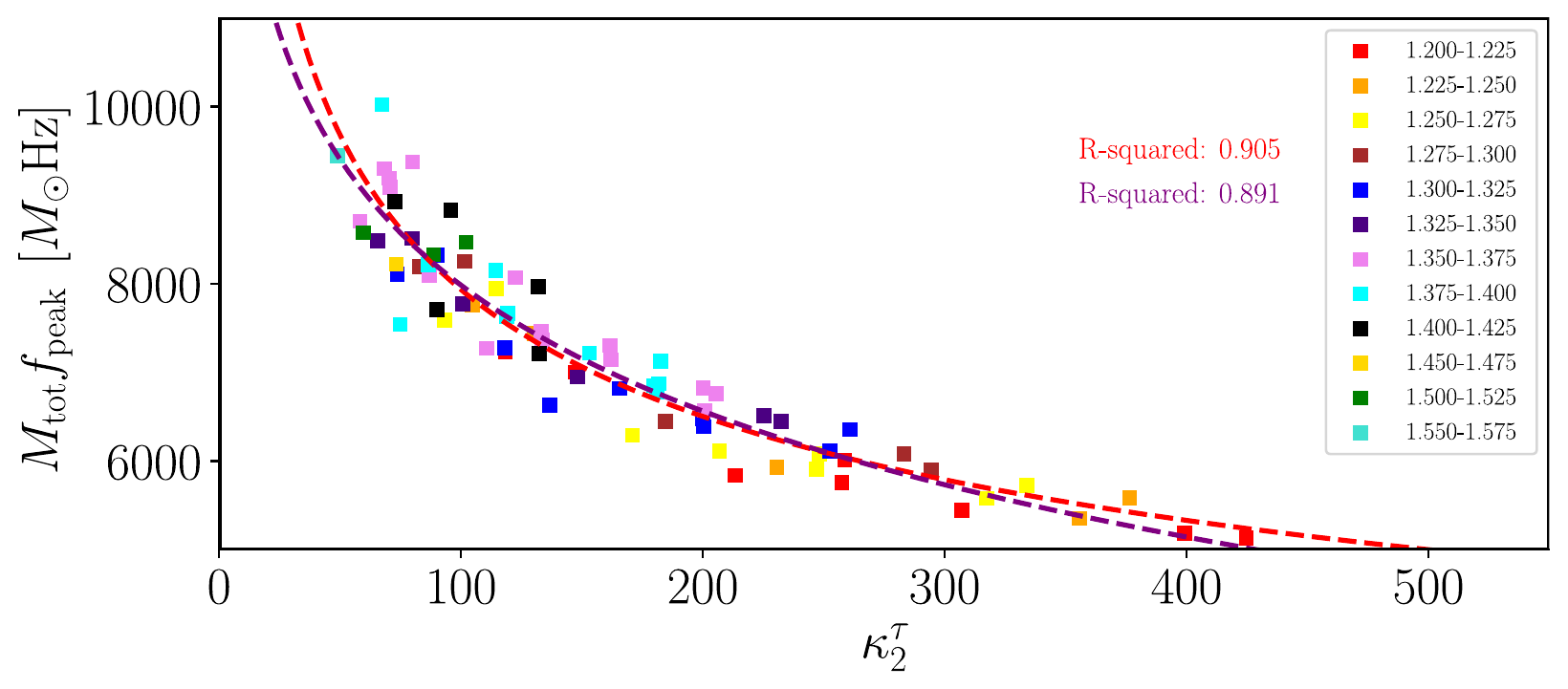}
        \end{minipage}
    \end{minipage}
	\caption{In these figures we depict the fitting of two conceptually different universal neutron star empirical relations, the purple one from Vretinaris et al. work \cite{vretinaris2020} used in \cite{Pesios2024}, and the other red from the  Chakravarti and Andersson work \cite{chakravarti2020} which is basically a theoretical power-law assumption. We color the points according to the EOS they belong to (left panels), and  according to their mass (right panels). The upper panels show that the predicted dataset of $M_\mathrm{tot}f_\mathrm{peak}$ versus dimensionless $\kappa_2^\tau$ agrees well with both empirical relations,  which validates the predictions produced by our ANN model. The lower panels show the corresponding ground truth data with respect to the two empirical relations. }
    \label{fig:Mf2_vs_kappa}
\end{figure*}

Figure \ref{fig:Mf2_vs_kappa} shows, in addition to the Vretinaris et al. \cite{vretinaris2020} empirical fit, the fit introduced by Chakravarti and Andersson \cite{chakravarti2020}. Its determination coefficient $R^2$ is equal to 0.782 for our predicted dataset compared to 0.905 for the ground truth dataset, slightly better than the 41st relation of \cite{vretinaris2020}.

\begin{figure*}[ht!]
    \begin{minipage}{\textwidth}
        \begin{minipage}{0.32\textwidth}
          \includegraphics[width=\linewidth,left]{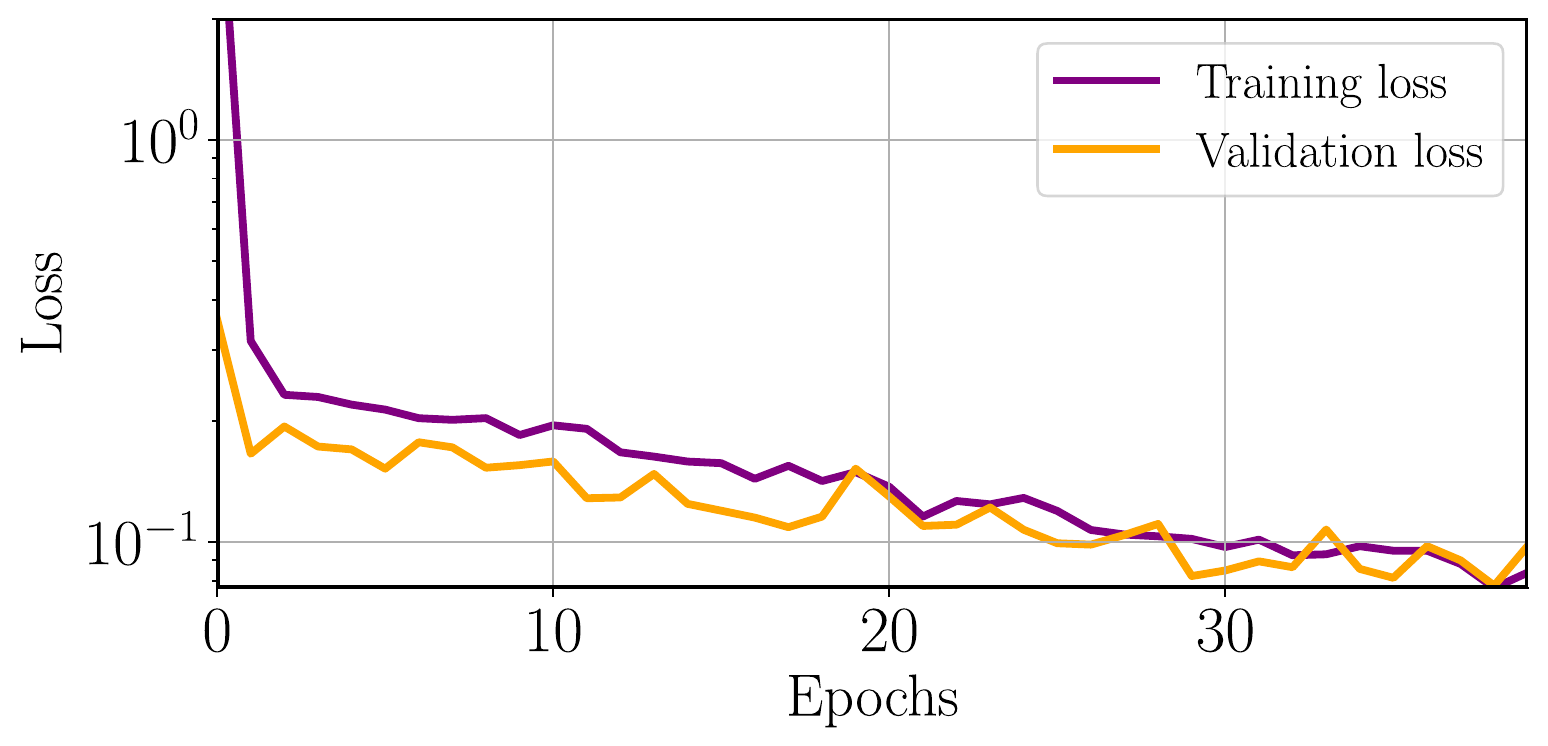}
        \end{minipage}
        \begin{minipage}{0.32\textwidth}
          \includegraphics[width=\linewidth,right]{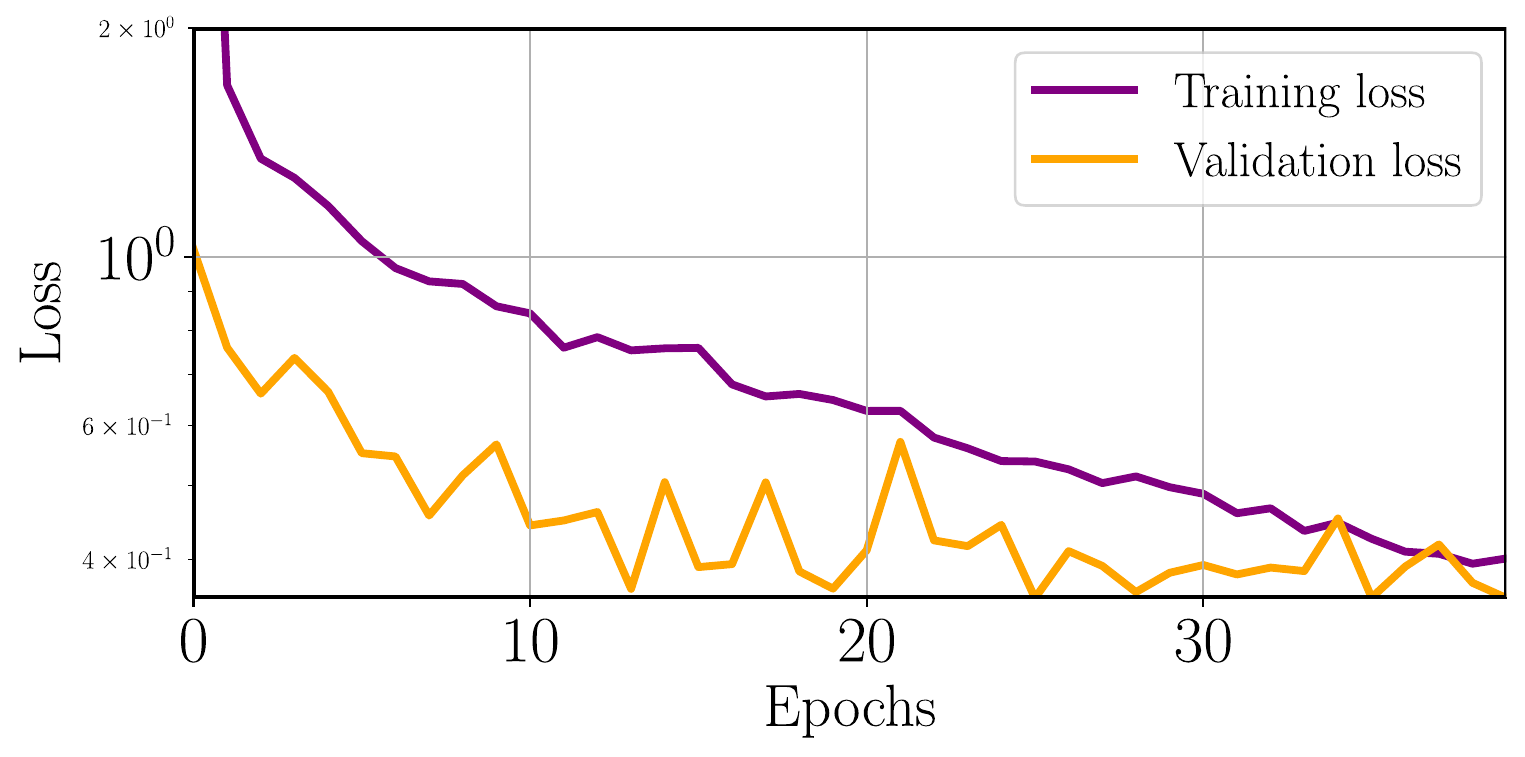}
        \end{minipage}%
        \begin{minipage}{0.32\textwidth}
          \includegraphics[width=\linewidth,right]{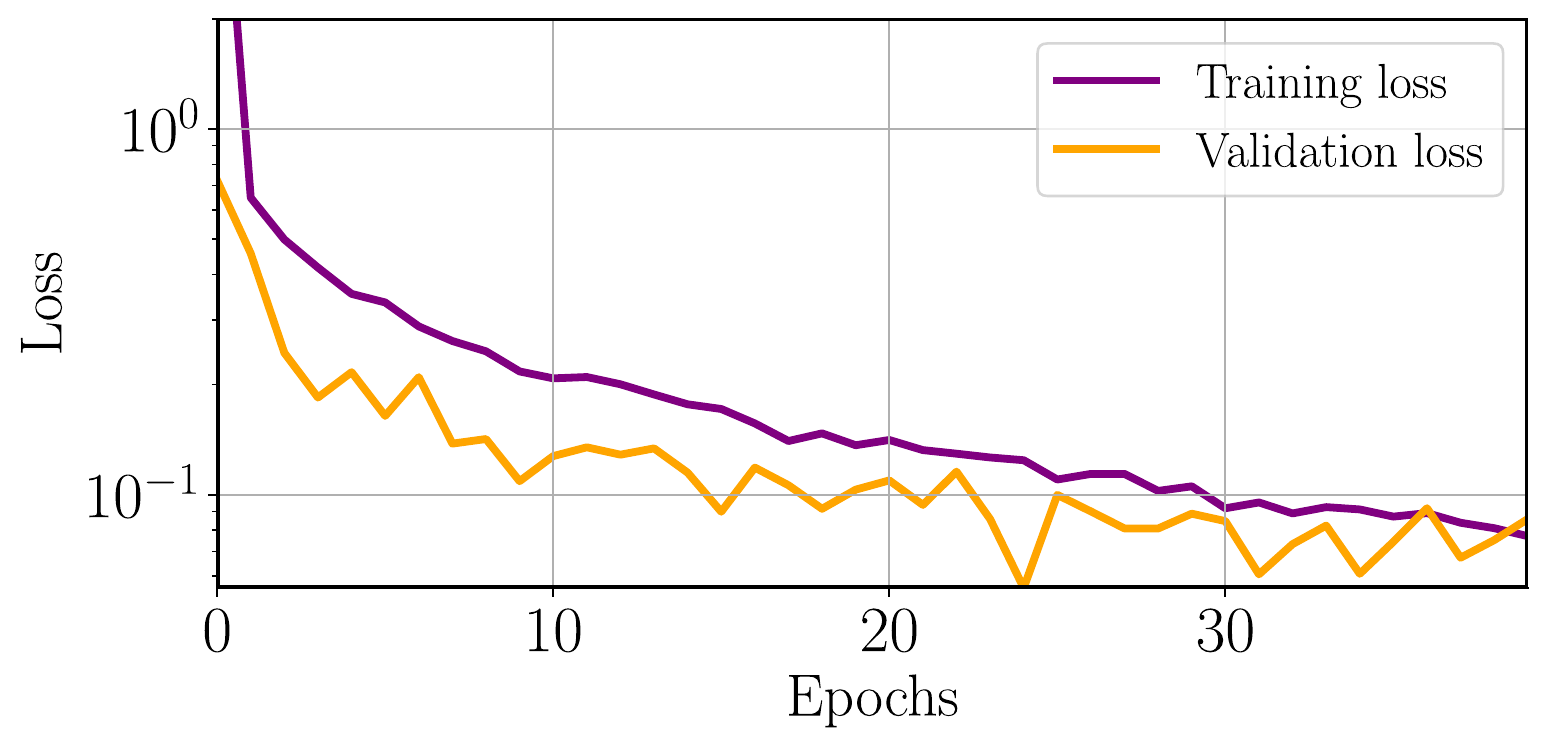}
        \end{minipage}%
    \end{minipage}
    \caption{Averaged validation loss curves and  averaged training loss curve, after performing LOO-CV among all folds. Training is halted when they are about to intersect each other to avoid overfitting effects.}
    \label{fig:loss_curves}
\end{figure*}

In Figure \ref{fig:loss_curves}, we show the {\it average loss curves} among 77 trained models following a LOO-CV evaluation process. We can see that the averaged validation loss curves are below the training loss, exhibiting a monotonic behavior,  for all three predicted neutron-star properties. Early stopping was enforced to avoid entering the overfitting regime. 
\newline

\bibliography{apssamp}

\providecommand{\noopsort}[1]{}\providecommand{\singleletter}[1]{#1}%
\begin{thebibliography}{122}%
\makeatletter
\providecommand \@ifxundefined [1]{%
 \@ifx{#1\undefined}
}%
\providecommand \@ifnum [1]{%
 \ifnum #1\expandafter \@firstoftwo
 \else \expandafter \@secondoftwo
 \fi
}%
\providecommand \@ifx [1]{%
 \ifx #1\expandafter \@firstoftwo
 \else \expandafter \@secondoftwo
 \fi
}%
\providecommand \natexlab [1]{#1}%
\providecommand \enquote  [1]{``#1''}%
\providecommand \bibnamefont  [1]{#1}%
\providecommand \bibfnamefont [1]{#1}%
\providecommand \citenamefont [1]{#1}%
\providecommand \href@noop [0]{\@secondoftwo}%
\providecommand \href [0]{\begingroup \@sanitize@url \@href}%
\providecommand \@href[1]{\@@startlink{#1}\@@href}%
\providecommand \@@href[1]{\endgroup#1\@@endlink}%
\providecommand \@sanitize@url [0]{\catcode `\\12\catcode `\$12\catcode
  `\&12\catcode `\#12\catcode `\^12\catcode `\_12\catcode `\%12\relax}%
\providecommand \@@startlink[1]{}%
\providecommand \@@endlink[0]{}%
\providecommand \url  [0]{\begingroup\@sanitize@url \@url }%
\providecommand \@url [1]{\endgroup\@href {#1}{\urlprefix }}%
\providecommand \urlprefix  [0]{URL }%
\providecommand \Eprint [0]{\href }%
\providecommand \doibase [0]{https://doi.org/}%
\providecommand \selectlanguage [0]{\@gobble}%
\providecommand \bibinfo  [0]{\@secondoftwo}%
\providecommand \bibfield  [0]{\@secondoftwo}%
\providecommand \translation [1]{[#1]}%
\providecommand \BibitemOpen [0]{}%
\providecommand \bibitemStop [0]{}%
\providecommand \bibitemNoStop [0]{.\EOS\space}%
\providecommand \EOS [0]{\spacefactor3000\relax}%
\providecommand \BibitemShut  [1]{\csname bibitem#1\endcsname}%
\let\auto@bib@innerbib\@empty
\bibitem [{\citenamefont {Haensel}\ \emph {et~al.}(2007)\citenamefont
  {Haensel}, \citenamefont {Potekhin},\ and\ \citenamefont
  {Yakovlev}}]{haensel_potekhin_yakovlev_2007}%
  \BibitemOpen
  \bibfield  {author} {\bibinfo {author} {\bibfnamefont {P.}~\bibnamefont
  {Haensel}}, \bibinfo {author} {\bibfnamefont {A.~Y.}\ \bibnamefont
  {Potekhin}},\ and\ \bibinfo {author} {\bibfnamefont {D.~G.}\ \bibnamefont
  {Yakovlev}},\ }\href {https://doi.org/10.1007/978-0-387-47301-7} {\emph
  {\bibinfo {title} {Neutron Stars 1: Equation of State and Structure}}},\
  \bibinfo {series} {Astrophysics and Space Science Library}, Vol.\ \bibinfo
  {volume} {326}\ (\bibinfo  {publisher} {Springer},\ \bibinfo {address} {New
  York},\ \bibinfo {year} {2007})\BibitemShut {NoStop}%
\bibitem [{\citenamefont {Abac}\ \emph {et~al.}(2025)\citenamefont {Abac},
  \citenamefont {Abramo}, \citenamefont {Albanesi}, \citenamefont {Albertini},
  \citenamefont {Agapito}, \citenamefont {Agathos}, \citenamefont {Albertus},
  \citenamefont {Andersson} \emph {et~al.}}]{abac2025scienceeinsteintelescope}%
  \BibitemOpen
  \bibfield  {author} {\bibinfo {author} {\bibfnamefont {A.}~\bibnamefont
  {Abac}}, \bibinfo {author} {\bibfnamefont {R.}~\bibnamefont {Abramo}},
  \bibinfo {author} {\bibfnamefont {S.}~\bibnamefont {Albanesi}}, \bibinfo
  {author} {\bibfnamefont {A.}~\bibnamefont {Albertini}}, \bibinfo {author}
  {\bibfnamefont {A.}~\bibnamefont {Agapito}}, \bibinfo {author} {\bibfnamefont
  {M.}~\bibnamefont {Agathos}}, \bibinfo {author} {\bibfnamefont
  {C.}~\bibnamefont {Albertus}}, \bibinfo {author} {\bibfnamefont
  {N.}~\bibnamefont {Andersson}}, \emph {et~al.},\ }\href
  {https://arxiv.org/abs/2503.12263} {\bibinfo {title} {The science of the
  einstein telescope}} (\bibinfo {year} {2025}),\ \Eprint
  {https://arxiv.org/abs/2503.12263} {arXiv:2503.12263 [gr-qc]} \BibitemShut
  {NoStop}%
\bibitem [{\citenamefont {Bauswein}\ \emph {et~al.}(2016)\citenamefont
  {Bauswein}, \citenamefont {Stergioulas},\ and\ \citenamefont
  {Janka}}]{bauswein_exploring_2016}%
  \BibitemOpen
  \bibfield  {author} {\bibinfo {author} {\bibfnamefont {A.}~\bibnamefont
  {Bauswein}}, \bibinfo {author} {\bibfnamefont {N.}~\bibnamefont
  {Stergioulas}},\ and\ \bibinfo {author} {\bibfnamefont {H.-T.}\ \bibnamefont
  {Janka}},\ }\href {https://doi.org/10.1140/epja/i2016-16056-7} {\bibfield
  {journal} {\bibinfo  {journal} {European Physical Journal A}\ }\textbf
  {\bibinfo {volume} {52}},\ \bibinfo {pages} {56} (\bibinfo {year}
  {2016})}\BibitemShut {NoStop}%
\bibitem [{\citenamefont {Dietrich}\ \emph {et~al.}(2021)\citenamefont
  {Dietrich}, \citenamefont {Hinderer},\ and\ \citenamefont
  {Samajdar}}]{dietrich2021interpreting}%
  \BibitemOpen
  \bibfield  {author} {\bibinfo {author} {\bibfnamefont {T.}~\bibnamefont
  {Dietrich}}, \bibinfo {author} {\bibfnamefont {T.}~\bibnamefont {Hinderer}},\
  and\ \bibinfo {author} {\bibfnamefont {A.}~\bibnamefont {Samajdar}},\
  }\href@noop {} {\bibfield  {journal} {\bibinfo  {journal} {Gen. Relativity
  Gravit.}\ }\textbf {\bibinfo {volume} {53}},\ \bibinfo {pages} {27} (\bibinfo
  {year} {2021})}\BibitemShut {NoStop}%
\bibitem [{\citenamefont {Abbott}\ \emph {et~al.}(2019)\citenamefont {Abbott}
  \emph {et~al.}}]{abbott2019properties}%
  \BibitemOpen
  \bibfield  {author} {\bibinfo {author} {\bibfnamefont {B.~P.}\ \bibnamefont
  {Abbott}} \emph {et~al.},\ }\href@noop {} {\bibfield  {journal} {\bibinfo
  {journal} {Phys. Rev. X}\ }\textbf {\bibinfo {volume} {9}},\ \bibinfo {pages}
  {011001} (\bibinfo {year} {2019})}\BibitemShut {NoStop}%
\bibitem [{\citenamefont {Abbott}\ \emph {et~al.}(2018)\citenamefont {Abbott}
  \emph {et~al.}}]{abbott2018gw170817}%
  \BibitemOpen
  \bibfield  {author} {\bibinfo {author} {\bibfnamefont {B.~P.}\ \bibnamefont
  {Abbott}} \emph {et~al.},\ }\href@noop {} {\bibfield  {journal} {\bibinfo
  {journal} {Phys. Rev. Lett.}\ }\textbf {\bibinfo {volume} {121}},\ \bibinfo
  {pages} {161101} (\bibinfo {year} {2018})}\BibitemShut {NoStop}%
\bibitem [{\citenamefont {Iacovelli}\ \emph
  {et~al.}(2023{\natexlab{a}})\citenamefont {Iacovelli}, \citenamefont
  {Mancarella}, \citenamefont {Mondal}, \citenamefont {Puecher}, \citenamefont
  {Dietrich}, \citenamefont {Gulminelli}, \citenamefont {Maggiore},\ and\
  \citenamefont {Oertel}}]{iacovelli2023nuclear}%
  \BibitemOpen
  \bibfield  {author} {\bibinfo {author} {\bibfnamefont {F.}~\bibnamefont
  {Iacovelli}}, \bibinfo {author} {\bibfnamefont {M.}~\bibnamefont
  {Mancarella}}, \bibinfo {author} {\bibfnamefont {C.}~\bibnamefont {Mondal}},
  \bibinfo {author} {\bibfnamefont {A.}~\bibnamefont {Puecher}}, \bibinfo
  {author} {\bibfnamefont {T.}~\bibnamefont {Dietrich}}, \bibinfo {author}
  {\bibfnamefont {F.}~\bibnamefont {Gulminelli}}, \bibinfo {author}
  {\bibfnamefont {M.}~\bibnamefont {Maggiore}},\ and\ \bibinfo {author}
  {\bibfnamefont {M.}~\bibnamefont {Oertel}},\ }\href@noop {} {\bibfield
  {journal} {\bibinfo  {journal} {Phys. Rev. D}\ }\textbf {\bibinfo {volume}
  {108}},\ \bibinfo {pages} {122006} (\bibinfo {year}
  {2023}{\natexlab{a}})}\BibitemShut {NoStop}%
\bibitem [{\citenamefont {Friedman}\ and\ \citenamefont
  {Stergioulas}(2020)}]{friedman_astrophysical_2020}%
  \BibitemOpen
  \bibfield  {author} {\bibinfo {author} {\bibfnamefont {J.~L.}\ \bibnamefont
  {Friedman}}\ and\ \bibinfo {author} {\bibfnamefont {N.}~\bibnamefont
  {Stergioulas}},\ }\href {https://doi.org/10.1142/s0218271820410151}
  {\bibfield  {journal} {\bibinfo  {journal} {International Journal of Modern
  Physics D}\ }\textbf {\bibinfo {volume} {29}},\ \bibinfo {pages} {2041015}
  (\bibinfo {year} {2020})}\BibitemShut {NoStop}%
\bibitem [{\citenamefont {Baiotti}\ and\ \citenamefont
  {Rezzolla}(2017)}]{baiotti_binary_2017}%
  \BibitemOpen
  \bibfield  {author} {\bibinfo {author} {\bibfnamefont {L.}~\bibnamefont
  {Baiotti}}\ and\ \bibinfo {author} {\bibfnamefont {L.}~\bibnamefont
  {Rezzolla}},\ }\href {https://doi.org/10.1088/1361-6633/aa67bb} {\bibfield
  {journal} {\bibinfo  {journal} {Reports on Progress in Physics}\ }\textbf
  {\bibinfo {volume} {80}},\ \bibinfo {pages} {096901} (\bibinfo {year}
  {2017})},\ \bibinfo {note} {publisher: IOP Publishing}\BibitemShut {NoStop}%
\bibitem [{\citenamefont {Baiotti}(2019)}]{baiotti_gravitational_2019}%
  \BibitemOpen
  \bibfield  {author} {\bibinfo {author} {\bibfnamefont {L.}~\bibnamefont
  {Baiotti}},\ }\href {https://doi.org/10.1016/j.ppnp.2019.103714} {\bibfield
  {journal} {\bibinfo  {journal} {Progress in Particle and Nuclear Physics}\
  }\textbf {\bibinfo {volume} {109}},\ \bibinfo {pages} {103714} (\bibinfo
  {year} {2019})},\ \bibinfo {note} {\_eprint: 1907.08534}\BibitemShut
  {NoStop}%
\bibitem [{\citenamefont {{Hild S., et
  al.}}(2011)}]{hild_s_et_al_sensitivity_2011}%
  \BibitemOpen
  \bibfield  {author} {\bibinfo {author} {\bibnamefont {{Hild S., et al.}}},\
  }\href {https://doi.org/10.1088/0264-9381/28/9/094013} {\bibfield  {journal}
  {\bibinfo  {journal} {Classical and Quantum Gravity}\ }\textbf {\bibinfo
  {volume} {28}},\ \bibinfo {pages} {094013} (\bibinfo {year}
  {2011})}\BibitemShut {NoStop}%
\bibitem [{\citenamefont {Ascenzi}\ \emph {et~al.}(2024)\citenamefont
  {Ascenzi}, \citenamefont {Graber},\ and\ \citenamefont
  {Rea}}]{ascenzi_neutron-star_2024}%
  \BibitemOpen
  \bibfield  {author} {\bibinfo {author} {\bibfnamefont {S.}~\bibnamefont
  {Ascenzi}}, \bibinfo {author} {\bibfnamefont {V.}~\bibnamefont {Graber}},\
  and\ \bibinfo {author} {\bibfnamefont {N.}~\bibnamefont {Rea}},\ }\href
  {https://doi.org/10.1016/j.astropartphys.2024.102935} {\bibfield  {journal}
  {\bibinfo  {journal} {Astroparticle Physics}\ }\textbf {\bibinfo {volume}
  {158}},\ \bibinfo {pages} {102935} (\bibinfo {year} {2024})},\ \bibinfo
  {note} {\_eprint: 2401.14930}\BibitemShut {NoStop}%
\bibitem [{\citenamefont {Bauswein}\ \emph {et~al.}(2020)\citenamefont
  {Bauswein}, \citenamefont {Blacker}, \citenamefont {Vijayan}, \citenamefont
  {Stergioulas}, \citenamefont {Chatziioannou}, \citenamefont {Clark},
  \citenamefont {Bastian}, \citenamefont {Blaschke}, \citenamefont {Cierniak},\
  and\ \citenamefont {Fischer}}]{bauswein_equation_2020}%
  \BibitemOpen
  \bibfield  {author} {\bibinfo {author} {\bibfnamefont {A.}~\bibnamefont
  {Bauswein}}, \bibinfo {author} {\bibfnamefont {S.}~\bibnamefont {Blacker}},
  \bibinfo {author} {\bibfnamefont {V.}~\bibnamefont {Vijayan}}, \bibinfo
  {author} {\bibfnamefont {N.}~\bibnamefont {Stergioulas}}, \bibinfo {author}
  {\bibfnamefont {K.}~\bibnamefont {Chatziioannou}}, \bibinfo {author}
  {\bibfnamefont {J.~A.}\ \bibnamefont {Clark}}, \bibinfo {author}
  {\bibfnamefont {N.-U.~F.}\ \bibnamefont {Bastian}}, \bibinfo {author}
  {\bibfnamefont {D.~B.}\ \bibnamefont {Blaschke}}, \bibinfo {author}
  {\bibfnamefont {M.}~\bibnamefont {Cierniak}},\ and\ \bibinfo {author}
  {\bibfnamefont {T.}~\bibnamefont {Fischer}},\ }\bibfield  {journal} {\bibinfo
   {journal} {Physical Review Letters}\ }\textbf {\bibinfo {volume} {125}},\
  \href {https://doi.org/10.1103/physrevlett.125.141103}
  {10.1103/physrevlett.125.141103} (\bibinfo {year} {2020}),\ \bibinfo {note}
  {publisher: American Physical Society (APS)}\BibitemShut {NoStop}%
\bibitem [{\citenamefont {Haster}\ \emph {et~al.}(2020)\citenamefont {Haster},
  \citenamefont {Chatziioannou}, \citenamefont {Bauswein},\ and\ \citenamefont
  {Clark}}]{haster_inference_2020}%
  \BibitemOpen
  \bibfield  {author} {\bibinfo {author} {\bibfnamefont {C.-J.}\ \bibnamefont
  {Haster}}, \bibinfo {author} {\bibfnamefont {K.}~\bibnamefont
  {Chatziioannou}}, \bibinfo {author} {\bibfnamefont {A.}~\bibnamefont
  {Bauswein}},\ and\ \bibinfo {author} {\bibfnamefont {J.~A.}\ \bibnamefont
  {Clark}},\ }\bibfield  {journal} {\bibinfo  {journal} {Physical Review
  Letters}\ }\textbf {\bibinfo {volume} {125}},\ \href
  {https://doi.org/10.1103/physrevlett.125.261101}
  {10.1103/physrevlett.125.261101} (\bibinfo {year} {2020}),\ \bibinfo {note}
  {publisher: American Physical Society (APS)}\BibitemShut {NoStop}%
\bibitem [{\citenamefont {Abbott}\ and\ \citenamefont
  {{others}}(2016)}]{abbott_observation_2016}%
  \BibitemOpen
  \bibfield  {author} {\bibinfo {author} {\bibfnamefont {B.~P.}\ \bibnamefont
  {Abbott}}\ and\ \bibinfo {author} {\bibnamefont {{others}}},\ }\href
  {https://doi.org/10.1103/PhysRevLett.116.061102} {\bibfield  {journal}
  {\bibinfo  {journal} {Phys. Rev. Lett.}\ }\textbf {\bibinfo {volume} {116}},\
  \bibinfo {pages} {061102} (\bibinfo {year} {2016})},\ \bibinfo {note}
  {\_eprint: 1602.03837}\BibitemShut {NoStop}%
\bibitem [{\citenamefont {L{"u}ck}\ \emph {et~al.}(2022)\citenamefont
  {L{"u}ck}, \citenamefont {Smith},\ and\ \citenamefont
  {Punturo}}]{luck2022third}%
  \BibitemOpen
  \bibfield  {author} {\bibinfo {author} {\bibfnamefont {H.}~\bibnamefont
  {L{"u}ck}}, \bibinfo {author} {\bibfnamefont {J.}~\bibnamefont {Smith}},\
  and\ \bibinfo {author} {\bibfnamefont {M.}~\bibnamefont {Punturo}},\ }in\
  \href@noop {} {\emph {\bibinfo {booktitle} {Handbook of Gravitational Wave
  Astronomy}}}\ (\bibinfo  {publisher} {Springer},\ \bibinfo {year} {2022})\
  pp.\ \bibinfo {pages} {1--18}\BibitemShut {NoStop}%
\bibitem [{\citenamefont {Collaboration}\ and\ \citenamefont
  {Harms}(2017)}]{Abbott_2017}%
  \BibitemOpen
  \bibfield  {author} {\bibinfo {author} {\bibfnamefont {L.~S.}\ \bibnamefont
  {Collaboration}}\ and\ \bibinfo {author} {\bibfnamefont {J.}~\bibnamefont
  {Harms}},\ }\href {https://doi.org/10.1088/1361-6382/aa51f4} {\bibfield
  {journal} {\bibinfo  {journal} {Class. Quantum Gravity}\ }\textbf {\bibinfo
  {volume} {34}},\ \bibinfo {pages} {044001} (\bibinfo {year}
  {2017})}\BibitemShut {NoStop}%
\bibitem [{\citenamefont {Reitze}\ \emph {et~al.}(2019)\citenamefont {Reitze},
  \citenamefont {Adhikari}, \citenamefont {Ballmer}, \citenamefont {Barish},
  \citenamefont {Barsotti}, \citenamefont {Billingsley}, \citenamefont {Brown},
  \citenamefont {Chen}, \citenamefont {Coyne}, \citenamefont {Eisenstein} \emph
  {et~al.}}]{reitze2019cosmic}%
  \BibitemOpen
  \bibfield  {author} {\bibinfo {author} {\bibfnamefont {D.}~\bibnamefont
  {Reitze}}, \bibinfo {author} {\bibfnamefont {R.~X.}\ \bibnamefont
  {Adhikari}}, \bibinfo {author} {\bibfnamefont {S.}~\bibnamefont {Ballmer}},
  \bibinfo {author} {\bibfnamefont {B.}~\bibnamefont {Barish}}, \bibinfo
  {author} {\bibfnamefont {L.}~\bibnamefont {Barsotti}}, \bibinfo {author}
  {\bibfnamefont {G.}~\bibnamefont {Billingsley}}, \bibinfo {author}
  {\bibfnamefont {D.~A.}\ \bibnamefont {Brown}}, \bibinfo {author}
  {\bibfnamefont {Y.}~\bibnamefont {Chen}}, \bibinfo {author} {\bibfnamefont
  {D.}~\bibnamefont {Coyne}}, \bibinfo {author} {\bibfnamefont
  {R.}~\bibnamefont {Eisenstein}}, \emph {et~al.},\ }\href@noop {} {\bibfield
  {journal} {\bibinfo  {journal} {arXiv preprint}\ } (\bibinfo {year}
  {2019})},\ \Eprint {https://arxiv.org/abs/arXiv:1907.04833}
  {arXiv:1907.04833} \BibitemShut {NoStop}%
\bibitem [{\citenamefont {Zhang}\ \emph {et~al.}(2022)\citenamefont {Zhang},
  \citenamefont {Yang}, \citenamefont {Martynov}, \citenamefont {Schmidt},\
  and\ \citenamefont {Miao}}]{zhang_gravitational_2022}%
  \BibitemOpen
  \bibfield  {author} {\bibinfo {author} {\bibfnamefont {T.}~\bibnamefont
  {Zhang}}, \bibinfo {author} {\bibfnamefont {H.}~\bibnamefont {Yang}},
  \bibinfo {author} {\bibfnamefont {D.}~\bibnamefont {Martynov}}, \bibinfo
  {author} {\bibfnamefont {P.}~\bibnamefont {Schmidt}},\ and\ \bibinfo {author}
  {\bibfnamefont {H.}~\bibnamefont {Miao}},\ }\href
  {https://doi.org/10.48550/ARXIV.2212.12144} {\bibinfo {title} {A
  {Gravitational} {Wave} {Detector} for {Post} {Merger} {Neutron} {Stars}:
  {Beyond} the {Quantum} {Loss} {Limit} of {Michelson} {Fabry} {Perot}
  {Interferometer}}} (\bibinfo {year} {2022})\BibitemShut {NoStop}%
\bibitem [{\citenamefont {{Evans M., et
  al.}}(2021)}]{evans_m_et_al_horizon_2021}%
  \BibitemOpen
  \bibfield  {author} {\bibinfo {author} {\bibnamefont {{Evans M., et al.}}},\
  }\href {https://doi.org/10.48550/ARXIV.2109.09882} {\bibinfo {title} {A
  {Horizon} {Study} for {Cosmic} {Explorer}: {Science}, {Observatories}, and
  {Community}}} (\bibinfo {year} {2021})\BibitemShut {NoStop}%
\bibitem [{\citenamefont {Borhanian}\ and\ \citenamefont
  {Sathyaprakash}(2024)}]{borhanian_listening_2024}%
  \BibitemOpen
  \bibfield  {author} {\bibinfo {author} {\bibfnamefont {S.}~\bibnamefont
  {Borhanian}}\ and\ \bibinfo {author} {\bibfnamefont {B.~S.}\ \bibnamefont
  {Sathyaprakash}},\ }\href {https://doi.org/10.1103/PhysRevD.110.083040}
  {\bibfield  {journal} {\bibinfo  {journal} {{\textbackslash}prd}\ }\textbf
  {\bibinfo {volume} {110}},\ \bibinfo {pages} {083040} (\bibinfo {year}
  {2024})},\ \bibinfo {note} {\_eprint: 2202.11048}\BibitemShut {NoStop}%
\bibitem [{\citenamefont {{Ackley K., et
  al.}}(2020)}]{ackley_k_et_al_neutron_2020}%
  \BibitemOpen
  \bibfield  {author} {\bibinfo {author} {\bibnamefont {{Ackley K., et al.}}},\
  }\bibfield  {journal} {\bibinfo  {journal} {Publications of the Astronomical
  Society of Australia}\ }\textbf {\bibinfo {volume} {37}},\ \href
  {https://doi.org/10.1017/pasa.2020.39} {10.1017/pasa.2020.39} (\bibinfo
  {year} {2020}),\ \bibinfo {note} {publisher: Cambridge University Press
  (CUP)}\BibitemShut {NoStop}%
\bibitem [{\citenamefont {{Branchesi M., et
  al.}}(2023)}]{branchesi_m_et_al_science_2023}%
  \BibitemOpen
  \bibfield  {author} {\bibinfo {author} {\bibnamefont {{Branchesi M., et
  al.}}},\ }\href {https://doi.org/10.1088/1475-7516/2023/07/068} {\bibfield
  {journal} {\bibinfo  {journal} {Journal of Cosmology and Astroparticle
  Physics}\ }\textbf {\bibinfo {volume} {2023}}\bibfield  {number} {\bibinfo
  {number} { (07)},\ \bibinfo {pages} {068}},\ }\bibinfo {note} {publisher: IOP
  Publishing}\BibitemShut {NoStop}%
\bibitem [{\citenamefont {Takami}\ \emph {et~al.}(2014)\citenamefont {Takami},
  \citenamefont {Rezzolla},\ and\ \citenamefont
  {Baiotti}}]{takami_constraining_2014}%
  \BibitemOpen
  \bibfield  {author} {\bibinfo {author} {\bibfnamefont {K.}~\bibnamefont
  {Takami}}, \bibinfo {author} {\bibfnamefont {L.}~\bibnamefont {Rezzolla}},\
  and\ \bibinfo {author} {\bibfnamefont {L.}~\bibnamefont {Baiotti}},\ }\href
  {https://doi.org/10.1103/PhysRevLett.113.091104} {\bibfield  {journal}
  {\bibinfo  {journal} {Phys. Rev. Lett.}\ }\textbf {\bibinfo {volume} {113}},\
  \bibinfo {pages} {091104} (\bibinfo {year} {2014})},\ \bibinfo {note}
  {publisher: American Physical Society}\BibitemShut {NoStop}%
\bibitem [{\citenamefont {Bernuzzi}\ \emph {et~al.}(2025)\citenamefont
  {Bernuzzi}, \citenamefont {Magistrelli}, \citenamefont {Jacobi},
  \citenamefont {Logoteta}, \citenamefont {Perego},\ and\ \citenamefont
  {Radice}}]{10.1093_mnras_staf1147}%
  \BibitemOpen
  \bibfield  {author} {\bibinfo {author} {\bibfnamefont {S.}~\bibnamefont
  {Bernuzzi}}, \bibinfo {author} {\bibfnamefont {F.}~\bibnamefont
  {Magistrelli}}, \bibinfo {author} {\bibfnamefont {M.}~\bibnamefont {Jacobi}},
  \bibinfo {author} {\bibfnamefont {D.}~\bibnamefont {Logoteta}}, \bibinfo
  {author} {\bibfnamefont {A.}~\bibnamefont {Perego}},\ and\ \bibinfo {author}
  {\bibfnamefont {D.}~\bibnamefont {Radice}},\ }\href
  {https://doi.org/10.1093/mnras/staf1147} {\bibfield  {journal} {\bibinfo
  {journal} {Monthly Notices of the Royal Astronomical Society}\ }\textbf
  {\bibinfo {volume} {542}},\ \bibinfo {pages} {256} (\bibinfo {year}
  {2025})},\ \Eprint
  {https://arxiv.org/abs/https://academic.oup.com/mnras/article-pdf/542/1/256/63841147/staf1147.pdf}
  {https://academic.oup.com/mnras/article-pdf/542/1/256/63841147/staf1147.pdf}
  \BibitemShut {NoStop}%
\bibitem [{\citenamefont {Shibata}(2005)}]{shibata_constraining_2005}%
  \BibitemOpen
  \bibfield  {author} {\bibinfo {author} {\bibfnamefont {M.}~\bibnamefont
  {Shibata}},\ }\bibfield  {journal} {\bibinfo  {journal} {Physical Review
  Letters}\ }\textbf {\bibinfo {volume} {94}},\ \href
  {https://doi.org/10.1103/physrevlett.94.201101}
  {10.1103/physrevlett.94.201101} (\bibinfo {year} {2005}),\ \bibinfo {note}
  {publisher: American Physical Society (APS)}\BibitemShut {NoStop}%
\bibitem [{\citenamefont {Breschi}\ \emph {et~al.}(2022)\citenamefont
  {Breschi}, \citenamefont {Bernuzzi}, \citenamefont {Godzieba}, \citenamefont
  {Perego},\ and\ \citenamefont {Radice}}]{breschi_constraints_2022}%
  \BibitemOpen
  \bibfield  {author} {\bibinfo {author} {\bibfnamefont {M.}~\bibnamefont
  {Breschi}}, \bibinfo {author} {\bibfnamefont {S.}~\bibnamefont {Bernuzzi}},
  \bibinfo {author} {\bibfnamefont {D.}~\bibnamefont {Godzieba}}, \bibinfo
  {author} {\bibfnamefont {A.}~\bibnamefont {Perego}},\ and\ \bibinfo {author}
  {\bibfnamefont {D.}~\bibnamefont {Radice}},\ }\bibfield  {journal} {\bibinfo
  {journal} {Physical Review Letters}\ }\textbf {\bibinfo {volume} {128}},\
  \href {https://doi.org/10.1103/physrevlett.128.161102}
  {10.1103/physrevlett.128.161102} (\bibinfo {year} {2022}),\ \bibinfo {note}
  {publisher: American Physical Society (APS)}\BibitemShut {NoStop}%
\bibitem [{\citenamefont {Easter}\ \emph
  {et~al.}(2020{\natexlab{a}})\citenamefont {Easter}, \citenamefont {Ghonge},
  \citenamefont {Lasky}, \citenamefont {Casey}, \citenamefont {Clark},
  \citenamefont {Hernandez~Vivanco},\ and\ \citenamefont
  {Chatziioannou}}]{PhysRevD.102.043011}%
  \BibitemOpen
  \bibfield  {author} {\bibinfo {author} {\bibfnamefont {P.~J.}\ \bibnamefont
  {Easter}}, \bibinfo {author} {\bibfnamefont {S.}~\bibnamefont {Ghonge}},
  \bibinfo {author} {\bibfnamefont {P.~D.}\ \bibnamefont {Lasky}}, \bibinfo
  {author} {\bibfnamefont {A.~R.}\ \bibnamefont {Casey}}, \bibinfo {author}
  {\bibfnamefont {J.~A.}\ \bibnamefont {Clark}}, \bibinfo {author}
  {\bibfnamefont {F.}~\bibnamefont {Hernandez~Vivanco}},\ and\ \bibinfo
  {author} {\bibfnamefont {K.}~\bibnamefont {Chatziioannou}},\ }\href
  {https://doi.org/10.1103/PhysRevD.102.043011} {\bibfield  {journal} {\bibinfo
   {journal} {Phys. Rev. D}\ }\textbf {\bibinfo {volume} {102}},\ \bibinfo
  {pages} {043011} (\bibinfo {year} {2020}{\natexlab{a}})}\BibitemShut
  {NoStop}%
\bibitem [{\citenamefont {Wijngaarden}\ \emph
  {et~al.}(2022{\natexlab{a}})\citenamefont {Wijngaarden}, \citenamefont
  {Hinderer}, \citenamefont {Schwenk} \emph {et~al.}}]{Wijngaarden_2022}%
  \BibitemOpen
  \bibfield  {author} {\bibinfo {author} {\bibfnamefont {M.}~\bibnamefont
  {Wijngaarden}}, \bibinfo {author} {\bibfnamefont {T.}~\bibnamefont
  {Hinderer}}, \bibinfo {author} {\bibfnamefont {A.}~\bibnamefont {Schwenk}},
  \emph {et~al.},\ }\href@noop {} {\bibfield  {journal} {\bibinfo  {journal}
  {Phys. Rev. D}\ }\textbf {\bibinfo {volume} {105}},\ \bibinfo {pages}
  {044015} (\bibinfo {year} {2022}{\natexlab{a}})}\BibitemShut {NoStop}%
\bibitem [{\citenamefont {Clark}\ \emph {et~al.}(2016)\citenamefont {Clark},
  \citenamefont {Bauswein}, \citenamefont {Stergioulas},\ and\ \citenamefont
  {Shoemaker}}]{clark_observing_2016}%
  \BibitemOpen
  \bibfield  {author} {\bibinfo {author} {\bibfnamefont {J.~A.}\ \bibnamefont
  {Clark}}, \bibinfo {author} {\bibfnamefont {A.}~\bibnamefont {Bauswein}},
  \bibinfo {author} {\bibfnamefont {N.}~\bibnamefont {Stergioulas}},\ and\
  \bibinfo {author} {\bibfnamefont {D.}~\bibnamefont {Shoemaker}},\ }\href
  {https://doi.org/10.1088/0264-9381/33/8/085003} {\bibfield  {journal}
  {\bibinfo  {journal} {Classical and Quantum Gravity}\ }\textbf {\bibinfo
  {volume} {33}},\ \bibinfo {pages} {085003} (\bibinfo {year} {2016})},\
  \bibinfo {note} {publisher: IOP Publishing}\BibitemShut {NoStop}%
\bibitem [{\citenamefont {Iacovelli}\ \emph
  {et~al.}(2023{\natexlab{b}})\citenamefont {Iacovelli}, \citenamefont
  {Mancarella}, \citenamefont {Mondal}, \citenamefont {Puecher}, \citenamefont
  {Dietrich}, \citenamefont {Gulminelli}, \citenamefont {Maggiore},\ and\
  \citenamefont {Oertel}}]{iacovelli_nuclear_2023}%
  \BibitemOpen
  \bibfield  {author} {\bibinfo {author} {\bibfnamefont {F.}~\bibnamefont
  {Iacovelli}}, \bibinfo {author} {\bibfnamefont {M.}~\bibnamefont
  {Mancarella}}, \bibinfo {author} {\bibfnamefont {C.}~\bibnamefont {Mondal}},
  \bibinfo {author} {\bibfnamefont {A.}~\bibnamefont {Puecher}}, \bibinfo
  {author} {\bibfnamefont {T.}~\bibnamefont {Dietrich}}, \bibinfo {author}
  {\bibfnamefont {F.}~\bibnamefont {Gulminelli}}, \bibinfo {author}
  {\bibfnamefont {M.}~\bibnamefont {Maggiore}},\ and\ \bibinfo {author}
  {\bibfnamefont {M.}~\bibnamefont {Oertel}},\ }\bibfield  {journal} {\bibinfo
  {journal} {Physical Review D}\ }\textbf {\bibinfo {volume} {108}},\ \href
  {https://doi.org/10.1103/physrevd.108.122006} {10.1103/physrevd.108.122006}
  (\bibinfo {year} {2023}{\natexlab{b}}),\ \bibinfo {note} {publisher: American
  Physical Society (APS)}\BibitemShut {NoStop}%
\bibitem [{\citenamefont {Vretinaris}\ \emph {et~al.}(2025)\citenamefont
  {Vretinaris}, \citenamefont {Vretinaris}, \citenamefont {Mermigkas},
  \citenamefont {Karamanis},\ and\ \citenamefont
  {Stergioulas}}]{vretinaris2025robustfastparameterestimation}%
  \BibitemOpen
  \bibfield  {author} {\bibinfo {author} {\bibfnamefont {S.}~\bibnamefont
  {Vretinaris}}, \bibinfo {author} {\bibfnamefont {G.}~\bibnamefont
  {Vretinaris}}, \bibinfo {author} {\bibfnamefont {C.}~\bibnamefont
  {Mermigkas}}, \bibinfo {author} {\bibfnamefont {M.}~\bibnamefont
  {Karamanis}},\ and\ \bibinfo {author} {\bibfnamefont {N.}~\bibnamefont
  {Stergioulas}},\ }\href {https://arxiv.org/abs/2501.11518} {\bibinfo {title}
  {Robust and fast parameter estimation for gravitational waves from binary
  neutron star merger remnants}} (\bibinfo {year} {2025}),\ \Eprint
  {https://arxiv.org/abs/2501.11518} {arXiv:2501.11518 [gr-qc]} \BibitemShut
  {NoStop}%
\bibitem [{\citenamefont {Banagiri}\ \emph {et~al.}(2020)\citenamefont
  {Banagiri}, \citenamefont {Coughlin}, \citenamefont {Clark}, \citenamefont
  {Lasky}, \citenamefont {Bizouard}, \citenamefont {Talbot}, \citenamefont
  {Thrane},\ and\ \citenamefont {Mandic}}]{Banagiri_2020}%
  \BibitemOpen
  \bibfield  {author} {\bibinfo {author} {\bibfnamefont {S.}~\bibnamefont
  {Banagiri}}, \bibinfo {author} {\bibfnamefont {M.~W.}\ \bibnamefont
  {Coughlin}}, \bibinfo {author} {\bibfnamefont {J.}~\bibnamefont {Clark}},
  \bibinfo {author} {\bibfnamefont {P.~D.}\ \bibnamefont {Lasky}}, \bibinfo
  {author} {\bibfnamefont {M.~A.}\ \bibnamefont {Bizouard}}, \bibinfo {author}
  {\bibfnamefont {C.}~\bibnamefont {Talbot}}, \bibinfo {author} {\bibfnamefont
  {E.}~\bibnamefont {Thrane}},\ and\ \bibinfo {author} {\bibfnamefont
  {V.}~\bibnamefont {Mandic}},\ }\href {https://doi.org/10.1093/mnras/staa181}
  {\bibfield  {journal} {\bibinfo  {journal} {Monthly Notices of the Royal
  Astronomical Society}\ }\textbf {\bibinfo {volume} {492}},\ \bibinfo {pages}
  {4945–4951} (\bibinfo {year} {2020})}\BibitemShut {NoStop}%
\bibitem [{\citenamefont {Jain}\ and\ \citenamefont
  {Agathos}(2024)}]{jain_improving_2024}%
  \BibitemOpen
  \bibfield  {author} {\bibinfo {author} {\bibfnamefont {T.}~\bibnamefont
  {Jain}}\ and\ \bibinfo {author} {\bibfnamefont {M.}~\bibnamefont {Agathos}},\
  }\href {https://doi.org/10.48550/arXiv.2404.12126} {\bibfield  {journal}
  {\bibinfo  {journal} {arXiv e-prints}\ ,\ \bibinfo {pages}
  {arXiv:2404.12126}} (\bibinfo {year} {2024})},\ \bibinfo {note} {\_eprint:
  2404.12126}\BibitemShut {NoStop}%
\bibitem [{\citenamefont {Criswell}\ \emph {et~al.}(2023)\citenamefont
  {Criswell}, \citenamefont {Miller}, \citenamefont {Woldemariam},
  \citenamefont {Soultanis}, \citenamefont {Bauswein}, \citenamefont
  {Chatziioannou}, \citenamefont {Coughlin}, \citenamefont {Jones},\ and\
  \citenamefont {Mandic}}]{criswell_hierarchical_2023}%
  \BibitemOpen
  \bibfield  {author} {\bibinfo {author} {\bibfnamefont {A.~W.}\ \bibnamefont
  {Criswell}}, \bibinfo {author} {\bibfnamefont {J.}~\bibnamefont {Miller}},
  \bibinfo {author} {\bibfnamefont {N.}~\bibnamefont {Woldemariam}}, \bibinfo
  {author} {\bibfnamefont {T.}~\bibnamefont {Soultanis}}, \bibinfo {author}
  {\bibfnamefont {A.}~\bibnamefont {Bauswein}}, \bibinfo {author}
  {\bibfnamefont {K.}~\bibnamefont {Chatziioannou}}, \bibinfo {author}
  {\bibfnamefont {M.~W.}\ \bibnamefont {Coughlin}}, \bibinfo {author}
  {\bibfnamefont {G.}~\bibnamefont {Jones}},\ and\ \bibinfo {author}
  {\bibfnamefont {V.}~\bibnamefont {Mandic}},\ }\bibfield  {journal} {\bibinfo
  {journal} {Physical Review D}\ }\textbf {\bibinfo {volume} {107}},\ \href
  {https://doi.org/10.1103/physrevd.107.043021} {10.1103/physrevd.107.043021}
  (\bibinfo {year} {2023}),\ \bibinfo {note} {publisher: American Physical
  Society (APS)}\BibitemShut {NoStop}%
\bibitem [{\citenamefont {Huez}\ \emph {et~al.}(2026)\citenamefont {Huez},
  \citenamefont {Bernuzzi}, \citenamefont {Breschi},\ and\ \citenamefont
  {Gamba}}]{Huez_2026}%
  \BibitemOpen
  \bibfield  {author} {\bibinfo {author} {\bibfnamefont {G.}~\bibnamefont
  {Huez}}, \bibinfo {author} {\bibfnamefont {S.}~\bibnamefont {Bernuzzi}},
  \bibinfo {author} {\bibfnamefont {M.}~\bibnamefont {Breschi}},\ and\ \bibinfo
  {author} {\bibfnamefont {R.}~\bibnamefont {Gamba}},\ }\bibfield  {journal}
  {\bibinfo  {journal} {Physical Review D}\ }\textbf {\bibinfo {volume}
  {113}},\ \href {https://doi.org/10.1103/2jxk-1bf1} {10.1103/2jxk-1bf1}
  (\bibinfo {year} {2026})\BibitemShut {NoStop}%
\bibitem [{\citenamefont {Mitra}\ \emph {et~al.}(2025)\citenamefont {Mitra},
  \citenamefont {Tiwari},\ and\ \citenamefont
  {Pai}}]{mitra2025prospectconstrainingeosneutron}%
  \BibitemOpen
  \bibfield  {author} {\bibinfo {author} {\bibfnamefont {S.}~\bibnamefont
  {Mitra}}, \bibinfo {author} {\bibfnamefont {P.}~\bibnamefont {Tiwari}},\ and\
  \bibinfo {author} {\bibfnamefont {A.}~\bibnamefont {Pai}},\ }\href
  {https://arxiv.org/abs/2505.21667} {\bibinfo {title} {Prospect of
  constraining the eos of neutron stars using post-merger signals}} (\bibinfo
  {year} {2025}),\ \Eprint {https://arxiv.org/abs/2505.21667} {arXiv:2505.21667
  [astro-ph.HE]} \BibitemShut {NoStop}%
\bibitem [{\citenamefont {Panther}\ and\ \citenamefont
  {Lasky}(2025)}]{panther2025carelesswhisperspopulationsubthreshold}%
  \BibitemOpen
  \bibfield  {author} {\bibinfo {author} {\bibfnamefont {F.~H.}\ \bibnamefont
  {Panther}}\ and\ \bibinfo {author} {\bibfnamefont {P.~D.}\ \bibnamefont
  {Lasky}},\ }\href {https://arxiv.org/abs/2512.08497} {\bibinfo {title}
  {Careless whispers: A population of sub-threshold post-merger gravitational
  waves constrains the hot nuclear equation of state}} (\bibinfo {year}
  {2025}),\ \Eprint {https://arxiv.org/abs/2512.08497} {arXiv:2512.08497
  [gr-qc]} \BibitemShut {NoStop}%
\bibitem [{\citenamefont {Christensen}\ and\ \citenamefont
  {Meyer}(2022)}]{christensen2022parameter}%
  \BibitemOpen
  \bibfield  {author} {\bibinfo {author} {\bibfnamefont {N.}~\bibnamefont
  {Christensen}}\ and\ \bibinfo {author} {\bibfnamefont {R.}~\bibnamefont
  {Meyer}},\ }\href@noop {} {\bibfield  {journal} {\bibinfo  {journal} {Rev.
  Mod. Phys.}\ }\textbf {\bibinfo {volume} {94}},\ \bibinfo {pages} {025001}
  (\bibinfo {year} {2022})}\BibitemShut {NoStop}%
\bibitem [{\citenamefont {Carson}\ \emph {et~al.}(2019)\citenamefont {Carson},
  \citenamefont {Steiner},\ and\ \citenamefont {Yagi}}]{carson_future_2019}%
  \BibitemOpen
  \bibfield  {author} {\bibinfo {author} {\bibfnamefont {Z.}~\bibnamefont
  {Carson}}, \bibinfo {author} {\bibfnamefont {A.~W.}\ \bibnamefont
  {Steiner}},\ and\ \bibinfo {author} {\bibfnamefont {K.}~\bibnamefont
  {Yagi}},\ }\href {https://doi.org/10.1103/PhysRevD.100.023012} {\bibfield
  {journal} {\bibinfo  {journal} {{\textbackslash}prd}\ }\textbf {\bibinfo
  {volume} {100}},\ \bibinfo {pages} {023012} (\bibinfo {year} {2019})},\
  \bibinfo {note} {\_eprint: 1906.05978}\BibitemShut {NoStop}%
\bibitem [{\citenamefont {Baiotti}(2022)}]{baiotti_gravitational_2022}%
  \BibitemOpen
  \bibfield  {author} {\bibinfo {author} {\bibfnamefont {L.}~\bibnamefont
  {Baiotti}},\ }\href {https://doi.org/10.1007/s40065-021-00357-7} {\bibfield
  {journal} {\bibinfo  {journal} {Arabian Journal of Mathematics}\ }\textbf
  {\bibinfo {volume} {11}},\ \bibinfo {pages} {105} (\bibinfo {year}
  {2022})}\BibitemShut {NoStop}%
\bibitem [{\citenamefont {Wijngaarden}\ \emph
  {et~al.}(2022{\natexlab{b}})\citenamefont {Wijngaarden}, \citenamefont
  {Chatziioannou}, \citenamefont {Bauswein}, \citenamefont {Clark},\ and\
  \citenamefont {Cornish}}]{wijngaarden_probing_2022}%
  \BibitemOpen
  \bibfield  {author} {\bibinfo {author} {\bibfnamefont {M.}~\bibnamefont
  {Wijngaarden}}, \bibinfo {author} {\bibfnamefont {K.}~\bibnamefont
  {Chatziioannou}}, \bibinfo {author} {\bibfnamefont {A.}~\bibnamefont
  {Bauswein}}, \bibinfo {author} {\bibfnamefont {J.~A.}\ \bibnamefont
  {Clark}},\ and\ \bibinfo {author} {\bibfnamefont {N.~J.}\ \bibnamefont
  {Cornish}},\ }\href {https://doi.org/10.1103/PhysRevD.105.104019} {\bibfield
  {journal} {\bibinfo  {journal} {{\textbackslash}prd}\ }\textbf {\bibinfo
  {volume} {105}},\ \bibinfo {pages} {104019} (\bibinfo {year}
  {2022}{\natexlab{b}})},\ \bibinfo {note} {\_eprint: 2202.09382}\BibitemShut
  {NoStop}%
\bibitem [{\citenamefont {Puecher}\ \emph {et~al.}(2023)\citenamefont
  {Puecher}, \citenamefont {Dietrich}, \citenamefont {Tsang}, \citenamefont
  {Kalaghatgi}, \citenamefont {Roy}, \citenamefont {Setyawati},\ and\
  \citenamefont {Van Den~Broeck}}]{puecher_unraveling_2023}%
  \BibitemOpen
  \bibfield  {author} {\bibinfo {author} {\bibfnamefont {A.}~\bibnamefont
  {Puecher}}, \bibinfo {author} {\bibfnamefont {T.}~\bibnamefont {Dietrich}},
  \bibinfo {author} {\bibfnamefont {K.~W.}\ \bibnamefont {Tsang}}, \bibinfo
  {author} {\bibfnamefont {C.}~\bibnamefont {Kalaghatgi}}, \bibinfo {author}
  {\bibfnamefont {S.}~\bibnamefont {Roy}}, \bibinfo {author} {\bibfnamefont
  {Y.}~\bibnamefont {Setyawati}},\ and\ \bibinfo {author} {\bibfnamefont
  {C.}~\bibnamefont {Van Den~Broeck}},\ }\href
  {https://doi.org/10.1103/PhysRevD.107.124009} {\bibfield  {journal} {\bibinfo
   {journal} {{\textbackslash}prd}\ }\textbf {\bibinfo {volume} {107}},\
  \bibinfo {pages} {124009} (\bibinfo {year} {2023})},\ \bibinfo {note}
  {\_eprint: 2210.09259}\BibitemShut {NoStop}%
\bibitem [{\citenamefont {Miller}\ and\ \citenamefont
  {Yunes}(2019)}]{miller2019new}%
  \BibitemOpen
  \bibfield  {author} {\bibinfo {author} {\bibfnamefont {M.~C.}\ \bibnamefont
  {Miller}}\ and\ \bibinfo {author} {\bibfnamefont {N.}~\bibnamefont {Yunes}},\
  }\href@noop {} {\bibfield  {journal} {\bibinfo  {journal} {Nature}\ }\textbf
  {\bibinfo {volume} {568}},\ \bibinfo {pages} {469} (\bibinfo {year}
  {2019})}\BibitemShut {NoStop}%
\bibitem [{\citenamefont {Arimoto}\ and\ \citenamefont
  {{others}}(2021)}]{arimoto_gravitational_2021}%
  \BibitemOpen
  \bibfield  {author} {\bibinfo {author} {\bibfnamefont {M.}~\bibnamefont
  {Arimoto}}\ and\ \bibinfo {author} {\bibnamefont {{others}}},\ }\href
  {https://doi.org/10.1093/ptep/ptab042} {\bibfield  {journal} {\bibinfo
  {journal} {Progress of Theoretical and Experimental Physics}\ }\textbf
  {\bibinfo {volume} {2023}},\ \bibinfo {pages} {10A103} (\bibinfo {year}
  {2021})}\BibitemShut {NoStop}%
\bibitem [{\citenamefont {Eardley}\ \emph {et~al.}(1973)\citenamefont
  {Eardley}, \citenamefont {Lee},\ and\ \citenamefont
  {Lightman}}]{PhysRevD.8.3308}%
  \BibitemOpen
  \bibfield  {author} {\bibinfo {author} {\bibfnamefont {D.~M.}\ \bibnamefont
  {Eardley}}, \bibinfo {author} {\bibfnamefont {D.~L.}\ \bibnamefont {Lee}},\
  and\ \bibinfo {author} {\bibfnamefont {A.~P.}\ \bibnamefont {Lightman}},\
  }\href {https://doi.org/10.1103/PhysRevD.8.3308} {\bibfield  {journal}
  {\bibinfo  {journal} {Phys. Rev. D}\ }\textbf {\bibinfo {volume} {8}},\
  \bibinfo {pages} {3308} (\bibinfo {year} {1973})}\BibitemShut {NoStop}%
\bibitem [{\citenamefont {Abbott}\ \emph {et~al.}(2021)\citenamefont {Abbott}
  \emph {et~al.}}]{abbott2021tests}%
  \BibitemOpen
  \bibfield  {author} {\bibinfo {author} {\bibfnamefont {R.}~\bibnamefont
  {Abbott}} \emph {et~al.},\ }\href@noop {} {\bibfield  {journal} {\bibinfo
  {journal} {Phys. Rev. D}\ }\textbf {\bibinfo {volume} {103}},\ \bibinfo
  {pages} {122002} (\bibinfo {year} {2021})}\BibitemShut {NoStop}%
\bibitem [{\citenamefont {Will}(2014)}]{will2014confrontation}%
  \BibitemOpen
  \bibfield  {author} {\bibinfo {author} {\bibfnamefont {C.~M.}\ \bibnamefont
  {Will}},\ }\href@noop {} {\bibfield  {journal} {\bibinfo  {journal} {Living
  Rev. Relativity}\ }\textbf {\bibinfo {volume} {17}},\ \bibinfo {pages} {1}
  (\bibinfo {year} {2014})}\BibitemShut {NoStop}%
\bibitem [{\citenamefont {Sathyaprakash}\ and\ \citenamefont
  {Schutz}(2009)}]{sathyaprakash2009physics}%
  \BibitemOpen
  \bibfield  {author} {\bibinfo {author} {\bibfnamefont {B.~S.}\ \bibnamefont
  {Sathyaprakash}}\ and\ \bibinfo {author} {\bibfnamefont {B.~F.}\ \bibnamefont
  {Schutz}},\ }\href@noop {} {\bibfield  {journal} {\bibinfo  {journal} {Living
  Rev. Relativity}\ }\textbf {\bibinfo {volume} {12}},\ \bibinfo {pages} {2}
  (\bibinfo {year} {2009})}\BibitemShut {NoStop}%
\bibitem [{\citenamefont {Maggiore}(2000)}]{maggiore2000gravitational}%
  \BibitemOpen
  \bibfield  {author} {\bibinfo {author} {\bibfnamefont {M.}~\bibnamefont
  {Maggiore}},\ }\href@noop {} {\bibfield  {journal} {\bibinfo  {journal}
  {Phys. Rep.}\ }\textbf {\bibinfo {volume} {331}},\ \bibinfo {pages} {283}
  (\bibinfo {year} {2000})}\BibitemShut {NoStop}%
\bibitem [{\citenamefont {Huxford}\ \emph {et~al.}(2024)\citenamefont
  {Huxford}, \citenamefont {Kashyap}, \citenamefont {Borhanian}, \citenamefont
  {Dhani}, \citenamefont {Gupta},\ and\ \citenamefont
  {Sathyaprakash}}]{huxford_accuracy_2024}%
  \BibitemOpen
  \bibfield  {author} {\bibinfo {author} {\bibfnamefont {R.}~\bibnamefont
  {Huxford}}, \bibinfo {author} {\bibfnamefont {R.}~\bibnamefont {Kashyap}},
  \bibinfo {author} {\bibfnamefont {S.}~\bibnamefont {Borhanian}}, \bibinfo
  {author} {\bibfnamefont {A.}~\bibnamefont {Dhani}}, \bibinfo {author}
  {\bibfnamefont {I.}~\bibnamefont {Gupta}},\ and\ \bibinfo {author}
  {\bibfnamefont {B.~S.}\ \bibnamefont {Sathyaprakash}},\ }\href
  {https://doi.org/10.1103/PhysRevD.109.103035} {\bibfield  {journal} {\bibinfo
   {journal} {{\textbackslash}prd}\ }\textbf {\bibinfo {volume} {109}},\
  \bibinfo {pages} {103035} (\bibinfo {year} {2024})},\ \bibinfo {note}
  {\_eprint: 2307.05376}\BibitemShut {NoStop}%
\bibitem [{\citenamefont {Easter}\ \emph
  {et~al.}(2020{\natexlab{b}})\citenamefont {Easter}, \citenamefont {Ghonge},
  \citenamefont {Lasky}, \citenamefont {Casey}, \citenamefont {Clark},
  \citenamefont {Hernandez~Vivanco},\ and\ \citenamefont
  {Chatziioannou}}]{easter_detection_2020}%
  \BibitemOpen
  \bibfield  {author} {\bibinfo {author} {\bibfnamefont {P.~J.}\ \bibnamefont
  {Easter}}, \bibinfo {author} {\bibfnamefont {S.}~\bibnamefont {Ghonge}},
  \bibinfo {author} {\bibfnamefont {P.~D.}\ \bibnamefont {Lasky}}, \bibinfo
  {author} {\bibfnamefont {A.~R.}\ \bibnamefont {Casey}}, \bibinfo {author}
  {\bibfnamefont {J.~A.}\ \bibnamefont {Clark}}, \bibinfo {author}
  {\bibfnamefont {F.}~\bibnamefont {Hernandez~Vivanco}},\ and\ \bibinfo
  {author} {\bibfnamefont {K.}~\bibnamefont {Chatziioannou}},\ }\href
  {https://doi.org/10.1103/PhysRevD.102.043011} {\bibfield  {journal} {\bibinfo
   {journal} {Phys. Rev. D}\ }\textbf {\bibinfo {volume} {102}},\ \bibinfo
  {pages} {043011} (\bibinfo {year} {2020}{\natexlab{b}})},\ \bibinfo {note}
  {publisher: American Physical Society}\BibitemShut {NoStop}%
\bibitem [{\citenamefont {Ecker}\ \emph {et~al.}(2024)\citenamefont {Ecker},
  \citenamefont {Gorda}, \citenamefont {Kurkela},\ and\ \citenamefont
  {Rezzolla}}]{ecker_listening_2024}%
  \BibitemOpen
  \bibfield  {author} {\bibinfo {author} {\bibfnamefont {C.}~\bibnamefont
  {Ecker}}, \bibinfo {author} {\bibfnamefont {T.}~\bibnamefont {Gorda}},
  \bibinfo {author} {\bibfnamefont {A.}~\bibnamefont {Kurkela}},\ and\ \bibinfo
  {author} {\bibfnamefont {L.}~\bibnamefont {Rezzolla}},\ }\href
  {https://doi.org/10.48550/arXiv.2403.03246} {\bibfield  {journal} {\bibinfo
  {journal} {arXiv e-prints}\ ,\ \bibinfo {pages} {arXiv:2403.03246}} (\bibinfo
  {year} {2024})},\ \bibinfo {note} {\_eprint: 2403.03246}\BibitemShut
  {NoStop}%
\bibitem [{\citenamefont {Breschi}\ \emph {et~al.}(2019)\citenamefont
  {Breschi}, \citenamefont {Bernuzzi}, \citenamefont {Zappa}, \citenamefont
  {Agathos}, \citenamefont {Perego}, \citenamefont {Radice},\ and\
  \citenamefont {Nagar}}]{breschi_kilohertz_2019}%
  \BibitemOpen
  \bibfield  {author} {\bibinfo {author} {\bibfnamefont {M.}~\bibnamefont
  {Breschi}}, \bibinfo {author} {\bibfnamefont {S.}~\bibnamefont {Bernuzzi}},
  \bibinfo {author} {\bibfnamefont {F.}~\bibnamefont {Zappa}}, \bibinfo
  {author} {\bibfnamefont {M.}~\bibnamefont {Agathos}}, \bibinfo {author}
  {\bibfnamefont {A.}~\bibnamefont {Perego}}, \bibinfo {author} {\bibfnamefont
  {D.}~\bibnamefont {Radice}},\ and\ \bibinfo {author} {\bibfnamefont
  {A.}~\bibnamefont {Nagar}},\ }\href
  {https://doi.org/10.1103/PhysRevD.100.104029} {\bibfield  {journal} {\bibinfo
   {journal} {{\textbackslash}prd}\ }\textbf {\bibinfo {volume} {100}},\
  \bibinfo {pages} {104029} (\bibinfo {year} {2019})},\ \bibinfo {note}
  {\_eprint: 1908.11418}\BibitemShut {NoStop}%
\bibitem [{\citenamefont {Breschi}\ \emph {et~al.}(2024)\citenamefont
  {Breschi}, \citenamefont {Bernuzzi}, \citenamefont {Chakravarti},
  \citenamefont {Camilletti}, \citenamefont {Prakash},\ and\ \citenamefont
  {Perego}}]{breschi_kilohertz_2024}%
  \BibitemOpen
  \bibfield  {author} {\bibinfo {author} {\bibfnamefont {M.}~\bibnamefont
  {Breschi}}, \bibinfo {author} {\bibfnamefont {S.}~\bibnamefont {Bernuzzi}},
  \bibinfo {author} {\bibfnamefont {K.}~\bibnamefont {Chakravarti}}, \bibinfo
  {author} {\bibfnamefont {A.}~\bibnamefont {Camilletti}}, \bibinfo {author}
  {\bibfnamefont {A.}~\bibnamefont {Prakash}},\ and\ \bibinfo {author}
  {\bibfnamefont {A.}~\bibnamefont {Perego}},\ }\bibfield  {journal} {\bibinfo
  {journal} {Physical Review D}\ }\textbf {\bibinfo {volume} {109}},\ \href
  {https://doi.org/10.1103/physrevd.109.064009} {10.1103/physrevd.109.064009}
  (\bibinfo {year} {2024}),\ \bibinfo {note} {publisher: American Physical
  Society (APS)}\BibitemShut {NoStop}%
\bibitem [{\citenamefont {Tsokaros}\ \emph {et~al.}(2024)\citenamefont
  {Tsokaros}, \citenamefont {Bamber}, \citenamefont {Ruiz},\ and\ \citenamefont
  {Shapiro}}]{tsokaros_masking_2024}%
  \BibitemOpen
  \bibfield  {author} {\bibinfo {author} {\bibfnamefont {A.}~\bibnamefont
  {Tsokaros}}, \bibinfo {author} {\bibfnamefont {J.}~\bibnamefont {Bamber}},
  \bibinfo {author} {\bibfnamefont {M.}~\bibnamefont {Ruiz}},\ and\ \bibinfo
  {author} {\bibfnamefont {S.~L.}\ \bibnamefont {Shapiro}},\ }\href
  {https://doi.org/10.48550/arXiv.2411.00939} {\bibfield  {journal} {\bibinfo
  {journal} {arXiv e-prints}\ ,\ \bibinfo {pages} {arXiv:2411.00939}} (\bibinfo
  {year} {2024})},\ \bibinfo {note} {\_eprint: 2411.00939}\BibitemShut
  {NoStop}%
\bibitem [{\citenamefont {Soultanis}\ \emph {et~al.}(2022)\citenamefont
  {Soultanis}, \citenamefont {Bauswein},\ and\ \citenamefont
  {Stergioulas}}]{Soultanis_2022}%
  \BibitemOpen
  \bibfield  {author} {\bibinfo {author} {\bibfnamefont {T.}~\bibnamefont
  {Soultanis}}, \bibinfo {author} {\bibfnamefont {A.}~\bibnamefont
  {Bauswein}},\ and\ \bibinfo {author} {\bibfnamefont {N.}~\bibnamefont
  {Stergioulas}},\ }\bibfield  {journal} {\bibinfo  {journal} {Physical Review
  D}\ }\textbf {\bibinfo {volume} {105}},\ \href
  {https://doi.org/10.1103/physrevd.105.043020} {10.1103/physrevd.105.043020}
  (\bibinfo {year} {2022})\BibitemShut {NoStop}%
\bibitem [{\citenamefont {Tringali}\ \emph {et~al.}(2023)\citenamefont
  {Tringali}, \citenamefont {Puecher}, \citenamefont {Lazzaro}, \citenamefont
  {Ciolfi}, \citenamefont {Drago}, \citenamefont {Giacomazzo}, \citenamefont
  {Vedovato},\ and\ \citenamefont
  {Prodi}}]{tringali_morphology-independent_2023}%
  \BibitemOpen
  \bibfield  {author} {\bibinfo {author} {\bibfnamefont {M.~C.}\ \bibnamefont
  {Tringali}}, \bibinfo {author} {\bibfnamefont {A.}~\bibnamefont {Puecher}},
  \bibinfo {author} {\bibfnamefont {C.}~\bibnamefont {Lazzaro}}, \bibinfo
  {author} {\bibfnamefont {R.}~\bibnamefont {Ciolfi}}, \bibinfo {author}
  {\bibfnamefont {M.}~\bibnamefont {Drago}}, \bibinfo {author} {\bibfnamefont
  {B.}~\bibnamefont {Giacomazzo}}, \bibinfo {author} {\bibfnamefont
  {G.}~\bibnamefont {Vedovato}},\ and\ \bibinfo {author} {\bibfnamefont
  {G.~A.}\ \bibnamefont {Prodi}},\ }\href
  {https://doi.org/10.1088/1361-6382/acfc0d} {\bibfield  {journal} {\bibinfo
  {journal} {Classical and Quantum Gravity}\ }\textbf {\bibinfo {volume}
  {40}},\ \bibinfo {pages} {225008} (\bibinfo {year} {2023})},\ \bibinfo {note}
  {\_eprint: 2304.12831}\BibitemShut {NoStop}%
\bibitem [{\citenamefont {Kochankovski}\ \emph {et~al.}(2024)\citenamefont
  {Kochankovski}, \citenamefont {Ramos}, \citenamefont {Tolos}, \citenamefont
  {Blacker},\ and\ \citenamefont
  {Bauswein}}]{kochankovski2024hyperonsneutronstarmergers}%
  \BibitemOpen
  \bibfield  {author} {\bibinfo {author} {\bibfnamefont {H.}~\bibnamefont
  {Kochankovski}}, \bibinfo {author} {\bibfnamefont {A.}~\bibnamefont {Ramos}},
  \bibinfo {author} {\bibfnamefont {L.}~\bibnamefont {Tolos}}, \bibinfo
  {author} {\bibfnamefont {S.}~\bibnamefont {Blacker}},\ and\ \bibinfo {author}
  {\bibfnamefont {A.}~\bibnamefont {Bauswein}},\ }\href
  {https://arxiv.org/abs/2411.14978} {\bibinfo {title} {Hyperons in neutron
  star mergers}} (\bibinfo {year} {2024}),\ \Eprint
  {https://arxiv.org/abs/2411.14978} {arXiv:2411.14978 [hep-ph]} \BibitemShut
  {NoStop}%
\bibitem [{\citenamefont {Blacker}\ \emph {et~al.}(2024)\citenamefont
  {Blacker}, \citenamefont {Kochankovski}, \citenamefont {Bauswein},
  \citenamefont {Ramos},\ and\ \citenamefont {Tolos}}]{Blacker_2024}%
  \BibitemOpen
  \bibfield  {author} {\bibinfo {author} {\bibfnamefont {S.}~\bibnamefont
  {Blacker}}, \bibinfo {author} {\bibfnamefont {H.}~\bibnamefont
  {Kochankovski}}, \bibinfo {author} {\bibfnamefont {A.}~\bibnamefont
  {Bauswein}}, \bibinfo {author} {\bibfnamefont {A.}~\bibnamefont {Ramos}},\
  and\ \bibinfo {author} {\bibfnamefont {L.}~\bibnamefont {Tolos}},\ }\href
  {https://doi.org/10.1103/PhysRevD.109.043015} {\bibfield  {journal} {\bibinfo
   {journal} {Phys. Rev. D}\ }\textbf {\bibinfo {volume} {109}},\ \bibinfo
  {pages} {043015} (\bibinfo {year} {2024})}\BibitemShut {NoStop}%
\bibitem [{\citenamefont {{De Pietri}}\ \emph {et~al.}(2018)\citenamefont {{De
  Pietri}}, \citenamefont {{Feo}}, \citenamefont {{Font}}, \citenamefont
  {{L{\"o}ffler}}, \citenamefont {{Maione}}, \citenamefont {{Pasquali}},\ and\
  \citenamefont {{Stergioulas}}}]{2018PhRvL.120v1101D}%
  \BibitemOpen
  \bibfield  {author} {\bibinfo {author} {\bibfnamefont {R.}~\bibnamefont {{De
  Pietri}}}, \bibinfo {author} {\bibfnamefont {A.}~\bibnamefont {{Feo}}},
  \bibinfo {author} {\bibfnamefont {J.~A.}\ \bibnamefont {{Font}}}, \bibinfo
  {author} {\bibfnamefont {F.}~\bibnamefont {{L{\"o}ffler}}}, \bibinfo {author}
  {\bibfnamefont {F.}~\bibnamefont {{Maione}}}, \bibinfo {author}
  {\bibfnamefont {M.}~\bibnamefont {{Pasquali}}},\ and\ \bibinfo {author}
  {\bibfnamefont {N.}~\bibnamefont {{Stergioulas}}},\ }\href
  {https://doi.org/10.1103/PhysRevLett.120.221101} {\bibfield  {journal}
  {\bibinfo  {journal} {\prl}\ }\textbf {\bibinfo {volume} {120}},\ \bibinfo
  {eid} {221101} (\bibinfo {year} {2018})},\ \Eprint
  {https://arxiv.org/abs/1802.03288} {arXiv:1802.03288 [gr-qc]} \BibitemShut
  {NoStop}%
\bibitem [{\citenamefont {{De Pietri}}\ \emph {et~al.}(2020)\citenamefont {{De
  Pietri}}, \citenamefont {{Feo}}, \citenamefont {{Font}}, \citenamefont
  {{L{\"o}ffler}}, \citenamefont {{Pasquali}},\ and\ \citenamefont
  {{Stergioulas}}}]{2020PhRvD.101f4052D}%
  \BibitemOpen
  \bibfield  {author} {\bibinfo {author} {\bibfnamefont {R.}~\bibnamefont {{De
  Pietri}}}, \bibinfo {author} {\bibfnamefont {A.}~\bibnamefont {{Feo}}},
  \bibinfo {author} {\bibfnamefont {J.~A.}\ \bibnamefont {{Font}}}, \bibinfo
  {author} {\bibfnamefont {F.}~\bibnamefont {{L{\"o}ffler}}}, \bibinfo {author}
  {\bibfnamefont {M.}~\bibnamefont {{Pasquali}}},\ and\ \bibinfo {author}
  {\bibfnamefont {N.}~\bibnamefont {{Stergioulas}}},\ }\href
  {https://doi.org/10.1103/PhysRevD.101.064052} {\bibfield  {journal} {\bibinfo
   {journal} {\prd}\ }\textbf {\bibinfo {volume} {101}},\ \bibinfo {eid}
  {064052} (\bibinfo {year} {2020})},\ \Eprint
  {https://arxiv.org/abs/1910.04036} {arXiv:1910.04036 [gr-qc]} \BibitemShut
  {NoStop}%
\bibitem [{\citenamefont {{Sasli}}\ \emph {et~al.}(2024)\citenamefont
  {{Sasli}}, \citenamefont {{Karnesis}},\ and\ \citenamefont
  {{Stergioulas}}}]{2024PhRvD.109d3045S}%
  \BibitemOpen
  \bibfield  {author} {\bibinfo {author} {\bibfnamefont {A.}~\bibnamefont
  {{Sasli}}}, \bibinfo {author} {\bibfnamefont {N.}~\bibnamefont
  {{Karnesis}}},\ and\ \bibinfo {author} {\bibfnamefont {N.}~\bibnamefont
  {{Stergioulas}}},\ }\href {https://doi.org/10.1103/PhysRevD.109.043045}
  {\bibfield  {journal} {\bibinfo  {journal} {\prd}\ }\textbf {\bibinfo
  {volume} {109}},\ \bibinfo {eid} {043045} (\bibinfo {year} {2024})},\ \Eprint
  {https://arxiv.org/abs/2311.10626} {arXiv:2311.10626 [gr-qc]} \BibitemShut
  {NoStop}%
\bibitem [{\citenamefont {Guerra}\ \emph {et~al.}(2026)\citenamefont {Guerra},
  \citenamefont {Ruiz}, \citenamefont {Pasquali}, \citenamefont
  {Cerdá-Durán}, \citenamefont {Rios},\ and\ \citenamefont
  {Font}}]{guerra2026treatmentthermaleffectsequation}%
  \BibitemOpen
  \bibfield  {author} {\bibinfo {author} {\bibfnamefont {D.}~\bibnamefont
  {Guerra}}, \bibinfo {author} {\bibfnamefont {M.}~\bibnamefont {Ruiz}},
  \bibinfo {author} {\bibfnamefont {M.}~\bibnamefont {Pasquali}}, \bibinfo
  {author} {\bibfnamefont {P.}~\bibnamefont {Cerdá-Durán}}, \bibinfo {author}
  {\bibfnamefont {A.}~\bibnamefont {Rios}},\ and\ \bibinfo {author}
  {\bibfnamefont {J.~A.}\ \bibnamefont {Font}},\ }\href
  {https://arxiv.org/abs/2512.05118} {\bibinfo {title} {On the treatment of
  thermal effects in the equation of state on neutron star merger remnants}}
  (\bibinfo {year} {2026}),\ \Eprint {https://arxiv.org/abs/2512.05118}
  {arXiv:2512.05118 [gr-qc]} \BibitemShut {NoStop}%
\bibitem [{\citenamefont {Puecher}\ and\ \citenamefont
  {Dietrich}(2024)}]{puecher_machine-learning_2024}%
  \BibitemOpen
  \bibfield  {author} {\bibinfo {author} {\bibfnamefont {A.}~\bibnamefont
  {Puecher}}\ and\ \bibinfo {author} {\bibfnamefont {T.}~\bibnamefont
  {Dietrich}},\ }\bibfield  {journal} {\bibinfo  {journal} {Physical Review D}\
  }\textbf {\bibinfo {volume} {110}},\ \href
  {https://doi.org/10.1103/physrevd.110.123038} {10.1103/physrevd.110.123038}
  (\bibinfo {year} {2024}),\ \bibinfo {note} {publisher: American Physical
  Society (APS)}\BibitemShut {NoStop}%
\bibitem [{\citenamefont {Whittaker}\ \emph {et~al.}(2022)\citenamefont
  {Whittaker}, \citenamefont {East}, \citenamefont {Green}, \citenamefont
  {Lehner},\ and\ \citenamefont {Yang}}]{whittaker_using_2022}%
  \BibitemOpen
  \bibfield  {author} {\bibinfo {author} {\bibfnamefont {T.}~\bibnamefont
  {Whittaker}}, \bibinfo {author} {\bibfnamefont {W.~E.}\ \bibnamefont {East}},
  \bibinfo {author} {\bibfnamefont {S.~R.}\ \bibnamefont {Green}}, \bibinfo
  {author} {\bibfnamefont {L.}~\bibnamefont {Lehner}},\ and\ \bibinfo {author}
  {\bibfnamefont {H.}~\bibnamefont {Yang}},\ }\href
  {https://doi.org/10.1103/PhysRevD.105.124021} {\bibfield  {journal} {\bibinfo
   {journal} {{\textbackslash}prd}\ }\textbf {\bibinfo {volume} {105}},\
  \bibinfo {pages} {124021} (\bibinfo {year} {2022})},\ \bibinfo {note}
  {\_eprint: 2201.06461}\BibitemShut {NoStop}%
\bibitem [{\citenamefont {Llorens-Monteagudo}\ \emph
  {et~al.}(2025)\citenamefont {Llorens-Monteagudo}, \citenamefont
  {Torres-Forné},\ and\ \citenamefont
  {Font}}]{llorensmonteagudo2025clawdiadictionarylearningframework}%
  \BibitemOpen
  \bibfield  {author} {\bibinfo {author} {\bibfnamefont {M.}~\bibnamefont
  {Llorens-Monteagudo}}, \bibinfo {author} {\bibfnamefont {A.}~\bibnamefont
  {Torres-Forné}},\ and\ \bibinfo {author} {\bibfnamefont {J.~A.}\
  \bibnamefont {Font}},\ }\href {https://arxiv.org/abs/2511.16750} {\bibinfo
  {title} {Clawdia: A dictionary learning framework for gravitational-wave data
  analysis}} (\bibinfo {year} {2025}),\ \Eprint
  {https://arxiv.org/abs/2511.16750} {arXiv:2511.16750 [astro-ph.IM]}
  \BibitemShut {NoStop}%
\bibitem [{\citenamefont {Soultanis}\ \emph {et~al.}(2024)\citenamefont
  {Soultanis}, \citenamefont {Maltsev}, \citenamefont {Bauswein}, \citenamefont
  {Chatziioannou}, \citenamefont {Roepke},\ and\ \citenamefont
  {Stergioulas}}]{soultanis2024gravitationalwavemodelneutronstar}%
  \BibitemOpen
  \bibfield  {author} {\bibinfo {author} {\bibfnamefont {T.}~\bibnamefont
  {Soultanis}}, \bibinfo {author} {\bibfnamefont {K.}~\bibnamefont {Maltsev}},
  \bibinfo {author} {\bibfnamefont {A.}~\bibnamefont {Bauswein}}, \bibinfo
  {author} {\bibfnamefont {K.}~\bibnamefont {Chatziioannou}}, \bibinfo {author}
  {\bibfnamefont {F.~K.}\ \bibnamefont {Roepke}},\ and\ \bibinfo {author}
  {\bibfnamefont {N.}~\bibnamefont {Stergioulas}},\ }\href
  {https://arxiv.org/abs/2405.09513} {\bibinfo {title} {Gravitational-wave
  model for neutron star merger remnants with supervised learning}} (\bibinfo
  {year} {2024}),\ \Eprint {https://arxiv.org/abs/2405.09513} {arXiv:2405.09513
  [astro-ph.HE]} \BibitemShut {NoStop}%
\bibitem [{\citenamefont {Pesios}\ \emph {et~al.}(2024)\citenamefont {Pesios},
  \citenamefont {Koutalios}, \citenamefont {Kugiumtzis},\ and\ \citenamefont
  {Stergioulas}}]{Pesios2024}%
  \BibitemOpen
  \bibfield  {author} {\bibinfo {author} {\bibfnamefont {D.}~\bibnamefont
  {Pesios}}, \bibinfo {author} {\bibfnamefont {I.}~\bibnamefont {Koutalios}},
  \bibinfo {author} {\bibfnamefont {D.}~\bibnamefont {Kugiumtzis}},\ and\
  \bibinfo {author} {\bibfnamefont {N.}~\bibnamefont {Stergioulas}},\ }\href
  {https://doi.org/10.1103/PhysRevD.110.063008} {\bibfield  {journal} {\bibinfo
   {journal} {Phys. Rev. D}\ }\textbf {\bibinfo {volume} {110}},\ \bibinfo
  {pages} {063008} (\bibinfo {year} {2024})}\BibitemShut {NoStop}%
\bibitem [{\citenamefont {Cuoco}\ \emph {et~al.}(2025)\citenamefont {Cuoco},
  \citenamefont {Cavaglià}, \citenamefont {Heng}, \citenamefont {Keitel},\
  and\ \citenamefont {Messenger}}]{Cuoco_2025}%
  \BibitemOpen
  \bibfield  {author} {\bibinfo {author} {\bibfnamefont {E.}~\bibnamefont
  {Cuoco}}, \bibinfo {author} {\bibfnamefont {M.}~\bibnamefont {Cavaglià}},
  \bibinfo {author} {\bibfnamefont {I.~S.}\ \bibnamefont {Heng}}, \bibinfo
  {author} {\bibfnamefont {D.}~\bibnamefont {Keitel}},\ and\ \bibinfo {author}
  {\bibfnamefont {C.}~\bibnamefont {Messenger}},\ }\bibfield  {journal}
  {\bibinfo  {journal} {Living Reviews in Relativity}\ }\textbf {\bibinfo
  {volume} {28}},\ \href {https://doi.org/10.1007/s41114-024-00055-8}
  {10.1007/s41114-024-00055-8} (\bibinfo {year} {2025})\BibitemShut {NoStop}%
\bibitem [{\citenamefont {Easter}\ \emph
  {et~al.}(2019{\natexlab{a}})\citenamefont {Easter}, \citenamefont {Lasky},
  \citenamefont {Casey}, \citenamefont {Rezzolla},\ and\ \citenamefont
  {Takami}}]{PhysRevD.100.043005}%
  \BibitemOpen
  \bibfield  {author} {\bibinfo {author} {\bibfnamefont {P.~J.}\ \bibnamefont
  {Easter}}, \bibinfo {author} {\bibfnamefont {P.~D.}\ \bibnamefont {Lasky}},
  \bibinfo {author} {\bibfnamefont {A.~R.}\ \bibnamefont {Casey}}, \bibinfo
  {author} {\bibfnamefont {L.}~\bibnamefont {Rezzolla}},\ and\ \bibinfo
  {author} {\bibfnamefont {K.}~\bibnamefont {Takami}},\ }\href
  {https://doi.org/10.1103/PhysRevD.100.043005} {\bibfield  {journal} {\bibinfo
   {journal} {Phys. Rev. D}\ }\textbf {\bibinfo {volume} {100}},\ \bibinfo
  {pages} {043005} (\bibinfo {year} {2019}{\natexlab{a}})}\BibitemShut
  {NoStop}%
\bibitem [{\citenamefont {Vretinaris}\ \emph {et~al.}(2020)\citenamefont
  {Vretinaris}, \citenamefont {Stergioulas},\ and\ \citenamefont
  {Bauswein}}]{vretinaris2020}%
  \BibitemOpen
  \bibfield  {author} {\bibinfo {author} {\bibfnamefont {S.}~\bibnamefont
  {Vretinaris}}, \bibinfo {author} {\bibfnamefont {N.}~\bibnamefont
  {Stergioulas}},\ and\ \bibinfo {author} {\bibfnamefont {A.}~\bibnamefont
  {Bauswein}},\ }\href {https://doi.org/10.1103/PhysRevD.101.084039} {\bibfield
   {journal} {\bibinfo  {journal} {Phys. Rev. D}\ }\textbf {\bibinfo {volume}
  {101}},\ \bibinfo {pages} {084039} (\bibinfo {year} {2020})}\BibitemShut
  {NoStop}%
\bibitem [{\citenamefont {Chakravarti}\ and\ \citenamefont
  {Andersson}(2020)}]{chakravarti2020}%
  \BibitemOpen
  \bibfield  {author} {\bibinfo {author} {\bibfnamefont {K.}~\bibnamefont
  {Chakravarti}}\ and\ \bibinfo {author} {\bibfnamefont {N.}~\bibnamefont
  {Andersson}},\ }\href {https://doi.org/10.1093/mnras/staa2342} {\bibfield
  {journal} {\bibinfo  {journal} {Monthly Notices of the Royal Astronomical
  Society}\ }\textbf {\bibinfo {volume} {497}},\ \bibinfo {pages} {5480}
  (\bibinfo {year} {2020})},\ \Eprint
  {https://arxiv.org/abs/https://academic.oup.com/mnras/article-pdf/497/4/5480/33695826/staa2342.pdf}
  {https://academic.oup.com/mnras/article-pdf/497/4/5480/33695826/staa2342.pdf}
  \BibitemShut {NoStop}%
\bibitem [{\citenamefont {Rezzolla}\ and\ \citenamefont
  {Takami}(2016)}]{PhysRevD.93.124051}%
  \BibitemOpen
  \bibfield  {author} {\bibinfo {author} {\bibfnamefont {L.}~\bibnamefont
  {Rezzolla}}\ and\ \bibinfo {author} {\bibfnamefont {K.}~\bibnamefont
  {Takami}},\ }\href {https://doi.org/10.1103/PhysRevD.93.124051} {\bibfield
  {journal} {\bibinfo  {journal} {Phys. Rev. D}\ }\textbf {\bibinfo {volume}
  {93}},\ \bibinfo {pages} {124051} (\bibinfo {year} {2016})}\BibitemShut
  {NoStop}%
\bibitem [{\citenamefont {Gonzalez}\ \emph {et~al.}(2022)\citenamefont
  {Gonzalez} \emph {et~al.}}]{Gonzalez_2022mgo}%
  \BibitemOpen
  \bibfield  {author} {\bibinfo {author} {\bibfnamefont {A.}~\bibnamefont
  {Gonzalez}} \emph {et~al.},\ }\href@noop {} {\  (\bibinfo {year} {2022})},\
  \Eprint {https://arxiv.org/abs/2210.16366} {arXiv:2210.16366 [gr-qc]}
  \BibitemShut {NoStop}%
\bibitem [{\citenamefont {Izenman}(2008)}]{Izenman2008}%
  \BibitemOpen
  \bibfield  {author} {\bibinfo {author} {\bibfnamefont {A.~J.}\ \bibnamefont
  {Izenman}},\ }\bibinfo {title} {Multivariate regression},\ in\ \href
  {https://doi.org/10.1007/978-0-387-78189-1_6} {\emph {\bibinfo {booktitle}
  {Modern Multivariate Statistical Techniques: Regression, Classification, and
  Manifold Learning}}}\ (\bibinfo  {publisher} {Springer New York},\ \bibinfo
  {address} {New York, NY},\ \bibinfo {year} {2008})\ pp.\ \bibinfo {pages}
  {159--194}\BibitemShut {NoStop}%
\bibitem [{\citenamefont {Fox}\ and\ \citenamefont
  {Weisberg}(2019)}]{foxweisberg}%
  \BibitemOpen
  \bibfield  {author} {\bibinfo {author} {\bibfnamefont {J.}~\bibnamefont
  {Fox}}\ and\ \bibinfo {author} {\bibfnamefont {S.}~\bibnamefont {Weisberg}},\
  }\href {https://socialsciences.mcmaster.ca/jfox/Books/Companion/index.html}
  {\emph {\bibinfo {title} {An R Companion to Applied Regression}}},\ \bibinfo
  {edition} {3rd}\ ed.\ (\bibinfo  {publisher} {Sage},\ \bibinfo {address}
  {Thousand Oaks CA},\ \bibinfo {year} {2019})\BibitemShut {NoStop}%
\bibitem [{\citenamefont {Hornik}\ \emph {et~al.}(1989)\citenamefont {Hornik},
  \citenamefont {Stinchcombe},\ and\ \citenamefont {White}}]{HORNIK1989359}%
  \BibitemOpen
  \bibfield  {author} {\bibinfo {author} {\bibfnamefont {K.}~\bibnamefont
  {Hornik}}, \bibinfo {author} {\bibfnamefont {M.}~\bibnamefont
  {Stinchcombe}},\ and\ \bibinfo {author} {\bibfnamefont {H.}~\bibnamefont
  {White}},\ }\href
  {https://doi.org/https://doi.org/10.1016/0893-6080(89)90020-8} {\bibfield
  {journal} {\bibinfo  {journal} {Neural Networks}\ }\textbf {\bibinfo {volume}
  {2}},\ \bibinfo {pages} {359} (\bibinfo {year} {1989})}\BibitemShut {NoStop}%
\bibitem [{\citenamefont {Hornik}(1991)}]{HORNIK1991251}%
  \BibitemOpen
  \bibfield  {author} {\bibinfo {author} {\bibfnamefont {K.}~\bibnamefont
  {Hornik}},\ }\href
  {https://doi.org/https://doi.org/10.1016/0893-6080(91)90009-T} {\bibfield
  {journal} {\bibinfo  {journal} {Neural Networks}\ }\textbf {\bibinfo {volume}
  {4}},\ \bibinfo {pages} {251} (\bibinfo {year} {1991})}\BibitemShut {NoStop}%
\bibitem [{\citenamefont {Mu}\ and\ \citenamefont {Lin}(2025)}]{mu2025}%
  \BibitemOpen
  \bibfield  {author} {\bibinfo {author} {\bibfnamefont {S.}~\bibnamefont
  {Mu}}\ and\ \bibinfo {author} {\bibfnamefont {S.}~\bibnamefont {Lin}},\
  }\href {https://arxiv.org/abs/2503.07137} {\bibinfo {title} {A comprehensive
  survey of mixture-of-experts: Algorithms, theory, and applications}}
  (\bibinfo {year} {2025}),\ \Eprint {https://arxiv.org/abs/2503.07137}
  {arXiv:2503.07137 [cs.LG]} \BibitemShut {NoStop}%
\bibitem [{\citenamefont {Royer}\ \emph {et~al.}(2023)\citenamefont {Royer},
  \citenamefont {Karmanov}, \citenamefont {Skliar}, \citenamefont {Bejnordi},\
  and\ \citenamefont {Blankevoort}}]{royer2023}%
  \BibitemOpen
  \bibfield  {author} {\bibinfo {author} {\bibfnamefont {A.}~\bibnamefont
  {Royer}}, \bibinfo {author} {\bibfnamefont {I.}~\bibnamefont {Karmanov}},
  \bibinfo {author} {\bibfnamefont {A.}~\bibnamefont {Skliar}}, \bibinfo
  {author} {\bibfnamefont {B.~E.}\ \bibnamefont {Bejnordi}},\ and\ \bibinfo
  {author} {\bibfnamefont {T.}~\bibnamefont {Blankevoort}},\ }\href
  {https://arxiv.org/abs/2304.05497} {\bibinfo {title} {Revisiting single-gated
  mixtures of experts}} (\bibinfo {year} {2023}),\ \Eprint
  {https://arxiv.org/abs/2304.05497} {arXiv:2304.05497 [cs.CV]} \BibitemShut
  {NoStop}%
\bibitem [{\citenamefont {Thorne}(1987)}]{Thorne1987}%
  \BibitemOpen
  \bibfield  {author} {\bibinfo {author} {\bibfnamefont {K.~S.}\ \bibnamefont
  {Thorne}},\ }in\ \href@noop {} {\emph {\bibinfo {booktitle} {Three Hundred
  Years of Gravitation}}},\ \bibinfo {editor} {edited by\ \bibinfo {editor}
  {\bibfnamefont {S.~W.}\ \bibnamefont {Hawking}}\ and\ \bibinfo {editor}
  {\bibfnamefont {W.}~\bibnamefont {Israel}}}\ (\bibinfo  {publisher}
  {Cambridge University Press},\ \bibinfo {address} {Cambridge, UK},\ \bibinfo
  {year} {1987})\ pp.\ \bibinfo {pages} {330--450}\BibitemShut {NoStop}%
\bibitem [{\citenamefont {Green}(1984)}]{Green1984}%
  \BibitemOpen
  \bibfield  {author} {\bibinfo {author} {\bibfnamefont {P.~J.}\ \bibnamefont
  {Green}},\ }\href {https://doi.org/10.1111/j.2517-6161.1984.tb01288.x}
  {\bibfield  {journal} {\bibinfo  {journal} {Journal of the Royal Statistical
  Society, Series B (Methodological)}\ }\textbf {\bibinfo {volume} {46}},\
  \bibinfo {pages} {149} (\bibinfo {year} {1984})}\BibitemShut {NoStop}%
\bibitem [{\citenamefont {Rusiecki}(2012)}]{Rusiecki2012}%
  \BibitemOpen
  \bibfield  {author} {\bibinfo {author} {\bibfnamefont {A.}~\bibnamefont
  {Rusiecki}},\ }\href {https://doi.org/10.1007/s11063-012-9227-z} {\bibfield
  {journal} {\bibinfo  {journal} {Neural Processing Letters}\ }\textbf
  {\bibinfo {volume} {36}},\ \bibinfo {pages} {145} (\bibinfo {year}
  {2012})}\BibitemShut {NoStop}%
\bibitem [{\citenamefont {Abadi}\ \emph {et~al.}(2015)\citenamefont {Abadi},
  \citenamefont {Agarwal}, \citenamefont {Barham}, \citenamefont {Brevdo},
  \citenamefont {Chen}, \citenamefont {Citro}, \citenamefont {Corrado},
  \citenamefont {Davis}, \citenamefont {Dean}, \citenamefont {Devin},
  \citenamefont {Ghemawat}, \citenamefont {Goodfellow}, \citenamefont {Harp},
  \citenamefont {Irving}, \citenamefont {Isard}, \citenamefont {Jia},
  \citenamefont {Jozefowicz}, \citenamefont {Kaiser}, \citenamefont {Kudlur},
  \citenamefont {Levenberg}, \citenamefont {Man\'{e}}, \citenamefont {Monga},
  \citenamefont {Moore}, \citenamefont {Murray}, \citenamefont {Olah},
  \citenamefont {Schuster}, \citenamefont {Shlens}, \citenamefont {Steiner},
  \citenamefont {Sutskever}, \citenamefont {Talwar}, \citenamefont {Tucker},
  \citenamefont {Vanhoucke}, \citenamefont {Vasudevan}, \citenamefont
  {Vi\'{e}gas}, \citenamefont {Vinyals}, \citenamefont {Warden}, \citenamefont
  {Wattenberg}, \citenamefont {Wicke}, \citenamefont {Yu},\ and\ \citenamefont
  {Zheng}}]{tensorflow2015-whitepaper}%
  \BibitemOpen
  \bibfield  {author} {\bibinfo {author} {\bibfnamefont {M.}~\bibnamefont
  {Abadi}}, \bibinfo {author} {\bibfnamefont {A.}~\bibnamefont {Agarwal}},
  \bibinfo {author} {\bibfnamefont {P.}~\bibnamefont {Barham}}, \bibinfo
  {author} {\bibfnamefont {E.}~\bibnamefont {Brevdo}}, \bibinfo {author}
  {\bibfnamefont {Z.}~\bibnamefont {Chen}}, \bibinfo {author} {\bibfnamefont
  {C.}~\bibnamefont {Citro}}, \bibinfo {author} {\bibfnamefont {G.~S.}\
  \bibnamefont {Corrado}}, \bibinfo {author} {\bibfnamefont {A.}~\bibnamefont
  {Davis}}, \bibinfo {author} {\bibfnamefont {J.}~\bibnamefont {Dean}},
  \bibinfo {author} {\bibfnamefont {M.}~\bibnamefont {Devin}}, \bibinfo
  {author} {\bibfnamefont {S.}~\bibnamefont {Ghemawat}}, \bibinfo {author}
  {\bibfnamefont {I.}~\bibnamefont {Goodfellow}}, \bibinfo {author}
  {\bibfnamefont {A.}~\bibnamefont {Harp}}, \bibinfo {author} {\bibfnamefont
  {G.}~\bibnamefont {Irving}}, \bibinfo {author} {\bibfnamefont
  {M.}~\bibnamefont {Isard}}, \bibinfo {author} {\bibfnamefont
  {Y.}~\bibnamefont {Jia}}, \bibinfo {author} {\bibfnamefont {R.}~\bibnamefont
  {Jozefowicz}}, \bibinfo {author} {\bibfnamefont {L.}~\bibnamefont {Kaiser}},
  \bibinfo {author} {\bibfnamefont {M.}~\bibnamefont {Kudlur}}, \bibinfo
  {author} {\bibfnamefont {J.}~\bibnamefont {Levenberg}}, \bibinfo {author}
  {\bibfnamefont {D.}~\bibnamefont {Man\'{e}}}, \bibinfo {author}
  {\bibfnamefont {R.}~\bibnamefont {Monga}}, \bibinfo {author} {\bibfnamefont
  {S.}~\bibnamefont {Moore}}, \bibinfo {author} {\bibfnamefont
  {D.}~\bibnamefont {Murray}}, \bibinfo {author} {\bibfnamefont
  {C.}~\bibnamefont {Olah}}, \bibinfo {author} {\bibfnamefont {M.}~\bibnamefont
  {Schuster}}, \bibinfo {author} {\bibfnamefont {J.}~\bibnamefont {Shlens}},
  \bibinfo {author} {\bibfnamefont {B.}~\bibnamefont {Steiner}}, \bibinfo
  {author} {\bibfnamefont {I.}~\bibnamefont {Sutskever}}, \bibinfo {author}
  {\bibfnamefont {K.}~\bibnamefont {Talwar}}, \bibinfo {author} {\bibfnamefont
  {P.}~\bibnamefont {Tucker}}, \bibinfo {author} {\bibfnamefont
  {V.}~\bibnamefont {Vanhoucke}}, \bibinfo {author} {\bibfnamefont
  {V.}~\bibnamefont {Vasudevan}}, \bibinfo {author} {\bibfnamefont
  {F.}~\bibnamefont {Vi\'{e}gas}}, \bibinfo {author} {\bibfnamefont
  {O.}~\bibnamefont {Vinyals}}, \bibinfo {author} {\bibfnamefont
  {P.}~\bibnamefont {Warden}}, \bibinfo {author} {\bibfnamefont
  {M.}~\bibnamefont {Wattenberg}}, \bibinfo {author} {\bibfnamefont
  {M.}~\bibnamefont {Wicke}}, \bibinfo {author} {\bibfnamefont
  {Y.}~\bibnamefont {Yu}},\ and\ \bibinfo {author} {\bibfnamefont
  {X.}~\bibnamefont {Zheng}},\ }\href {https://www.tensorflow.org/} {\bibinfo
  {title} {{TensorFlow}: Large-scale machine learning on heterogeneous
  systems}} (\bibinfo {year} {2015}),\ \bibinfo {note} {software available from
  tensorflow.org}\BibitemShut {NoStop}%
\bibitem [{\citenamefont {Easter}\ \emph
  {et~al.}(2019{\natexlab{b}})\citenamefont {Easter}, \citenamefont {Lasky},
  \citenamefont {Casey}, \citenamefont {Rezzolla},\ and\ \citenamefont
  {Takami}}]{Easter2019}%
  \BibitemOpen
  \bibfield  {author} {\bibinfo {author} {\bibfnamefont {P.~J.}\ \bibnamefont
  {Easter}}, \bibinfo {author} {\bibfnamefont {P.~D.}\ \bibnamefont {Lasky}},
  \bibinfo {author} {\bibfnamefont {A.~R.}\ \bibnamefont {Casey}}, \bibinfo
  {author} {\bibfnamefont {L.}~\bibnamefont {Rezzolla}},\ and\ \bibinfo
  {author} {\bibfnamefont {K.}~\bibnamefont {Takami}},\ }\href
  {https://doi.org/10.1103/PhysRevD.100.043005} {\bibfield  {journal} {\bibinfo
   {journal} {Phys. Rev. D}\ }\textbf {\bibinfo {volume} {100}},\ \bibinfo
  {pages} {043005} (\bibinfo {year} {2019}{\natexlab{b}})}\BibitemShut
  {NoStop}%
\bibitem [{\citenamefont {Kiuchi}\ \emph {et~al.}(2018)\citenamefont {Kiuchi},
  \citenamefont {Kyutoku}, \citenamefont {Sekiguchi},\ and\ \citenamefont
  {Shibata}}]{Kiuchi:2017zzg}%
  \BibitemOpen
  \bibfield  {author} {\bibinfo {author} {\bibfnamefont {K.}~\bibnamefont
  {Kiuchi}}, \bibinfo {author} {\bibfnamefont {K.}~\bibnamefont {Kyutoku}},
  \bibinfo {author} {\bibfnamefont {Y.}~\bibnamefont {Sekiguchi}},\ and\
  \bibinfo {author} {\bibfnamefont {M.}~\bibnamefont {Shibata}},\ }\href
  {https://doi.org/10.1103/PhysRevD.97.124039} {\bibfield  {journal} {\bibinfo
  {journal} {Phys. Rev. D}\ }\textbf {\bibinfo {volume} {97}},\ \bibinfo
  {pages} {124039} (\bibinfo {year} {2018})}\BibitemShut {NoStop}%
\bibitem [{\citenamefont {Kiuchi}\ \emph {et~al.}(2015)\citenamefont {Kiuchi},
  \citenamefont {Cerd\'a-Dur\'an}, \citenamefont {Kyutoku}, \citenamefont
  {Sekiguchi},\ and\ \citenamefont {Shibata}}]{Kiuchi:2015sga}%
  \BibitemOpen
  \bibfield  {author} {\bibinfo {author} {\bibfnamefont {K.}~\bibnamefont
  {Kiuchi}}, \bibinfo {author} {\bibfnamefont {P.}~\bibnamefont
  {Cerd\'a-Dur\'an}}, \bibinfo {author} {\bibfnamefont {K.}~\bibnamefont
  {Kyutoku}}, \bibinfo {author} {\bibfnamefont {Y.}~\bibnamefont {Sekiguchi}},\
  and\ \bibinfo {author} {\bibfnamefont {M.}~\bibnamefont {Shibata}},\ }\href
  {https://doi.org/10.1103/PhysRevD.92.124034} {\bibfield  {journal} {\bibinfo
  {journal} {Phys. Rev. D}\ }\textbf {\bibinfo {volume} {92}},\ \bibinfo
  {pages} {124034} (\bibinfo {year} {2015})}\BibitemShut {NoStop}%
\bibitem [{\citenamefont {Shibata}\ and\ \citenamefont
  {Kiuchi}(2017)}]{Shibata:2017xht}%
  \BibitemOpen
  \bibfield  {author} {\bibinfo {author} {\bibfnamefont {M.}~\bibnamefont
  {Shibata}}\ and\ \bibinfo {author} {\bibfnamefont {K.}~\bibnamefont
  {Kiuchi}},\ }\href {https://doi.org/10.1103/PhysRevD.95.123003} {\bibfield
  {journal} {\bibinfo  {journal} {Phys. Rev. D}\ }\textbf {\bibinfo {volume}
  {95}},\ \bibinfo {pages} {123003} (\bibinfo {year} {2017})},\ \Eprint
  {https://arxiv.org/abs/1705.06142} {arXiv:1705.06142 [astro-ph.HE]}
  \BibitemShut {NoStop}%
\bibitem [{\citenamefont {Ciolfi}\ \emph {et~al.}(2019)\citenamefont {Ciolfi},
  \citenamefont {Kastaun}, \citenamefont {Kalinani},\ and\ \citenamefont
  {Giacomazzo}}]{Ciolfi:2019fie}%
  \BibitemOpen
  \bibfield  {author} {\bibinfo {author} {\bibfnamefont {R.}~\bibnamefont
  {Ciolfi}}, \bibinfo {author} {\bibfnamefont {W.}~\bibnamefont {Kastaun}},
  \bibinfo {author} {\bibfnamefont {J.~V.}\ \bibnamefont {Kalinani}},\ and\
  \bibinfo {author} {\bibfnamefont {B.}~\bibnamefont {Giacomazzo}},\ }\href
  {https://doi.org/10.1103/PhysRevD.100.023005} {\bibfield  {journal} {\bibinfo
   {journal} {Phys. Rev. D}\ }\textbf {\bibinfo {volume} {100}},\ \bibinfo
  {pages} {023005} (\bibinfo {year} {2019})}\BibitemShut {NoStop}%
\bibitem [{\citenamefont {Radice}(2017)}]{Radice:2017zta}%
  \BibitemOpen
  \bibfield  {author} {\bibinfo {author} {\bibfnamefont {D.}~\bibnamefont
  {Radice}},\ }\href {https://doi.org/10.3847/2041-8213/aa6483} {\bibfield
  {journal} {\bibinfo  {journal} {Astrophys. J. Lett.}\ }\textbf {\bibinfo
  {volume} {838}},\ \bibinfo {pages} {L2} (\bibinfo {year} {2017})}\BibitemShut
  {NoStop}%
\bibitem [{\citenamefont {Giacomazzo}\ \emph {et~al.}(2015)\citenamefont
  {Giacomazzo}, \citenamefont {Zrake}, \citenamefont {Duffell}, \citenamefont
  {MacFadyen},\ and\ \citenamefont {Perna}}]{Giacomazzo:2014qba}%
  \BibitemOpen
  \bibfield  {author} {\bibinfo {author} {\bibfnamefont {B.}~\bibnamefont
  {Giacomazzo}}, \bibinfo {author} {\bibfnamefont {J.}~\bibnamefont {Zrake}},
  \bibinfo {author} {\bibfnamefont {P.}~\bibnamefont {Duffell}}, \bibinfo
  {author} {\bibfnamefont {A.~I.}\ \bibnamefont {MacFadyen}},\ and\ \bibinfo
  {author} {\bibfnamefont {R.}~\bibnamefont {Perna}},\ }\href
  {https://doi.org/10.1088/0004-637X/809/1/39} {\bibfield  {journal} {\bibinfo
  {journal} {Astrophys. J.}\ }\textbf {\bibinfo {volume} {809}},\ \bibinfo
  {pages} {39} (\bibinfo {year} {2015})},\ \Eprint
  {https://arxiv.org/abs/1410.0013} {arXiv:1410.0013 [astro-ph.HE]}
  \BibitemShut {NoStop}%
\bibitem [{\citenamefont {Kiuchi}\ \emph {et~al.}(2014)\citenamefont {Kiuchi},
  \citenamefont {Kyutoku}, \citenamefont {Sekiguchi}, \citenamefont {Shibata},\
  and\ \citenamefont {Wada}}]{Kiuchi:2014hja}%
  \BibitemOpen
  \bibfield  {author} {\bibinfo {author} {\bibfnamefont {K.}~\bibnamefont
  {Kiuchi}}, \bibinfo {author} {\bibfnamefont {K.}~\bibnamefont {Kyutoku}},
  \bibinfo {author} {\bibfnamefont {Y.}~\bibnamefont {Sekiguchi}}, \bibinfo
  {author} {\bibfnamefont {M.}~\bibnamefont {Shibata}},\ and\ \bibinfo {author}
  {\bibfnamefont {T.}~\bibnamefont {Wada}},\ }\href
  {https://doi.org/10.1103/PhysRevD.90.041502} {\bibfield  {journal} {\bibinfo
  {journal} {Phys. Rev. D}\ }\textbf {\bibinfo {volume} {90}},\ \bibinfo
  {pages} {041502} (\bibinfo {year} {2014})}\BibitemShut {NoStop}%
\bibitem [{\citenamefont {Sagunski}\ \emph
  {et~al.}(2018{\natexlab{a}})\citenamefont {Sagunski}, \citenamefont {Zhang},
  \citenamefont {Johnson}, \citenamefont {Lehner}, \citenamefont
  {Sakellariadou}, \citenamefont {Liebling}, \citenamefont {Palenzuela},\ and\
  \citenamefont {Neilsen}}]{PhysRevD.97.064016}%
  \BibitemOpen
  \bibfield  {author} {\bibinfo {author} {\bibfnamefont {L.}~\bibnamefont
  {Sagunski}}, \bibinfo {author} {\bibfnamefont {J.}~\bibnamefont {Zhang}},
  \bibinfo {author} {\bibfnamefont {M.~C.}\ \bibnamefont {Johnson}}, \bibinfo
  {author} {\bibfnamefont {L.}~\bibnamefont {Lehner}}, \bibinfo {author}
  {\bibfnamefont {M.}~\bibnamefont {Sakellariadou}}, \bibinfo {author}
  {\bibfnamefont {S.~L.}\ \bibnamefont {Liebling}}, \bibinfo {author}
  {\bibfnamefont {C.}~\bibnamefont {Palenzuela}},\ and\ \bibinfo {author}
  {\bibfnamefont {D.}~\bibnamefont {Neilsen}},\ }\href
  {https://doi.org/10.1103/PhysRevD.97.064016} {\bibfield  {journal} {\bibinfo
  {journal} {Phys. Rev. D}\ }\textbf {\bibinfo {volume} {97}},\ \bibinfo
  {pages} {064016} (\bibinfo {year} {2018}{\natexlab{a}})}\BibitemShut
  {NoStop}%
\bibitem [{\citenamefont {Blacker}\ \emph {et~al.}(2020)\citenamefont
  {Blacker}, \citenamefont {Bastian}, \citenamefont {Bauswein}, \citenamefont
  {Blaschke}, \citenamefont {Fischer}, \citenamefont {Oertel}, \citenamefont
  {Soultanis},\ and\ \citenamefont {Typel}}]{blacker_constraining_2020}%
  \BibitemOpen
  \bibfield  {author} {\bibinfo {author} {\bibfnamefont {S.}~\bibnamefont
  {Blacker}}, \bibinfo {author} {\bibfnamefont {N.-U.~F.}\ \bibnamefont
  {Bastian}}, \bibinfo {author} {\bibfnamefont {A.}~\bibnamefont {Bauswein}},
  \bibinfo {author} {\bibfnamefont {D.~B.}\ \bibnamefont {Blaschke}}, \bibinfo
  {author} {\bibfnamefont {T.}~\bibnamefont {Fischer}}, \bibinfo {author}
  {\bibfnamefont {M.}~\bibnamefont {Oertel}}, \bibinfo {author} {\bibfnamefont
  {T.}~\bibnamefont {Soultanis}},\ and\ \bibinfo {author} {\bibfnamefont
  {S.}~\bibnamefont {Typel}},\ }\href
  {https://doi.org/10.1103/PhysRevD.102.123023} {\bibfield  {journal} {\bibinfo
   {journal} {{\textbackslash}prd}\ }\textbf {\bibinfo {volume} {102}},\
  \bibinfo {pages} {123023} (\bibinfo {year} {2020})},\ \bibinfo {note}
  {\_eprint: 2006.03789}\BibitemShut {NoStop}%
\bibitem [{\citenamefont {Raithel}\ and\ \citenamefont
  {Paschalidis}(2024)}]{raithel_detectability_2024}%
  \BibitemOpen
  \bibfield  {author} {\bibinfo {author} {\bibfnamefont {C.~A.}\ \bibnamefont
  {Raithel}}\ and\ \bibinfo {author} {\bibfnamefont {V.}~\bibnamefont
  {Paschalidis}},\ }\href {https://doi.org/10.1103/PhysRevD.110.043002}
  {\bibfield  {journal} {\bibinfo  {journal} {{\textbackslash}prd}\ }\textbf
  {\bibinfo {volume} {110}},\ \bibinfo {pages} {043002} (\bibinfo {year}
  {2024})},\ \bibinfo {note} {\_eprint: 2312.14046}\BibitemShut {NoStop}%
\bibitem [{\citenamefont {Prakash}\ \emph {et~al.}(2024)\citenamefont
  {Prakash}, \citenamefont {Gupta}, \citenamefont {Breschi}, \citenamefont
  {Kashyap}, \citenamefont {Radice}, \citenamefont {Bernuzzi}, \citenamefont
  {Logoteta},\ and\ \citenamefont
  {Sathyaprakash}}]{prakash_detectability_2024}%
  \BibitemOpen
  \bibfield  {author} {\bibinfo {author} {\bibfnamefont {A.}~\bibnamefont
  {Prakash}}, \bibinfo {author} {\bibfnamefont {I.}~\bibnamefont {Gupta}},
  \bibinfo {author} {\bibfnamefont {M.}~\bibnamefont {Breschi}}, \bibinfo
  {author} {\bibfnamefont {R.}~\bibnamefont {Kashyap}}, \bibinfo {author}
  {\bibfnamefont {D.}~\bibnamefont {Radice}}, \bibinfo {author} {\bibfnamefont
  {S.}~\bibnamefont {Bernuzzi}}, \bibinfo {author} {\bibfnamefont
  {D.}~\bibnamefont {Logoteta}},\ and\ \bibinfo {author} {\bibfnamefont
  {B.~S.}\ \bibnamefont {Sathyaprakash}},\ }\href
  {https://doi.org/10.1103/PhysRevD.109.103008} {\bibfield  {journal} {\bibinfo
   {journal} {{\textbackslash}prd}\ }\textbf {\bibinfo {volume} {109}},\
  \bibinfo {pages} {103008} (\bibinfo {year} {2024})},\ \bibinfo {note}
  {\_eprint: 2310.06025}\BibitemShut {NoStop}%
\bibitem [{\citenamefont {Jacobi}\ \emph {et~al.}(2023)\citenamefont {Jacobi},
  \citenamefont {Guercilena}, \citenamefont {Huth}, \citenamefont {Ricigliano},
  \citenamefont {Arcones},\ and\ \citenamefont
  {Schwenk}}]{jacobi_effects_2023}%
  \BibitemOpen
  \bibfield  {author} {\bibinfo {author} {\bibfnamefont {M.}~\bibnamefont
  {Jacobi}}, \bibinfo {author} {\bibfnamefont {F.~M.}\ \bibnamefont
  {Guercilena}}, \bibinfo {author} {\bibfnamefont {S.}~\bibnamefont {Huth}},
  \bibinfo {author} {\bibfnamefont {G.}~\bibnamefont {Ricigliano}}, \bibinfo
  {author} {\bibfnamefont {A.}~\bibnamefont {Arcones}},\ and\ \bibinfo {author}
  {\bibfnamefont {A.}~\bibnamefont {Schwenk}},\ }\href
  {https://doi.org/10.1093/mnras/stad3738} {\bibfield  {journal} {\bibinfo
  {journal} {Monthly Notices of the Royal Astronomical Society}\ }\textbf
  {\bibinfo {volume} {527}},\ \bibinfo {pages} {8812} (\bibinfo {year}
  {2023})},\ \bibinfo {note} {publisher: Oxford University Press
  (OUP)}\BibitemShut {NoStop}%
\bibitem [{\citenamefont {Most}\ \emph {et~al.}(2024)\citenamefont {Most},
  \citenamefont {Haber}, \citenamefont {Harris}, \citenamefont {Zhang},
  \citenamefont {Alford},\ and\ \citenamefont {Noronha}}]{most_emergence_2024}%
  \BibitemOpen
  \bibfield  {author} {\bibinfo {author} {\bibfnamefont {E.~R.}\ \bibnamefont
  {Most}}, \bibinfo {author} {\bibfnamefont {A.}~\bibnamefont {Haber}},
  \bibinfo {author} {\bibfnamefont {S.~P.}\ \bibnamefont {Harris}}, \bibinfo
  {author} {\bibfnamefont {Z.}~\bibnamefont {Zhang}}, \bibinfo {author}
  {\bibfnamefont {M.~G.}\ \bibnamefont {Alford}},\ and\ \bibinfo {author}
  {\bibfnamefont {J.}~\bibnamefont {Noronha}},\ }\href
  {https://doi.org/10.3847/2041-8213/ad454f} {\bibfield  {journal} {\bibinfo
  {journal} {Astrophys. J. Lett.}\ }\textbf {\bibinfo {volume} {967}},\
  \bibinfo {pages} {L14} (\bibinfo {year} {2024})},\ \bibinfo {note} {\_eprint:
  2207.00442}\BibitemShut {NoStop}%
\bibitem [{\citenamefont {Suárez-Fontanella}\ \emph
  {et~al.}(2024)\citenamefont {Suárez-Fontanella}, \citenamefont
  {Barba-González}, \citenamefont {Albertus},\ and\ \citenamefont {Ángeles
  Pérez-García}}]{suarez-fontanella_gravitational_2024}%
  \BibitemOpen
  \bibfield  {author} {\bibinfo {author} {\bibfnamefont {D.}~\bibnamefont
  {Suárez-Fontanella}}, \bibinfo {author} {\bibfnamefont {D.}~\bibnamefont
  {Barba-González}}, \bibinfo {author} {\bibfnamefont {C.}~\bibnamefont
  {Albertus}},\ and\ \bibinfo {author} {\bibfnamefont {M.}~\bibnamefont
  {Ángeles Pérez-García}},\ }\href
  {https://doi.org/10.48550/arXiv.2408.05226} {\bibfield  {journal} {\bibinfo
  {journal} {arXiv e-prints}\ ,\ \bibinfo {pages} {arXiv:2408.05226}} (\bibinfo
  {year} {2024})},\ \bibinfo {note} {\_eprint: 2408.05226}\BibitemShut
  {NoStop}%
\bibitem [{\citenamefont {Fujimoto}\ \emph {et~al.}(2023)\citenamefont
  {Fujimoto}, \citenamefont {Fukushima}, \citenamefont {Hotokezaka},\ and\
  \citenamefont {Kyutoku}}]{fujimoto_gravitational_2023}%
  \BibitemOpen
  \bibfield  {author} {\bibinfo {author} {\bibfnamefont {Y.}~\bibnamefont
  {Fujimoto}}, \bibinfo {author} {\bibfnamefont {K.}~\bibnamefont {Fukushima}},
  \bibinfo {author} {\bibfnamefont {K.}~\bibnamefont {Hotokezaka}},\ and\
  \bibinfo {author} {\bibfnamefont {K.}~\bibnamefont {Kyutoku}},\ }\href
  {https://doi.org/10.1103/PhysRevLett.130.091404} {\bibfield  {journal}
  {\bibinfo  {journal} {Phys. Rev. Lett.}\ }\textbf {\bibinfo {volume} {130}},\
  \bibinfo {pages} {091404} (\bibinfo {year} {2023})},\ \bibinfo {note}
  {\_eprint: 2205.03882}\BibitemShut {NoStop}%
\bibitem [{\citenamefont {Rivieccio}\ \emph {et~al.}(2024)\citenamefont
  {Rivieccio}, \citenamefont {Guerra}, \citenamefont {Ruiz},\ and\
  \citenamefont {Font}}]{rivieccio_gravitational-wave_2024}%
  \BibitemOpen
  \bibfield  {author} {\bibinfo {author} {\bibfnamefont {G.}~\bibnamefont
  {Rivieccio}}, \bibinfo {author} {\bibfnamefont {D.}~\bibnamefont {Guerra}},
  \bibinfo {author} {\bibfnamefont {M.}~\bibnamefont {Ruiz}},\ and\ \bibinfo
  {author} {\bibfnamefont {J.~A.}\ \bibnamefont {Font}},\ }\href
  {https://doi.org/10.1103/PhysRevD.109.064032} {\bibfield  {journal} {\bibinfo
   {journal} {{\textbackslash}prd}\ }\textbf {\bibinfo {volume} {109}},\
  \bibinfo {pages} {064032} (\bibinfo {year} {2024})},\ \bibinfo {note}
  {\_eprint: 2401.06849}\BibitemShut {NoStop}%
\bibitem [{\citenamefont {Bauswein}\ \emph {et~al.}(2019)\citenamefont
  {Bauswein}, \citenamefont {Bastian}, \citenamefont {Blaschke}, \citenamefont
  {Chatziioannou}, \citenamefont {Clark}, \citenamefont {Fischer},\ and\
  \citenamefont {Oertel}}]{bauswein_identifying_2019}%
  \BibitemOpen
  \bibfield  {author} {\bibinfo {author} {\bibfnamefont {A.}~\bibnamefont
  {Bauswein}}, \bibinfo {author} {\bibfnamefont {N.-U.~F.}\ \bibnamefont
  {Bastian}}, \bibinfo {author} {\bibfnamefont {D.~B.}\ \bibnamefont
  {Blaschke}}, \bibinfo {author} {\bibfnamefont {K.}~\bibnamefont
  {Chatziioannou}}, \bibinfo {author} {\bibfnamefont {J.~A.}\ \bibnamefont
  {Clark}}, \bibinfo {author} {\bibfnamefont {T.}~\bibnamefont {Fischer}},\
  and\ \bibinfo {author} {\bibfnamefont {M.}~\bibnamefont {Oertel}},\ }\href
  {https://doi.org/10.1103/PhysRevLett.122.061102} {\bibfield  {journal}
  {\bibinfo  {journal} {Phys. Rev. Lett.}\ }\textbf {\bibinfo {volume} {122}},\
  \bibinfo {pages} {061102} (\bibinfo {year} {2019})},\ \bibinfo {note}
  {publisher: American Physical Society}\BibitemShut {NoStop}%
\bibitem [{\citenamefont {Miravet-Tenés}\ \emph {et~al.}(2024)\citenamefont
  {Miravet-Tenés}, \citenamefont {Guerra}, \citenamefont {Ruiz}, \citenamefont
  {Cerdá-Durán},\ and\ \citenamefont
  {Font}}]{miravet-tenes_identifying_2024}%
  \BibitemOpen
  \bibfield  {author} {\bibinfo {author} {\bibfnamefont {M.}~\bibnamefont
  {Miravet-Tenés}}, \bibinfo {author} {\bibfnamefont {D.}~\bibnamefont
  {Guerra}}, \bibinfo {author} {\bibfnamefont {M.}~\bibnamefont {Ruiz}},
  \bibinfo {author} {\bibfnamefont {P.}~\bibnamefont {Cerdá-Durán}},\ and\
  \bibinfo {author} {\bibfnamefont {J.~A.}\ \bibnamefont {Font}},\ }\href
  {https://doi.org/10.48550/arXiv.2401.02493} {\bibfield  {journal} {\bibinfo
  {journal} {arXiv e-prints}\ ,\ \bibinfo {pages} {arXiv:2401.02493}} (\bibinfo
  {year} {2024})},\ \bibinfo {note} {\_eprint: 2401.02493}\BibitemShut
  {NoStop}%
\bibitem [{\citenamefont {Vijayan}\ \emph {et~al.}(2023)\citenamefont
  {Vijayan}, \citenamefont {Rahman}, \citenamefont {Bauswein}, \citenamefont
  {Martínez-Pinedo},\ and\ \citenamefont {Arbina}}]{vijayan_impact_2023}%
  \BibitemOpen
  \bibfield  {author} {\bibinfo {author} {\bibfnamefont {V.}~\bibnamefont
  {Vijayan}}, \bibinfo {author} {\bibfnamefont {N.}~\bibnamefont {Rahman}},
  \bibinfo {author} {\bibfnamefont {A.}~\bibnamefont {Bauswein}}, \bibinfo
  {author} {\bibfnamefont {G.}~\bibnamefont {Martínez-Pinedo}},\ and\ \bibinfo
  {author} {\bibfnamefont {I.~L.}\ \bibnamefont {Arbina}},\ }\href
  {https://doi.org/10.1103/PhysRevD.108.023020} {\bibfield  {journal} {\bibinfo
   {journal} {{\textbackslash}prd}\ }\textbf {\bibinfo {volume} {108}},\
  \bibinfo {pages} {023020} (\bibinfo {year} {2023})},\ \bibinfo {note}
  {\_eprint: 2302.12055}\BibitemShut {NoStop}%
\bibitem [{\citenamefont {Raithel}\ and\ \citenamefont
  {Paschalidis}(2023)}]{raithel_influence_2023}%
  \BibitemOpen
  \bibfield  {author} {\bibinfo {author} {\bibfnamefont {C.~A.}\ \bibnamefont
  {Raithel}}\ and\ \bibinfo {author} {\bibfnamefont {V.}~\bibnamefont
  {Paschalidis}},\ }\bibfield  {journal} {\bibinfo  {journal} {Physical Review
  D}\ }\textbf {\bibinfo {volume} {108}},\ \href
  {https://doi.org/10.1103/physrevd.108.083029} {10.1103/physrevd.108.083029}
  (\bibinfo {year} {2023}),\ \bibinfo {note} {publisher: American Physical
  Society (APS)}\BibitemShut {NoStop}%
\bibitem [{\citenamefont {Most}\ \emph {et~al.}(2020)\citenamefont {Most},
  \citenamefont {Jens~Papenfort}, \citenamefont {Dexheimer}, \citenamefont
  {Hanauske}, \citenamefont {Stoecker},\ and\ \citenamefont
  {Rezzolla}}]{most_deconfinement_2020}%
  \BibitemOpen
  \bibfield  {author} {\bibinfo {author} {\bibfnamefont {E.~R.}\ \bibnamefont
  {Most}}, \bibinfo {author} {\bibfnamefont {L.}~\bibnamefont
  {Jens~Papenfort}}, \bibinfo {author} {\bibfnamefont {V.}~\bibnamefont
  {Dexheimer}}, \bibinfo {author} {\bibfnamefont {M.}~\bibnamefont {Hanauske}},
  \bibinfo {author} {\bibfnamefont {H.}~\bibnamefont {Stoecker}},\ and\
  \bibinfo {author} {\bibfnamefont {L.}~\bibnamefont {Rezzolla}},\ }\href
  {https://doi.org/10.1140/epja/s10050-020-00073-4} {\bibfield  {journal}
  {\bibinfo  {journal} {The European Physical Journal A}\ }\textbf {\bibinfo
  {volume} {56}},\ \bibinfo {pages} {59} (\bibinfo {year} {2020})},\ \bibinfo
  {note} {iSBN: 1434-601X}\BibitemShut {NoStop}%
\bibitem [{\citenamefont {Bamber}\ \emph {et~al.}(2024)\citenamefont {Bamber},
  \citenamefont {Tsokaros}, \citenamefont {Ruiz},\ and\ \citenamefont
  {Shapiro}}]{bamber_post-merger_2024}%
  \BibitemOpen
  \bibfield  {author} {\bibinfo {author} {\bibfnamefont {J.}~\bibnamefont
  {Bamber}}, \bibinfo {author} {\bibfnamefont {A.}~\bibnamefont {Tsokaros}},
  \bibinfo {author} {\bibfnamefont {M.}~\bibnamefont {Ruiz}},\ and\ \bibinfo
  {author} {\bibfnamefont {S.~L.}\ \bibnamefont {Shapiro}},\ }\href
  {https://doi.org/10.48550/arXiv.2411.00943} {\bibfield  {journal} {\bibinfo
  {journal} {arXiv e-prints}\ ,\ \bibinfo {pages} {arXiv:2411.00943}} (\bibinfo
  {year} {2024})},\ \bibinfo {note} {\_eprint: 2411.00943}\BibitemShut
  {NoStop}%
\bibitem [{\citenamefont {Espino}\ \emph {et~al.}(2024)\citenamefont {Espino},
  \citenamefont {Prakash}, \citenamefont {Radice},\ and\ \citenamefont
  {Logoteta}}]{espino_revealing_2024}%
  \BibitemOpen
  \bibfield  {author} {\bibinfo {author} {\bibfnamefont {P.~L.}\ \bibnamefont
  {Espino}}, \bibinfo {author} {\bibfnamefont {A.}~\bibnamefont {Prakash}},
  \bibinfo {author} {\bibfnamefont {D.}~\bibnamefont {Radice}},\ and\ \bibinfo
  {author} {\bibfnamefont {D.}~\bibnamefont {Logoteta}},\ }\bibfield  {journal}
  {\bibinfo  {journal} {Physical Review D}\ }\textbf {\bibinfo {volume}
  {109}},\ \href {https://doi.org/10.1103/physrevd.109.123009}
  {10.1103/physrevd.109.123009} (\bibinfo {year} {2024}),\ \bibinfo {note}
  {publisher: American Physical Society (APS)}\BibitemShut {NoStop}%
\bibitem [{\citenamefont {Prakash}\ \emph {et~al.}(2021)\citenamefont
  {Prakash}, \citenamefont {Radice}, \citenamefont {Logoteta}, \citenamefont
  {Perego}, \citenamefont {Nedora}, \citenamefont {Bombaci}, \citenamefont
  {Kashyap}, \citenamefont {Bernuzzi},\ and\ \citenamefont
  {Endrizzi}}]{prakash_signatures_2021}%
  \BibitemOpen
  \bibfield  {author} {\bibinfo {author} {\bibfnamefont {A.}~\bibnamefont
  {Prakash}}, \bibinfo {author} {\bibfnamefont {D.}~\bibnamefont {Radice}},
  \bibinfo {author} {\bibfnamefont {D.}~\bibnamefont {Logoteta}}, \bibinfo
  {author} {\bibfnamefont {A.}~\bibnamefont {Perego}}, \bibinfo {author}
  {\bibfnamefont {V.}~\bibnamefont {Nedora}}, \bibinfo {author} {\bibfnamefont
  {I.}~\bibnamefont {Bombaci}}, \bibinfo {author} {\bibfnamefont
  {R.}~\bibnamefont {Kashyap}}, \bibinfo {author} {\bibfnamefont
  {S.}~\bibnamefont {Bernuzzi}},\ and\ \bibinfo {author} {\bibfnamefont
  {A.}~\bibnamefont {Endrizzi}},\ }\href
  {https://doi.org/10.1103/PhysRevD.104.083029} {\bibfield  {journal} {\bibinfo
   {journal} {Phys. Rev. D}\ }\textbf {\bibinfo {volume} {104}},\ \bibinfo
  {pages} {083029} (\bibinfo {year} {2021})},\ \bibinfo {note} {publisher:
  American Physical Society}\BibitemShut {NoStop}%
\bibitem [{\citenamefont {Most}\ \emph {et~al.}(2019)\citenamefont {Most},
  \citenamefont {Papenfort}, \citenamefont {Dexheimer}, \citenamefont
  {Hanauske}, \citenamefont {Schramm}, \citenamefont {Stöcker},\ and\
  \citenamefont {Rezzolla}}]{most_signatures_2019}%
  \BibitemOpen
  \bibfield  {author} {\bibinfo {author} {\bibfnamefont {E.~R.}\ \bibnamefont
  {Most}}, \bibinfo {author} {\bibfnamefont {L.~J.}\ \bibnamefont {Papenfort}},
  \bibinfo {author} {\bibfnamefont {V.}~\bibnamefont {Dexheimer}}, \bibinfo
  {author} {\bibfnamefont {M.}~\bibnamefont {Hanauske}}, \bibinfo {author}
  {\bibfnamefont {S.}~\bibnamefont {Schramm}}, \bibinfo {author} {\bibfnamefont
  {H.}~\bibnamefont {Stöcker}},\ and\ \bibinfo {author} {\bibfnamefont
  {L.}~\bibnamefont {Rezzolla}},\ }\href
  {https://doi.org/10.1103/PhysRevLett.122.061101} {\bibfield  {journal}
  {\bibinfo  {journal} {Phys. Rev. Lett.}\ }\textbf {\bibinfo {volume} {122}},\
  \bibinfo {pages} {061101} (\bibinfo {year} {2019})},\ \bibinfo {note}
  {publisher: American Physical Society}\BibitemShut {NoStop}%
\bibitem [{\citenamefont {Fields}\ \emph {et~al.}(2023)\citenamefont {Fields},
  \citenamefont {Prakash}, \citenamefont {Breschi}, \citenamefont {Radice},
  \citenamefont {Bernuzzi},\ and\ \citenamefont
  {Schneider}}]{fields_thermal_2023}%
  \BibitemOpen
  \bibfield  {author} {\bibinfo {author} {\bibfnamefont {J.}~\bibnamefont
  {Fields}}, \bibinfo {author} {\bibfnamefont {A.}~\bibnamefont {Prakash}},
  \bibinfo {author} {\bibfnamefont {M.}~\bibnamefont {Breschi}}, \bibinfo
  {author} {\bibfnamefont {D.}~\bibnamefont {Radice}}, \bibinfo {author}
  {\bibfnamefont {S.}~\bibnamefont {Bernuzzi}},\ and\ \bibinfo {author}
  {\bibfnamefont {A.~d.~S.}\ \bibnamefont {Schneider}},\ }\href
  {https://doi.org/10.3847/2041-8213/ace5b2} {\bibfield  {journal} {\bibinfo
  {journal} {Astrophys. J. Lett.}\ }\textbf {\bibinfo {volume} {952}},\
  \bibinfo {pages} {L36} (\bibinfo {year} {2023})},\ \bibinfo {note} {\_eprint:
  2302.11359}\BibitemShut {NoStop}%
\bibitem [{\citenamefont {Kedia}\ \emph {et~al.}(2022)\citenamefont {Kedia},
  \citenamefont {Kim}, \citenamefont {Suh},\ and\ \citenamefont
  {Mathews}}]{kedia_binary_2022}%
  \BibitemOpen
  \bibfield  {author} {\bibinfo {author} {\bibfnamefont {A.}~\bibnamefont
  {Kedia}}, \bibinfo {author} {\bibfnamefont {H.~I.}\ \bibnamefont {Kim}},
  \bibinfo {author} {\bibfnamefont {I.-S.}\ \bibnamefont {Suh}},\ and\ \bibinfo
  {author} {\bibfnamefont {G.~J.}\ \bibnamefont {Mathews}},\ }\href
  {https://doi.org/10.1103/PhysRevD.106.103027} {\bibfield  {journal} {\bibinfo
   {journal} {Phys. Rev. D}\ }\textbf {\bibinfo {volume} {106}},\ \bibinfo
  {pages} {103027} (\bibinfo {year} {2022})},\ \bibinfo {note} {\_eprint:
  2203.05461}\BibitemShut {NoStop}%
\bibitem [{\citenamefont {East}\ and\ \citenamefont
  {Pretorius}(2022)}]{east_binary_2022}%
  \BibitemOpen
  \bibfield  {author} {\bibinfo {author} {\bibfnamefont {W.~E.}\ \bibnamefont
  {East}}\ and\ \bibinfo {author} {\bibfnamefont {F.}~\bibnamefont
  {Pretorius}},\ }\href {https://doi.org/10.1103/PhysRevD.106.104055}
  {\bibfield  {journal} {\bibinfo  {journal} {{\textbackslash}prd}\ }\textbf
  {\bibinfo {volume} {106}},\ \bibinfo {pages} {104055} (\bibinfo {year}
  {2022})},\ \bibinfo {note} {\_eprint: 2208.09488}\BibitemShut {NoStop}%
\bibitem [{\citenamefont {Lam}\ \emph {et~al.}(2024)\citenamefont {Lam},
  \citenamefont {Kuan}, \citenamefont {Shibata}, \citenamefont {Van~Aelst},\
  and\ \citenamefont {Kiuchi}}]{lam_binary_2024}%
  \BibitemOpen
  \bibfield  {author} {\bibinfo {author} {\bibfnamefont {A.~T.-L.}\
  \bibnamefont {Lam}}, \bibinfo {author} {\bibfnamefont {H.-J.}\ \bibnamefont
  {Kuan}}, \bibinfo {author} {\bibfnamefont {M.}~\bibnamefont {Shibata}},
  \bibinfo {author} {\bibfnamefont {K.}~\bibnamefont {Van~Aelst}},\ and\
  \bibinfo {author} {\bibfnamefont {K.}~\bibnamefont {Kiuchi}},\ }\href
  {https://doi.org/10.1103/PhysRevD.110.104018} {\bibfield  {journal} {\bibinfo
   {journal} {{\textbackslash}prd}\ }\textbf {\bibinfo {volume} {110}},\
  \bibinfo {pages} {104018} (\bibinfo {year} {2024})},\ \bibinfo {note}
  {\_eprint: 2406.05211}\BibitemShut {NoStop}%
\bibitem [{\citenamefont {Kuan}\ \emph
  {et~al.}(2023{\natexlab{a}})\citenamefont {Kuan}, \citenamefont {Van~Aelst},
  \citenamefont {Lam},\ and\ \citenamefont {Shibata}}]{kuan_binary_2023}%
  \BibitemOpen
  \bibfield  {author} {\bibinfo {author} {\bibfnamefont {H.-J.}\ \bibnamefont
  {Kuan}}, \bibinfo {author} {\bibfnamefont {K.}~\bibnamefont {Van~Aelst}},
  \bibinfo {author} {\bibfnamefont {A.~T.-L.}\ \bibnamefont {Lam}},\ and\
  \bibinfo {author} {\bibfnamefont {M.}~\bibnamefont {Shibata}},\ }\href
  {https://doi.org/10.1103/PhysRevD.108.064057} {\bibfield  {journal} {\bibinfo
   {journal} {Phys. Rev. D}\ }\textbf {\bibinfo {volume} {108}},\ \bibinfo
  {pages} {064057} (\bibinfo {year} {2023}{\natexlab{a}})},\ \bibinfo {note}
  {publisher: American Physical Society}\BibitemShut {NoStop}%
\bibitem [{\citenamefont {Shibata}\ \emph {et~al.}(2014)\citenamefont
  {Shibata}, \citenamefont {Taniguchi}, \citenamefont {Okawa},\ and\
  \citenamefont {Buonanno}}]{shibata_coalescence_2014}%
  \BibitemOpen
  \bibfield  {author} {\bibinfo {author} {\bibfnamefont {M.}~\bibnamefont
  {Shibata}}, \bibinfo {author} {\bibfnamefont {K.}~\bibnamefont {Taniguchi}},
  \bibinfo {author} {\bibfnamefont {H.}~\bibnamefont {Okawa}},\ and\ \bibinfo
  {author} {\bibfnamefont {A.}~\bibnamefont {Buonanno}},\ }\href
  {https://doi.org/10.1103/PhysRevD.89.084005} {\bibfield  {journal} {\bibinfo
  {journal} {Phys. Rev. D}\ }\textbf {\bibinfo {volume} {89}},\ \bibinfo
  {pages} {084005} (\bibinfo {year} {2014})},\ \bibinfo {note} {publisher:
  American Physical Society}\BibitemShut {NoStop}%
\bibitem [{\citenamefont {Staykov}\ \emph {et~al.}(2023)\citenamefont
  {Staykov}, \citenamefont {Doneva}, \citenamefont {Heisenberg}, \citenamefont
  {Stergioulas},\ and\ \citenamefont
  {Yazadjiev}}]{staykov_differentially_2023}%
  \BibitemOpen
  \bibfield  {author} {\bibinfo {author} {\bibfnamefont {K.~V.}\ \bibnamefont
  {Staykov}}, \bibinfo {author} {\bibfnamefont {D.~D.}\ \bibnamefont {Doneva}},
  \bibinfo {author} {\bibfnamefont {L.}~\bibnamefont {Heisenberg}}, \bibinfo
  {author} {\bibfnamefont {N.}~\bibnamefont {Stergioulas}},\ and\ \bibinfo
  {author} {\bibfnamefont {S.~S.}\ \bibnamefont {Yazadjiev}},\ }\href
  {https://doi.org/10.1103/PhysRevD.108.024058} {\bibfield  {journal} {\bibinfo
   {journal} {{\textbackslash}prd}\ }\textbf {\bibinfo {volume} {108}},\
  \bibinfo {pages} {024058} (\bibinfo {year} {2023})},\ \bibinfo {note}
  {\_eprint: 2303.07769}\BibitemShut {NoStop}%
\bibitem [{\citenamefont {Kuan}\ \emph
  {et~al.}(2023{\natexlab{b}})\citenamefont {Kuan}, \citenamefont {Lam},
  \citenamefont {Doneva}, \citenamefont {Yazadjiev}, \citenamefont {Shibata},\
  and\ \citenamefont {Kiuchi}}]{kuan_dynamical_2023}%
  \BibitemOpen
  \bibfield  {author} {\bibinfo {author} {\bibfnamefont {H.-J.}\ \bibnamefont
  {Kuan}}, \bibinfo {author} {\bibfnamefont {A.~T.-L.}\ \bibnamefont {Lam}},
  \bibinfo {author} {\bibfnamefont {D.~D.}\ \bibnamefont {Doneva}}, \bibinfo
  {author} {\bibfnamefont {S.~S.}\ \bibnamefont {Yazadjiev}}, \bibinfo {author}
  {\bibfnamefont {M.}~\bibnamefont {Shibata}},\ and\ \bibinfo {author}
  {\bibfnamefont {K.}~\bibnamefont {Kiuchi}},\ }\href
  {https://doi.org/10.1103/PhysRevD.108.063033} {\bibfield  {journal} {\bibinfo
   {journal} {Phys. Rev. D}\ }\textbf {\bibinfo {volume} {108}},\ \bibinfo
  {pages} {063033} (\bibinfo {year} {2023}{\natexlab{b}})},\ \bibinfo {note}
  {publisher: American Physical Society}\BibitemShut {NoStop}%
\bibitem [{\citenamefont {Palenzuela}\ \emph {et~al.}(2014)\citenamefont
  {Palenzuela}, \citenamefont {Barausse}, \citenamefont {Ponce},\ and\
  \citenamefont {Lehner}}]{palenzuela_dynamical_2014}%
  \BibitemOpen
  \bibfield  {author} {\bibinfo {author} {\bibfnamefont {C.}~\bibnamefont
  {Palenzuela}}, \bibinfo {author} {\bibfnamefont {E.}~\bibnamefont
  {Barausse}}, \bibinfo {author} {\bibfnamefont {M.}~\bibnamefont {Ponce}},\
  and\ \bibinfo {author} {\bibfnamefont {L.}~\bibnamefont {Lehner}},\ }\href
  {https://doi.org/10.1103/PhysRevD.89.044024} {\bibfield  {journal} {\bibinfo
  {journal} {Phys. Rev. D}\ }\textbf {\bibinfo {volume} {89}},\ \bibinfo
  {pages} {044024} (\bibinfo {year} {2014})},\ \bibinfo {note} {publisher:
  American Physical Society}\BibitemShut {NoStop}%
\bibitem [{\citenamefont {Sagunski}\ \emph
  {et~al.}(2018{\natexlab{b}})\citenamefont {Sagunski}, \citenamefont {Zhang},
  \citenamefont {Johnson}, \citenamefont {Lehner}, \citenamefont
  {Sakellariadou}, \citenamefont {Liebling}, \citenamefont {Palenzuela},\ and\
  \citenamefont {Neilsen}}]{sagunski_neutron_2018}%
  \BibitemOpen
  \bibfield  {author} {\bibinfo {author} {\bibfnamefont {L.}~\bibnamefont
  {Sagunski}}, \bibinfo {author} {\bibfnamefont {J.}~\bibnamefont {Zhang}},
  \bibinfo {author} {\bibfnamefont {M.~C.}\ \bibnamefont {Johnson}}, \bibinfo
  {author} {\bibfnamefont {L.}~\bibnamefont {Lehner}}, \bibinfo {author}
  {\bibfnamefont {M.}~\bibnamefont {Sakellariadou}}, \bibinfo {author}
  {\bibfnamefont {S.~L.}\ \bibnamefont {Liebling}}, \bibinfo {author}
  {\bibfnamefont {C.}~\bibnamefont {Palenzuela}},\ and\ \bibinfo {author}
  {\bibfnamefont {D.}~\bibnamefont {Neilsen}},\ }\href
  {https://doi.org/10.1103/PhysRevD.97.064016} {\bibfield  {journal} {\bibinfo
  {journal} {{\textbackslash}prd}\ }\textbf {\bibinfo {volume} {97}},\ \bibinfo
  {pages} {064016} (\bibinfo {year} {2018}{\natexlab{b}})},\ \bibinfo {note}
  {\_eprint: 1709.06634}\BibitemShut {NoStop}%
\bibitem [{\citenamefont {Barausse}\ \emph {et~al.}(2013)\citenamefont
  {Barausse}, \citenamefont {Palenzuela}, \citenamefont {Ponce},\ and\
  \citenamefont {Lehner}}]{barausse_neutron-star_2013}%
  \BibitemOpen
  \bibfield  {author} {\bibinfo {author} {\bibfnamefont {E.}~\bibnamefont
  {Barausse}}, \bibinfo {author} {\bibfnamefont {C.}~\bibnamefont
  {Palenzuela}}, \bibinfo {author} {\bibfnamefont {M.}~\bibnamefont {Ponce}},\
  and\ \bibinfo {author} {\bibfnamefont {L.}~\bibnamefont {Lehner}},\ }\href
  {https://doi.org/10.1103/PhysRevD.87.081506} {\bibfield  {journal} {\bibinfo
  {journal} {Phys. Rev. D}\ }\textbf {\bibinfo {volume} {87}},\ \bibinfo
  {pages} {081506} (\bibinfo {year} {2013})},\ \bibinfo {note} {publisher:
  American Physical Society}\BibitemShut {NoStop}%
\end{thebibliography}%

\pagebreak

\end{document}